\documentclass[aps,prd,reprint,groupedaddress,amsmath,amssymb,longbibliography,nofootinbib]{revtex4-2} 
\usepackage[]{hyperref} 
\usepackage[]{color} 
\usepackage[]{graphicx} 
\usepackage{afterpage} 
\usepackage{enumerate} 
 
\allowdisplaybreaks[2] 
\newcommand{\be}{\begin{equation}}\newcommand{\ee}{\end{equation}} 
\newcommand{\bes}{\begin{equation*}}\newcommand{\ees}{\end{equation*}} 
\newcommand{\bea}{\begin{eqnarray}}\newcommand{\eea}{\end{eqnarray}} 
\newcommand{\del}{\partial} 
\newcommand{\bvec}[1]{\mbox{\boldmath $#1$}}

\newcommand{\nn}{\nonumber} 
 
\newcommand{\eq}[1]{${\displaystyle{#1}}$}

\newcommand{\figonecolumn}{8.8cm} 
\newcommand{\figonecolumnsmall}{7cm} 
\newcommand{\figtwocolumn}{18cm} 
\newcommand{\figtwocolumnsmall}{16.8cm} 
 
\begin{document} 
\title{Finite-length flux tube in the dual Ginzburg-Landau theory on the dual lattice}{} 
 
\author{Yoshiaki Koma} 
\email{koma@numazu-ct.ac.jp} 
\affiliation{National Institute of Technology, Numazu College, Ooka 3600, Numazu 410-8501, Japan} 
 
\author{Miho Koma} 
\email{takayama@rcnp.osaka-u.ac.jp} 
\affiliation{Research Center for Nuclear Physics (RCNP), Osaka University, 
Mihogaoka 10-1, Ibaraki, Osaka 567-0047, Japan} 
 
\date{October 24, 2025 (v2)} 
 
\begin{abstract} 
The dual Ginzburg-Landau (DGL) theory is one of the nonperturbative effective field theories of quantum chromodynamics (QCD). The DGL theory describes the QCD vacuum as a dual superconductor and possesses electric flux-tube solutions via the dual Meissner effect, which applies to the quark confinement mechanism. We demonstrate a powerful numerical method for solving the field equations in the DGL theory with U(1) dual gauge symmetry. An essential aspect of our method is to formulate the DGL theory on the dual lattice, which enables us to investigate any system composed of finite-length flux tubes in a systematic manner. Taking full advantage of the dual lattice formulation, we investigate the finite-length flux-tube solution corresponding to the quark-antiquark system in detail, which exposes the significant terminal effects absent in the infinitely long flux-tube solution. We also study the flux-tube interaction in the two-flux-tube and multiflux-tube systems, providing new insights into the nonperturbative properties of QCD. 
\end{abstract} 
 
\maketitle 
 
\section{Introduction} 
\label{sec:intro} 
 
\par 
If the quantum chromodynamics (QCD) vacuum behaves as a dual superconductor~\cite{Nambu:1974zg, Mandelstam:1974pi, tHooft:1975krp}, the color-electric field emitted from a quark will be squeezed into a stringlike flux tube~\cite{Nielsen:1973cs} due to the dual Meissner effect until being absorbed by an antiquark or by another two quarks with different color charges. Then, the potential energy for the bound state of quarks increases linearly as the flux-tube length becomes longer, so that quarks can never be isolated with a finite amount of energy, which provides us with a possible mechanism of quark confinement. It is quite well known that the interquark potential computed by lattice QCD\footnote{We assume that ``lattice QCD" also includes compact U(1) and SU(2) lattice gauge theories, as these theories also exhibit the dual superconducting nature.} simulations contains a linearly rising part for the distance between quarks. 
 
\par 
The properties of an ordinary superconductor are known to be described by the Ginzburg-Landau theory, where its response to the magnetic field depends sensitively on the type of superconducting phases characterized by the Ginzburg-Landau parameter $\kappa$, defined by the ratio of the penetration depth to the coherence length. Likewise, the QCD vacuum can effectively be described by the dual Ginzburg-Landau (DGL) theory~\cite{Suzuki:1988yq, Maedan:1988yi, Suzuki:1988qa, Suganuma:1993ps, Sasaki:1994sa, Ichie:1996jr}. It is then natural to ask the type of dual superconducting phase of the QCD vacuum, since it must directly be responsible for the structure of hadrons. So far, many efforts have been made to identify the type of phase by investigating the field profile of the quark-antiquark system from lattice QCD simulations, but it seems that the results are not easily settled~\cite{Singh:1993jj, Cea:1995zt, Bali:1997cp, Gubarev:1999yp, Koma:2003gq, Koma:2003hv, Haymaker:2005py, Chernodub:2005gz, DAlessandro:2006hfn, Sekido:2007mp, Suzuki:2007jp, Suzuki:2009xy, Cea:2012qw, Shibata:2012ae, Kato:2014nka, Cea:2014uja, Nishino:2019bzb, Battelli:2019lkz}. 
 
\par 
Part of the reason may be due to the difficulty of reducing numerical errors in the lattice QCD simulation, while another part may stem from the unsatisfactory analyses of the lattice QCD results. For instance, the field profile of the quark-antiquark system obtained by the lattice QCD simulations has often been compared with exponential functions with a negative exponent for identifying the penetration depth or the coherence length. In most cases, however, despite the distance between the quark and the antiquark being always finite, the effect of finiteness is not taken into account seriously. Since the numerical errors of lattice simulations are getting smaller, thanks not only to the increase of computer resources but also to the efficient numerical algorithms~\cite{Luscher:2001up, Luscher:2002qv}, it is quite important to use appropriate model functions for the analysis. 
 
\par 
In this paper, we demonstrate a powerful numerical method for obtaining the finite-length flux-tube solution in the DGL theory corresponding to the quark-antiquark system in QCD, and clarify the properties of the solution in detail. In general, the flux-tube solution can be obtained by solving simultaneous nonlinear partial differential equations for the field variables with appropriate boundary conditions. An essential aspect of our method is to formulate the DGL theory on the dual lattice~\cite{Koma:2000hw}, which enables us to investigate any system composed of finite-length flux tubes in a systematic manner in contrast to normal discretization methods~\cite{Baker:1991bc}. Our method has been successfully applied to the analysis of the lattice QCD results~\cite{Koma:2003gq, Koma:2003hv, Koma:2003gi}. We describe all the details of the dual lattice formulation of the DGL theory, including several updates from our previous work~\cite{Koma:2000hw}. As further applications of our method, we also investigate the interaction properties of flux tubes in the two-flux-tube and multiflux-tube systems. 
 
\par 
This paper is organized as follows. In Sec.~\ref{sec:dgl}, we introduce the DGL theory with the U(1) dual gauge symmetry and summarize the basics of the flux-tube solution. In Sec.~\ref{sec:duallattice}, we describe the dual lattice formulation of the DGL theory, and demonstrate how to obtain the finite-length flux-tube solution. The methods of the Hodge decomposition and of identifying the location of the flux-tube core are also given, which are quite useful to reveal the structure of the flux tube. We then provide an optimal set of parameters for controlling systematic effects due to the lattice formulation. In Sec.~\ref{sec:results}, we present the numerical results on the finite-length flux-tube solution, and on the interaction properties of flux tubes. Section~\ref{sec:summary} is devoted to the summary of our findings and outlook. In Appendices, we present the basic properties of the lattice Green functions for the massless field used in our study, and how to obtain the flux-tube solution in the large $\kappa$ limit. 
 
\par 
Although we intend to apply our method to the analysis of lattice QCD results, or more widely, that of lattice gauge theories results, the presented method can obviously be used to investigate the ordinary superconductor by regarding the dual variables as the original ones, with introducing an appropriate physical scale~\cite{Abrikosov:1956sx}. It would also be useful to understand any physical phenomena that are expected to be described by the Abelian Higgs model~\cite{Higgs:1964ia, Nielsen:1973cs}, since its field-theoretical structure is the same as the DGL theory. 
 
\section{The flux-tube solution in the DGL theory} 
\label{sec:dgl} 

In this section, we introduce the DGL theory with U(1) dual gauge symmetry, and summarize the essential feature of the most popular flux-tube solution with infinite length. We clarify the role of the Dirac string for the flux-tube solution, which is necessary to define the electric charge in the dual world, both from analytical and numerical points of view. 
 
\subsection{The model} 

We consider a $D$-dimensional Euclidean space, whose coordinate is specified by $x=(x_{1},x_{2},...,x_{D})$. The DGL theory is then defined by the Lagrangian density 
\be 
{\cal L} = 
\frac{1}{4} {}^{*\!}F_{\mu\nu}^{2} 
\! +\! |(\del_{\mu} + i g B_{\mu})\chi |^{2} 
+ \lambda ( |\chi |^{2}\! -\! v^{2} )^{2} 
\; , 
\label{eqn:dgl-action} 
\ee 
where $B_{\mu}$ denotes the axial vector dual gauge field and $\chi$ the complex scalar monopole field, and 
\be 
{}^{*\!}F_{\mu\nu}= 
\del_{\mu}B_{\nu} -\del_{\nu}B_{\mu} - 
e\Sigma_{\mu\nu} 
\; 
\ee 
is the antisymmetric dual field strength tensor (the star represents the Hodge dual). The Greek indices $\mu$ and $\nu$ take values from 1 to $D$, and repeated indices are summed over. The monopole field carries the magnetic charge $g$, while the quark and antiquark fields, which are introduced as the endpoints of the Dirac string $\Sigma_{\mu\nu}$, carry the electric charge $e$ and $-e$. Between the magnetic and the electric charges, there is an important relation called the Dirac quantization condition, 
\be 
eg=2\pi \; . 
\label{eqn:charge-relation} 
\ee 
The Lagrangian density is apparently invariant under the U(1) dual gauge transformation, 
\be 
\chi \to \chi e^{i\xi }, \quad 
\chi^* \to \chi^* e^{-i\xi} ,\quad 
B_{\mu} \to B_{\mu} - \partial_{\mu}\xi 
\;, 
\label{eqn:u1symmetry} 
\ee 
for $\xi \in \Re$, where the star denotes the complex conjugate. This symmetry will be broken spontaneously once monopole condensation occurs $|\chi | \to v$ (vacuum expectation value), and then both the dual gauge field and the monopole field acquire masses, 
\be 
m_{B}=\sqrt{2}gv \;, \quad m_{\chi} =2\sqrt{\lambda} v \;, 
\ee 
respectively. The distances defined by the inverse of these masses correspond to the penetration depth and the coherence length, and the ratio of these masses, 
\be 
\kappa 
=\frac{m_{\chi}}{m_{B}} =\frac{\sqrt{2 \lambda}}{g} \;, 
\ee 
characterizes the type of the dual superconducting phase; type-I for $\kappa <1$ and type-II for $\kappa >1$. The border of the two phases, $\kappa =1$, is often called the Bogomol'nyi limit~\cite{Bogomolny:1975de}. We may call $\kappa$ the Ginzburg-Landau (GL) parameter as in the ordinary superconductor.\footnote{The border of the type-I and type-II phases is often defined at $\kappa=1/\sqrt{2}$. The difference from our definition of the border is just a matter of convention.} 
 
\par 
For the numerical analysis, it is useful to isolate the physical scale $v$ from the theory, which is recovered when needed. In practice, we may introduce dimensionless variables with the carets by 
\bea 
&& gB_{\mu} \equiv v\hat{B}_{\mu}\;,\quad 
\chi \equiv v\hat{\chi} =v (\hat{\chi}^{1} +i\hat{\chi}^{2})\;,\nn\\* 
&& 
\del_{\mu} \equiv v\hat{\del}_{\mu} 
\;, 
\quad 
x_{\mu} \equiv v^{-1}\hat{x}_{\mu} 
\;,\quad 
\Sigma_{\mu\nu} \equiv v^{2}\hat{\Sigma}_{\mu\nu}\;, \nn\\* 
&& m_{B} \equiv v \hat{m}_{B} \; ,\quad m_{\chi}\equiv v \hat{m}_{\chi} \;. 
\label{eqn:dimensionless} 
\eea 
Then, the Lagrangian density in Eq.~\eqref{eqn:dgl-action} is written as 
\be 
{\cal L}= 
\beta_{g} v^{4} 
\biggl [ 
\frac{1}{4} {}^{*\!} \hat{F}_{\mu\nu}^{2} 
+ \frac{\hat{m}_{B}^{2}}{2} ( \hat{D}_{\mu} \hat{\chi}^{\alpha} )^{2} + \frac{\hat{m}_{B}^{2} \hat{m}_{\chi}^{2}}{8}( \hat{\chi}^{\alpha\,2} -1)^{2} 
\biggr] 
, 
\label{eqn:dgl-dimensionless} 
\ee 
where 
\bea 
&& {}^{*\!}\hat{F}_{\mu\nu} = 
\hat{\del}_{\mu} \hat{B}_{\nu} - \hat{\del}_{\nu} \hat{B}_{\mu} 
- 2 \pi \hat{\Sigma}_{\mu\nu}\;, 
\label{eqn:dfs} \\* 
&& 
\hat{D}_{\mu} \hat{\chi}^{\alpha} =\hat{\del}_{\mu}\hat{\chi}^{\alpha}- \epsilon_{\alpha \beta} \hat{B}_{\mu} \hat{\chi}^{\beta} \;, 
\label{eqn:cdv} 
\eea 
and $\beta_{g} = 1/g^{2}$. The Greek indices $\alpha$ and $\beta$ take values from 1 to 2, and repeated indices are summed over, where $\epsilon_{\alpha\beta}$ in Eq.~\eqref{eqn:cdv} is the 2nd-rank antisymmetric tensor ($\epsilon_{12}=- \epsilon_{21}=1$, $\epsilon_{11}=\epsilon_{22}=0$). For simplicity, we omit all of the caret hereafter \textit{except for the masses}. As the coupling $\beta_{g}$ is just an overall factor, we always assume $\beta_{g} = 1$ when the numerical results are presented in the following. This value needs to be adjusted when the results are compared to the lattice QCD ones. 
 
\subsection{The analytical approach to the flux-tube solution} 
 
\par 
The DGL theory has a flux-tube solution when a chain of ${\Sigma}_{\mu\nu}({x}) \ne 0$ is put. For ${\Sigma}_{\mu\nu}({x}) \ne 0$, the dual gauge field can be decomposed into three parts by applying the Hodge decomposition,\footnote{Using codifferential $\delta$ and exterior derivative $d$, the Laplacian is defined by $\Delta =d \delta +\delta d$. Then, a 1-form field $B$ can be decomposed as $B=\Delta^{-1}\Delta B=\Delta^{-1}(d\delta +\delta d)B = \Delta^{-1}dB + \Delta^{-1}\delta dB = \Delta^{-1}d\delta B + \Delta^{-1}\delta (*F + 2\pi \Sigma) = \Delta^{-1} d\delta B + \Delta^{-1}\delta * F + 2\pi \Delta^{-1}\delta \Sigma = B^{\rm red} +B^{\rm reg} +B^{\rm sing}$, where $*$ denotes the Hodge star operator, and the definition of the 2-form dual field strength $*F =dB-2\pi \Sigma$ is used~\cite{Koma:2001pz}.} 
\be 
B_{\mu} = B_{\mu}^{\rm reg}+ B_{\mu}^{\rm sing} +B_{\mu}^{\rm red}. 
\label{eqn:hodge-decomposition} 
\ee 
The first term $B_{\mu}^{\rm reg}$ is, as demonstrated later, related to the regular monopole supercurrent and leads to a solenoidal structure of the electric field. The second term $B_{\mu}^{\rm sing}$ is related to the singular boundary condition of the system, which satisfies the relation~\cite{Baker:1991bc, Koma:2000wn}, 
\be 
{\del}_{\mu} B_{\nu}^{\rm sing} -{\del}_{\nu} B_{\mu}^{\rm sing} - 2\pi {\Sigma}_{\mu\nu} =2\pi C_{\mu\nu} \;, 
\label{eqn:bsing} 
\ee 
where ${C}_{\mu\nu}$ leads to the Coulombic electric field, and is related to the electric charge density by 
\be 
j_{0} (x) = - \frac{1}{2} \epsilon_{0\eta\lambda\sigma} \del_{\eta}\Sigma_{\lambda\sigma} = + \frac{1}{2} \epsilon_{0\eta\lambda\sigma} \del_{\eta}C_{\lambda\sigma}\;. 
\label{eqn:echarge} 
\ee 
The formal solution of Eq.~\eqref{eqn:echarge} with respect to ${C}_{\mu\nu}$ in $D$ dimensions is given by 
\be 
{C}_{\mu\nu}(x) = -\epsilon_{\mu\nu\sigma} \del_{\sigma} 
\int d^{D} x' G(x-x') j_{0} (x') \; , 
\ee 
where $G (x)$ denotes the massless Green function, which satisfies $\Delta G (x) = - \delta^{(D)} (x)$. Using the fact that $\del_{\nu} C_{\mu\nu}(x) =0$ and $\del_{\mu}B_{\mu}^{\rm sing}=0$, the formal solution of Eq.~\eqref{eqn:bsing} with respect to $B_{\mu}^{\rm sing}$ is expressed as 
\be 
B_{\mu}^{\rm sing} (x) = 2\pi \!\! \int \!\! d^{D}x' G (x-x') 
\del_{\nu}' \Sigma_{\mu\nu}(x') \; . 
\label{eqn:bsing-general} 
\ee 
The last term $B_{\mu}^{\rm red}$ in Eq.~\eqref{eqn:hodge-decomposition} is redundant, and is irrelevant to the electric field, as ${\del}_{\mu} B_{\nu}^{\rm red} -{\del}_{\nu} B_{\mu}^{\rm red} =0$. 
 
\par 
The existence of the flux-tube solution is demonstrated analytically by looking at the static $D=2$ dimensional system. We put a single Dirac string at the origin $x=0$, such as 
\be 
{\Sigma}_{\mu\nu}(x) = - \epsilon_{\mu\nu} N_{q} \delta^{(2)} (x) \;, 
\ee 
where $N_{q}\in \mathbb{Z}$ characterizes the scalar multiple of the electric charge $e$. Note that this system can also be regarded as the cross section of an infinitely long flux-tube system along the $x_{3}$-axis in $D=3$ dimensions. Therefore, there is no endpoint of the Dirac string, and $C_{\mu\nu}$ is absent. By inserting the $D=2$ dimensional Green function, 
\be 
G (x-x') = - \frac{1}{2\pi} \ln | x-x' | 
\ee 
into Eq.~\eqref{eqn:bsing-general}, we obtain an explicit form 
\be 
B_{\mu}^{\rm sing} (x) = 
\frac{N_{q} \epsilon_{\mu\nu} x_{\nu}}{|x|^{2}} \;. 
\label{eqn:bsing-explicit} 
\ee 
As the system becomes cylindrically symmetric around the origin, we may introduce polar coordinates, $x_{1}=\rho \cos\theta$ and $x_{2}=\rho \sin\theta$ with $\rho = \sqrt{x_{1}^{2}+x_{2}^{2}}$, and then, Eq.~\eqref{eqn:bsing-explicit} can be written as 
\be 
\vec{B}^{\rm sing} 
= - \frac{N_{q}}{\rho} (- \sin \theta \vec{e}_{1} +\cos\theta \vec{e}_{2} ) = - \frac{N_{q}}{\rho} \vec{e}_{\theta}\; 
\label{eqn:bsing-cylindrical} 
\ee 
with the unit vector $\vec{e}_{\theta} = - \sin \theta \vec{e}_{1} +\cos\theta \vec{e}_{2} $. Other field variables are also written as 
\be 
\vec{B}^{\rm reg} =B^{\rm reg} (\rho) \vec{e}_{\theta}=\frac{B_{\rho}(\rho)}{\rho}\vec{e}_{\theta}\;, 
\quad \chi^{1} =\chi^{2} = \frac{1}{\sqrt{2}}\phi(\rho) 
\; . 
\ee 
 
\par 
The functional forms of $B_{\rho}(\rho)$ and $\phi(\rho)$ are determined to minimize the spatial integration of the Lagrangian density in Eq.~\eqref{eqn:dgl-dimensionless} written in the same polar coordinates, 
\bea 
\sigma &=& 
\beta_{g} v^{2}\!\! \int_{0}^{\infty} \!\! (2 \pi \rho ) d \rho \biggl [ 
\frac{1}{2} \biggl (\frac{1}{\rho}\frac{d B_{\rho}}{d\rho} \biggr )^{2} 
\!\!\! + \!\frac{\hat{m}_{B}^{2}}{2}\Bigl \{ 
\biggl(\frac{d \phi}{d\rho}\biggr)^{2} 
\nn\\ 
&&\!+ \biggl(\frac{B_{\rho} -N_{q}}{\rho}\biggr)^{2}\!\! \phi^{2} 
\Bigr \} 
\!+ \!\frac{\hat{m}_{B}^{2}\hat{m}_{\chi}^{2}}{8}( \phi^{2} -1)^{2} 
\biggr] \; . 
\label{eqn:dgl-cylind} 
\eea 
From the extremum condition, $\del \sigma/\del B_{\rho} =0$ and $\del \sigma/\del \phi =0$, we obtain the field equations\footnote{Assuming that the phase of the monopole field $\chi$ is multivalued as $[\del_{\mu},\del_{\nu}]\eta =2\pi N_{q} \epsilon_{\mu\nu} \delta^{(2)}(x)$ for $\chi = \phi e^{i\eta}$~\cite{Nielsen:1973cs}, the expression of Eq.~\eqref{eqn:dgl-cylind} and the following field equations in Eqs.~\eqref{eqn:feq1} and \eqref{eqn:feq2} can be derived. However, this way of introducing $N_{q}$ obscures the relationship between $N_{q}$ and the electric charge, especially when the finite-length flux-tube solution is considered.} 
\be 
\frac{d^{2}B_{\rho}}{d\rho^{2}} -\frac{1}{\rho}\frac{d B_{\rho}}{d\rho} - \hat{m}_{B}^{2}(B_{\rho}-N_{q})\phi^{2} =0\;, 
\label{eqn:feq1} 
\ee 
\be 
\frac{d^{2}\phi}{d\rho^{2}} +\frac{1}{\rho}\frac{d\phi}{d\rho } 
- \biggl ( \frac{B_{\rho} -N_{q}}{\rho} \biggr )^{2}\phi - \frac{\hat{m}_{\chi}^{2}}{2}\phi ( \phi^{2}-1)=0\;. 
\label{eqn:feq2} 
\ee 
With the solution of $B_{\rho}$ and $\phi$ in Eqs.~\eqref{eqn:feq1} and \eqref{eqn:feq2}, which usually requires numerical methods, Eq.~\eqref{eqn:dgl-cylind} provides the energy of the flux tube per unit length, i.e., the string tension. 
 
\par 
Although the exact analytical solution of these simultaneous field equations is not known, it is possible to show the existence of the solution as follows. By expressing the dual gauge field as $B_{\rho}=N_{q}- \rho' K(\rho')$ with the rescaled radial coordinate $\rho' = \hat{m}_{B}\rho$, Eq.~\eqref{eqn:feq1} can be written as 
\be 
\frac{d^{2}K}{d{\rho'}^{2}} + \frac{1}{{\rho'}}\frac{dK}{d{\rho'}} - ( 1 +\frac{1}{{\rho'}^{2}})K \phi^{2} =0 \; . 
\label{eqn:bessel-k} 
\ee 
In the region of $\rho'$ where $\phi (\rho') \sim 1$ is satisfied, the modified Bessel function of the second kind, $K=K_{1}(\rho')\sim \sqrt{\pi/(2\rho')}e^{-\rho'}$ can be the solution of Eq.~\eqref{eqn:bessel-k}. Thus, the dual gauge field can be written as $B_{\rho}=N_{q}- \rho' K_{1}(\rho')$. By using the relation $K_{1}+\rho' \frac{d K_{1}}{d\rho'} =-\rho' K_{0}$, the electric field is given by $\vec{E}=\nabla \times \vec{B} = (\frac{1}{\rho}\frac{d B_{\rho}}{d\rho})\vec{e}_{3} = E_{\rho} (\rho) \; \vec{e}_{3}$ with 
\be 
E_{\rho} (\rho) 
\sim \hat{m}_{B}^{2} K_{0}(\hat{m}_{B}\rho) \, . 
\label{eqn:ele-besselk0} 
\ee 
This reads that the electric field is suppressed exponentially with the increase of $\rho$, indicating a confining tubelike structure of the electric field within $\hat{m}_{B} \rho <1$. The appearance of the electric field $\vec{E}$ is linked to that of the monopole supercurrent $\vec{k}$ via the dual Amp\`ere law, $\nabla \times \vec{E} = \vec{k}$, which signals the occurrence of the dual Meissner effect. In the cylindrical case, the monopole supercurrent can be written as $\vec{k} = k_{\rho} (\rho)\; \vec{e}_{\theta}$, where 
\be 
k_{\rho} (\rho) = - \hat{m}_{B}^{2} \frac{B_{\rho}-N_{q}}{\rho} \phi^{2} 
\sim \hat{m}_{B}^{3} K_{1}(\hat{m}_{B}\rho) 
\; . 
\label{eqn:k-besselk1} 
\ee 
This reads that the monopole supercurrent is circulating the electric flux tube. 
 
\par 
For the special case such that $\hat{m}_{B}=\hat{m}_{\chi}$ is satisfied ($\kappa=1$), it is possible to evaluate Eq.~\eqref{eqn:dgl-cylind} analytically without knowing the functional forms of $B_{\rho}$ and $\phi$. Following the idea of Bogomol'nyi~\cite{Bogomolny:1975de}, we may rewrite Eq.~\eqref{eqn:dgl-cylind} as 
\bea 
\sigma &=& 
\beta_{g} v^{2} \int_{0}^{\infty} (2 \pi \rho) d \rho 
\biggl [ 
\frac{1}{2} \biggl ( \frac{1}{\rho}\frac{d B_{\rho}}{d\rho} \pm \frac{\hat{m}_{B}^{2}}{2} ( \phi^{2} -1) \biggr )^{2}\nn\\ 
&& +\frac{\hat{m}_{B}^{2}}{2}\biggl (\frac{d \phi}{d\rho} \pm \biggl ( \frac{B_{\rho} - N_{q}}{\rho} \biggr )\phi \biggr)^{2}\nn\\ && + \frac{\hat{m}_{B}^{2} (\hat{m}_{\chi}^{2} - \hat{m}_{B}^{2})}{8}( \phi^{2} - 1)^{2} 
\biggr] \nn\\ 
&& 
\mp \beta_{g} v^{2} \int_{0}^{\infty} (2 \pi \rho) d \rho \frac{\hat{m}_{B}^{2} }{2\rho} 
\biggl [ 
\frac{d B_{\rho}}{d\rho} ( \phi^{2} - 1) 
\nn\\&& 
+\frac{d \phi^{2}}{d\rho} ( B_{\rho} - N_{q}) 
\biggr] \,, 
\label{eqn:sigma-border} 
\eea 
where the upper sign is for $N_{q}>0$ and the lower sign is for $N_{q}<0$. For $\hat{m}_{\chi}=\hat{m}_{B}$, the field equations are given by the first-order differential equations, 
\be 
\frac{1}{\rho}\frac{dB_{\rho}}{d\rho} \pm \frac{\hat{m}_{B}^{2}}{2} ( \phi^{2} - 1) = 0 \;, 
\label{eqn:firstordereq1} 
\ee 
\be 
\frac{d \phi}{d\rho} \pm \biggl ( \frac{B_{\rho} - N_{q}}{\rho}\!\biggr)\phi = 0 \;. 
\label{eqn:firstordereq2} 
\ee 
By taking into account the boundary conditions of the fields such as $B_{\rho}(0)=0$, $\phi(0)=0$, and $B_{\rho}(\infty) =N_{q}$, $\phi (\infty) =1$, Eq.~\eqref{eqn:sigma-border} is reduced to 
\be 
\sigma = 
\pi \beta_{g} | N_{q} | \hat{m}_{B}^{2} v^{2} 
= \pi \beta_{g} | N_{q} | m_{B}^{2} \;. 
\label{eqn:stringtension-bb} 
\ee 
Note that differentiating Eqs.~\eqref{eqn:firstordereq1} and ~\eqref{eqn:firstordereq2} with respect to $\rho$ recovers the second-order differential equations as in Eqs.~\eqref{eqn:feq1} and~\eqref{eqn:feq2}. 
 
\subsection{The numerical approach to the flux-tube solution} 
\label{sec:dgl-cylind-numerical} 

The field equations in Eqs.~\eqref{eqn:feq1} and~\eqref{eqn:feq2} can be solved numerically. We may write the radial coordinate as $\rho =nh$ ($n=0,1,2,...,n_{\max}$) with a small interval~$h$, and approximate derivatives with finite (central) differences as 
\be 
\frac{df(\rho)}{d\rho} \to \frac{f(n+1) -f(n-1)}{2h} 
\ee 
for an arbitrary function $f(\rho)$. Eq.~\eqref{eqn:dgl-cylind} is then expressed as 
\bea 
&&\sigma = 
\frac{ \beta_{g} v^{2}}{h^{2}}\sum_{n} 2 \pi n \biggl [ 
\frac{1}{2} \biggl ( \frac{B_{\rho}(n + 1) - B_{\rho}(n - 1)}{2 n } \biggr )^{2} 
\nn\\ 
&& + \frac{(\hat{m}_{B}h)^{2}}{2}\biggl \{ \biggl ( \frac{\phi(n + 1) - \phi(n - 1)}{2} \biggr )^{2}\nn\\ && + \biggl ( \frac{B_{\rho}(n) -N_{q}}{n} \biggr )^{2} \phi(n) ^{2} \biggr \} \nn\\ && + \frac{(\hat{m}_{B}h)^{2}(\hat{m}_{\chi}h)^{2}}{8}( \phi(n)^{2} - 1)^{2} \biggr] 
\;, 
\label{eqn:dgl-cylind-numerical} 
\eea 
and the field equations are given by 
\bea 
X (n) &\equiv& B_{\rho}(n + 1) +B_{\rho}(n - 1) - 2B_{\rho}(n)\nn\\ && - \frac{B_{\rho}(n + 1) - B_{\rho}(n - 1)}{2n}\nn\\ && - ( \hat{m}_{B}h)^{2}(B_{\rho}(n) - N_{q})\phi (n)^{2} =0 \;, \\ Y (n) & \equiv & 
\phi(n + 1) + \phi(n -1)- 2\phi(n) \nn\\ 
&& + \frac{\phi(n + 1) - \phi(n - 1)}{2n}\nn\\ && - \biggl ( \frac{B_{\rho}(n) - N_{q}}{n} \biggr )^{2} \phi (n)\nn\\ && - \frac{(\hat{m}_{\chi}h)^{2}}{2}\phi (n) ( \phi(n)^{2}-1) = 0 \;. 
\eea 
The simultaneous equations $X(n)=Y(n)=0$ can be solved, for instance, by applying the Newton-Raphson method; using derivatives of the field equations with respect to the field variable, 
\bea 
\delta_{B} X (n) \equiv 
\frac{\del X(n)}{\del B_{\rho}(n)} 
&=& -2- (\hat{m}_{B}h)^{2}\phi (n)^{2} \;,\\ 
\delta_{\phi} Y (n) 
\equiv \frac{\del Y(n)}{\del \phi (n)} 
&=& -2- \biggl (\frac{B_{\rho}(n) - N_{q}}{n} \biggr )^{2} 
\nn\\&& 
- \frac{(\hat{m}_{\chi}h)^{2}}{2} (3\phi(n)^{2}- 1 ) \; , 
\eea 
the field variables are iteratively updated by 
\bea 
B_{\rho}(n) &\to & B_{\rho}(n)-\frac{X(n)}{\delta_{B} X (n)} 
\;,\nn\\* 
\phi(n) &\to& \phi(n) -\frac{Y(n)}{\delta_{\phi} Y (n)} 
\;. 
\eea 
The iteration process is terminated when both $\max (|X \!(n)| )<\epsilon$ and $\max ( |Y \!(n)| )< \epsilon$ are satisfied simultaneously with a reasonably small value of $\epsilon$. The boundary conditions of the dual gauge field and the monopole field are set to be $B_{\rho}(0)=0$, $\phi (0)= 0$, and $B_{\rho}(n_{\rm max})=N_{q}$, $\phi (n_{\rm max})= 1$ for a large enough $n_{\max}$. 
 
\par 
The dimensionless interval $h$ should be small enough to avoid discretization effects, and at the same time, the size of $n_{\rm max}$ should be large enough to avoid finite size effects. Here, it is useful to pay attention to the fact that the interval $h$ is always accompanied by the mass parameters such as $\hat{m}_{B}h$ and $\hat{m}_{\chi}h$. This means that it is possible to take $h=1$ and to regard the $\hat{m}_{B}$ as the contraction percentage of the interval $h$. Then, the continuum limit can be defined by $\hat{m}_{B} \to 0$, while $\hat{m}_{\chi}$ should be adjusted to keep $\kappa =\hat{m}_{\chi}/\hat{m}_{B}$ intact. 
 
\par 
According to Eq.~\eqref{eqn:dimensionless}, the physical scale of a dimensionless distance can be recovered by multiplying it by 
\be 
a \equiv \frac{1}{v}=\frac{\hat{m}_{B}}{m_{B}} \;. 
\label{eqn:physical-lat-a} 
\ee 
Thus, the physical scale of the radial coordinate $\rho$ is given by $\rho a = nh a =na =n\hat{m}_{B}\;\mathrm{[1 /m_{B}]}$ for $n=0,1,2,...,n_{\rm max}$. Although the physical value of $m_{B}$ is not known \textit{a priori}, we can always compare the distances for various $\hat{m}_{B}$ values with each other in units of $1/m_{B}$, the penetration depth. The small $\hat{m}_{B}$ or large $\hat{m}_{B}$ corresponds to the computation with the fine or coarse discretization. 
 
\begin{figure}[!t] 
\centering\includegraphics[width=\figonecolumn]{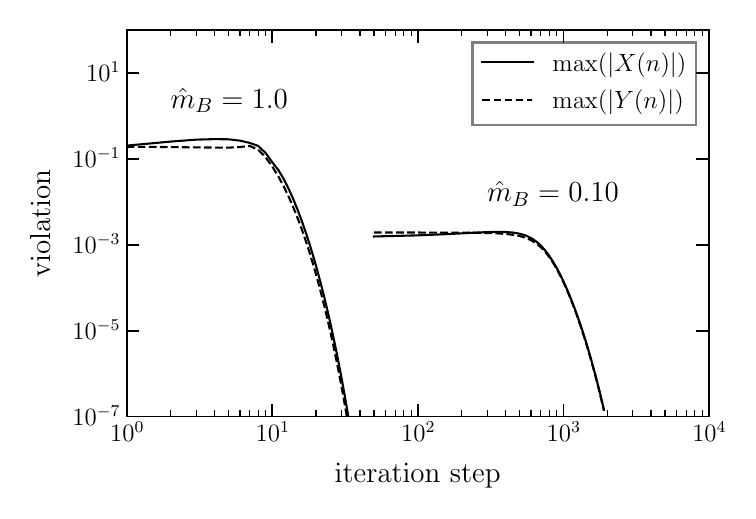} 
\caption{Histories of the maximum violation of the field equations $\max (|X \!(n)| )$ and $\max ( |Y \!(n)| )$ for $\kappa =1.0$ as a function of iteration step.} 
\label{fig:vio_vs_ite} 
\end{figure} 
 
\begin{figure}[!t] 
\centering\includegraphics[width=\figonecolumn]{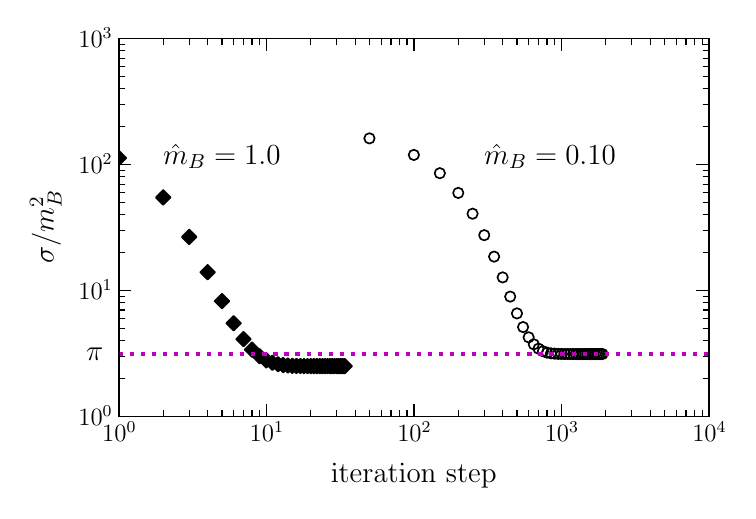} 
\caption{History of the value of $\sigma$ in Eq.~\eqref{eqn:dgl-cylind} for $\kappa =1.0$ as a function of iteration step. The dotted line denotes the analytical continuum value $\sigma/m_{B}^{2}=\pi$ in Eq.~\eqref{eqn:stringtension-bb}. } 
\label{fig:st_vs_ite} 
\end{figure} 
 
\par 
To examine the discretization effect, let us demonstrate two cases of a fundamental flux-tube system $N_{q}=1$ for the border $\kappa =1$ with $\hat{m}_{B}=1.0$ (coarse) and $0.10$ (fine). We set the maximum radial coordinate as $\rho_{\max}a =40\;\mathrm{[1 / m_{B}]}$, which means $n_{\max}=40$ for $\hat{m}_{B}=1.0$ and $n_{\max}=400$ for $\hat{m}_{B}=0.10$, respectively. In Figs.~\ref{fig:vio_vs_ite} and~\ref{fig:st_vs_ite}, we show the histories of the maximum violation $\max (|X \!(n)| )$ and $\max ( |Y \!(n)| )$, and of the value of $\sigma$ in Eq.~\eqref{eqn:dgl-cylind-numerical} both as a function of iteration steps. We find that if we set the convergence criterion $\epsilon$ such that the maximum violation should be smaller than $\epsilon = 10^{-7}$ for instance, this is achieved after about $300$ steps for $\hat{m}_{B} =1.0$, and about $2000$ steps for $\hat{m}_{B} =0.10$. The number of iteration steps for convergence depends on the initial guess of the solution. The examples are the case starting from the linear interpolation ansatz such that $B_{\rho}(n)= N_{q} n/n_{\rm max}$ and $\phi (n)= n/n_{\rm max}$. It is a general tendency that smaller $\hat{m}_{B}$ requires longer computation time, but it takes only 0.06 seconds for $\hat{m}_{B}=0.10$ with a single CPU on the Apple Mac mini 3\,GHz Intel Core i7 without any specific optimization of the numerical code. 

The converged values of $\sigma$ are found to be $\sigma/ m_{B}^{2}= 2.5137 $ for $\hat{m}_{B}=1.0$, and $\sigma /m_{B}^{2}= 3.1362$ for $\hat{m}_{B}=0.10$. Since the analytical continuum value is $\sigma /m_{B}^{2}= \pi$ according to Eq.~\eqref{eqn:stringtension-bb}, we find that the former reproduces the value with only 20 percent of accuracy, while the latter reproduces the value successfully within 0.2 percent of accuracy. Note that the difference from the analytical continuum value cannot be cured even if we simply use the smaller convergence criterion $\epsilon$. The final value is limited by the setting of $\rho_{\rm max}a$ and $\hat{m}_{B}$, the infrared and ultraviolet cutoffs, respectively. A detailed computation with larger $\rho_{\rm max}a$ with smaller $\hat{m}_{B}$ leads to more accurate results as presented in Table~\ref{tbl:comparison1D2D} in the next section. 
 
\par
\begin{figure}[!t] 
\centering\includegraphics[width=\figonecolumn]{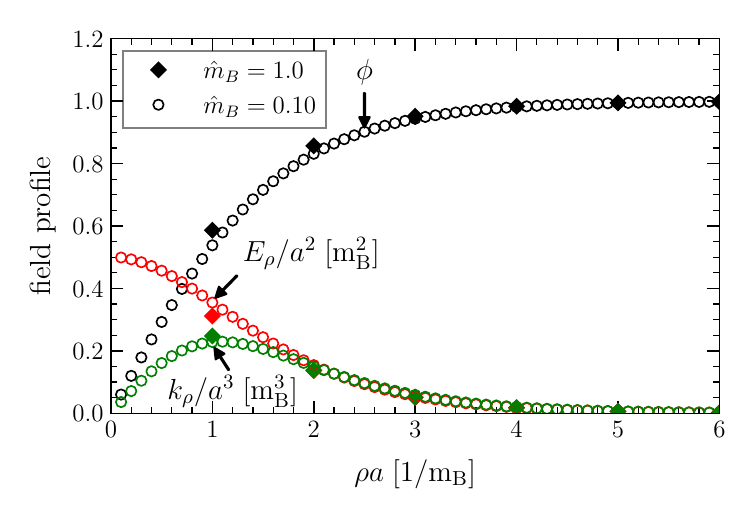} 
\caption{The field profiles for $\kappa =1.0$ as a function of the radial coordinate $\rho$. The circles with dotted lines are for $\hat{m}_{B}=1.0$ (coarse), and the solid lines are for $\hat{m}_{B}=0.10$ (fine).} 
\label{fig:fieldprofile} 
\end{figure}

 
In Fig.~\ref{fig:fieldprofile}, we plot the field profiles of the electric field 
\be 
E_{\rho}(n) =\frac{B_{\rho}(n+1)-B_{\rho}(n-1)}{2n } \;, 
\label{eqn:e-cylind} 
\ee 
and the monopole supercurrent 
\be 
k_{\rho}(n) = - \hat{m}_{B}^{2}\frac{B_{\rho}(n)-N_{q}}{n}\phi(n)^{2} 
\;, 
\label{eqn:k-cylind} 
\ee 
and the monopole field $\phi (n)$ as a function of $\rho$. In this plot, the physical scales of $E_{\rho}$ and $k_{\rho}$ are recovered by multiplying $1/a^{2}$ and $1/a^{3}$, respectively, to compare the results between two different discretization constants $a$ in Eq.~\eqref{eqn:physical-lat-a}. 
 
\par 
The flux-tube structure can be observed from both coarse and fine settings such that the electric field is distributed mainly in the region where the monopole field takes the value $0 \leq \phi < 1$. As expected, the finer setting $\hat{m}_{B}=0.10$ can capture the detailed $\rho$-dependence of the field variables; the electric field $E_{\rho}$ approaches a constant value as $\rho \to 0$, and the monopole supercurrent $k_{\rho}$ peaks around $\rho a \sim 1\;\mathrm{[1/m_{B}]}$. For a larger $\rho$, the difference between the coarse and finer settings is getting insignificant; regardless of the $\hat{m}_{B}$ values, the monopole fields are condensed as $\phi \sim 1$ for $\rho a > 5\;\mathrm{[1/m_{B}]}$, and the electric field and monopole supercurrent are similarly suppressed exponentially as already discussed. 
 
\par
\begin{figure}[!t] 
\centering\includegraphics[width=\figonecolumn]{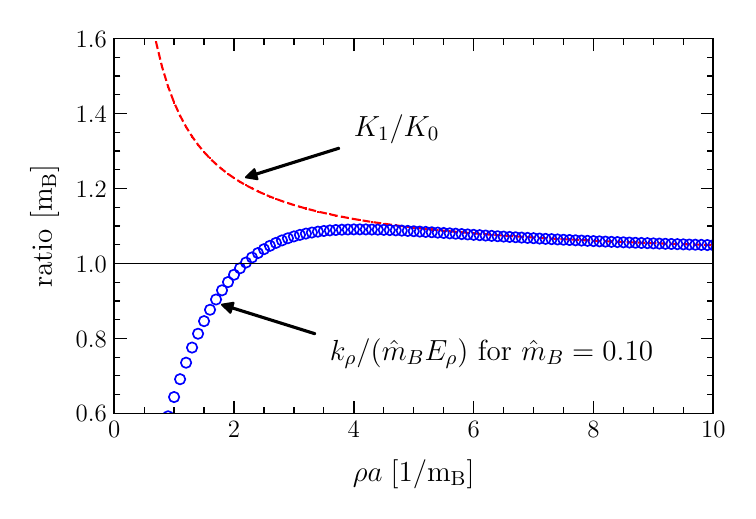} 
\caption{The ratio of the monopole supercurrent to the electric field with the normalization $\hat{m}_{B}$ for $\kappa =1.0$ as a function of the radial coordinate $\rho$. The ratio is compared to that of the modified Bessel functions $K_{1}/K_{0}$.} 
\label{fig:ratio} 
\end{figure}

 
In Fig.~\ref{fig:ratio}, we present the ratio of the monopole supercurrent $k_{\rho}$ to the electric field $E_{\rho}$ (multiplied by $\hat{m}_{B}$) as a function of $\rho$, where the vertical axis is normalized by $m_{B}$, and the ratio of the modified Bessel functions $K_{1}/K_{0}$ is also plotted. The result confirms the analytical expectation in Eqs.~\eqref{eqn:ele-besselk0} and ~\eqref{eqn:k-besselk1} for $\rho a > 5\;\mathrm{[1/m_{B}]}$, which gradually approaches $K_{1}/K_{0}\to 1$ for larger $\rho$, as $K_{1}$ and $K_{0}$ behave similarly at long distances. Note that this tendency offers a possibility of extracting the mass $m_{B}$ from the ratio between $k_{\rho}$ and $E_{\rho}$~\cite{Koma:2003gi}. 
 
\section{The DGL theory on the dual lattice} 
\label{sec:duallattice} 

It is straightforward to impose the boundary condition on the field variables as long as we consider a single flux tube in $D=2$ dimensions with cylindrical symmetry, as demonstrated in the previous section. However, a naive extension of such an approach to a finite-length \textit{open-ended} flux tube in $D=3$ dimensions makes the procedure complicated~\cite{Baker:1991bc}. Our approach is then to formulate the DGL theory on the dual lattice, and to solve the lattice version of the field equations by monitoring carefully the lattice effects caused by the finite cutoff and finite volume. The DGL theory on the dual lattice has been investigated previously~\cite{Koma:2000hw}. In this section, we complete the framework for a systematic numerical study. 
 
\subsection{The dual lattice formulation} 

We consider the DGL theory on the lattice in $D$ dimensions, such that the lattice volume is given by $L^{D}\equiv \prod_{\mu=1}^{D}L_{\mu}$ ($L_{\mu} \in \mathbb{N}$) and the lattice spacing by $a= 1/v =\hat{m}_{B}/m_{B}$ as in Eq.~\eqref{eqn:physical-lat-a}, where we set $a=1$ hereafter. The physical scale of $a$ will be recovered when the numerical results of different $\hat{m}_{B}$ are compared with each other. Although we use the argument $x=(x_{1},x_{2},...,x_{D})$ to express space coordinates as in the continuum theory, each component is discretized as $x_{\mu} \in [1,L_{\mu}]$. We impose periodic boundary conditions in all directions, as in the ordinary lattice QCD simulations. 

On the lattice, we may define the monopole field $\chi(x) =\chi^{1} (x)+i\chi^{2}(x)$ ($\chi^{\alpha} \in \Re$) as the {\em site} variables, the dual gauge field as the {\em link} variables, and the rotation of the dual gauge field $B_{\mu}(x)$ as the noncompact {\em plaquette} variables, 
\be 
(\del \wedge B)_{\mu\nu}(x) \equiv B_{\mu}(x) + B_{\nu} (x+\hat{\mu}) -B_{\mu}(x+\hat{\nu}) - B_{\nu}(x) \;, 
\ee 
where $\hat{\mu}$ is assumed to be a unit vector to the direction~$\mu$. The lattice action, the spatial sum of the lattice discretized version of the Lagrangian density in Eq.~\eqref{eqn:dgl-dimensionless}, is then given by 
\bea 
S &=& 
\beta_{g} 
\sum_{x} 
\Biggl [ \frac{1}{4} 
{}^{*\!} F_{\mu\nu} (x)^{2} + 
\frac{\hat{m}_B^2}{2} 
(D_{\mu}\chi^{\alpha}(x) )^{2} 
\nn\\* 
&& + 
\frac{\hat{m}_B^2 \hat{m}_\chi^2}{8} \left ( {\chi^{\alpha}(x)}^{2} - 1 \right )^2 
\Biggr ] \;, 
\label{eqn:action} 
\eea 
where 
\be 
{}^{*\!} F_{\mu\nu} (x)= (\del \wedge B)_{\mu\nu}(x) - 2 \pi \Sigma_{\mu\nu}(x) 
\label{eqn:dualgauge-fieldstrength} 
\ee 
is the dual field strength in Eq.~\eqref{eqn:dfs}, and 
\bea 
D_{\mu} \chi^{\alpha}(x) &=& \chi^{\alpha}(x) - \chi^{\alpha}(x+\hat{\mu}) \cos B_{\mu} (x)\nn\\ && + \epsilon_{\alpha\beta}\chi^{\beta}(x+\hat{\mu}) \sin B_{\mu}(x) 
\eea 
is the covariant derivative in Eq.~\eqref{eqn:cdv}. The field equations for the dual gauge field $B_{\mu}(x)$ and the monopole field $\chi^{\alpha}$ are given by 
\bea 
&& 
\frac{\del S}{\del B_{\mu}(x)} = \beta_{g} 
X_{\mu} (x) =0 \;, 
\label{eqn:feq-duallattice-b}\\ 
&& 
\frac{\del S}{\del \chi^{\alpha}(x)} =\beta_{g} \hat{m}_{B}^{2} Y^{\alpha} (x) =0 \;, 
\label{eqn:feq-duallattice-chi} 
\eea 
where 
\bea 
X_{\mu} (x) &= & 
\sum_{\nu \ne\mu} [ {}^{*\!} F_{\mu \nu}(x)+ {}^{*\!} F_{\nu \mu}(x-\hat{\nu}) ] \nn\\ 
&& +\hat{m}_{B}^{2} [ \chi^{\alpha}(x) \chi^{\alpha}(x+\hat{\mu})\sin B_{\mu}(x) \nn\\ &&+ \epsilon_{\alpha\beta}\chi^{\alpha}(x)\chi^{\beta}(x+\hat{\mu}) \cos B_{\mu} (x) ] \;, 
\label{eqn:feq-x} 
\eea 
\bea 
Y^{\alpha} (x) &=& 
\sum_{\mu} [ 
2 \chi^{\alpha}(x) - \chi^{\alpha}(x+\hat{\mu}) \cos B_{\mu} (x) \nn\\ && + \epsilon_{\alpha\beta}\chi^{\beta}(x+\hat{\mu}) \sin B_{\mu}(x) 
\nn\\&& 
-\chi^{\alpha}(x-\hat{\mu}) \cos B_{\mu} (x-\hat{\mu}) 
\nn\\&& 
- \epsilon_{\alpha\beta}\chi^{\beta}(x-\hat{\mu}) \sin B_{\mu}(x-\hat{\mu}) ] 
\nn\\&& 
+\frac{\hat{m}_{\chi}^{2}}{2} \chi^{\alpha} (x) ( {\chi^{\beta}(x)}^{2}-1) \; . 
\label{eqn:feq-y} 
\eea 
 
\par 
One needs to find the field variables that satisfy $X_{\mu}(x)=0$ ($\mu=1,2,...,D$) and $Y^{\alpha}(x)=0$ ($\alpha=1,2$) simultaneously, where the Newton-Raphson method can be used again for this purpose. In this method, after setting initial field variables, for instance $B_{\mu} (x)=0$, $\chi^{1} (x)=1$, and $\chi^{2} (x)=0$ everywhere (we may call this as the trivial initial condition), they are updated iteratively by 
\be 
B_{\mu}(x) \to B_{\mu}(x) - \frac{X_{\mu}(x)}{\delta_{B} X_{\mu}(x)} \;, 
\ee 
\be 
\left (\!\!\begin{array}{c} 
\chi^{1}(x) \\ 
\chi^{2}(x) \\ 
\end{array} 
\!\!\right ) 
\to 
\left (\!\!\begin{array}{c} 
\chi^{1}(x) \\ 
\chi^{2}(x) \\ 
\end{array} 
\!\!\right ) 
- 
\Biggl (\!\!\begin{array}{cc} 
\frac{Y^{1}(x)}{\delta_{\chi} Y^{1,1} (x)} &\frac{Y^{1}(x)}{\delta_{\chi} Y^{1,2} (x)} \\ 
\frac{Y^{2}(x)}{\delta_{\chi} Y^{2,1} (x)} &\frac{Y^{2}(x)}{\delta_{\chi} Y^{2,2} (x)}\\ 
\end{array} 
\!\!\Biggr )^{\!\!-1} 
\!\!\!\!\! 
\left (\!\!\begin{array}{c} 
\chi^{1}(x) \\ 
\chi^{2}(x) \\ 
\end{array} 
\!\!\right ) \;, 
\ee 
where 
\bea 
\delta_{B} X_{\mu}(x) &\equiv & \frac{\del X_{\mu}(x)}{\del B_{\mu}(x)} \nn\\* 
&=& 2 (D-1) + \hat{m}_{B}^{2} [ \chi^{\alpha}(x) \chi^{\alpha}(x+\hat{\mu})\cos B_{\mu} (x) \nn\\* &&- \epsilon_{\alpha\beta}\chi^{\alpha}(x)\chi^{\beta}(x+\hat{\mu}) \sin B_{\mu} (x) ] \;, \\ 
\delta_{\chi} Y^{\alpha,\beta}(x) &\equiv & 
\frac{\del Y^{\alpha} (x)}{\del \chi^{\beta}(x)} \nn\\* 
&=& 
\delta_{\alpha\beta} [ 2 D + \frac{ \hat{m}_\chi^2}{2} ( \chi^{\gamma} (x)^{2} - 1 ) ] 
\nn\\* 
&& +\hat{m}_\chi^2 \chi^{\alpha} (x) \chi^{\beta} (x) \; , 
\eea 
where $\delta_{\alpha\beta}$ denotes the Kronecker delta. As in the cylindrical case, the iteration process is terminated when all of $\max (|X_{\mu} (x)| )<\epsilon$ ($\mu=1,2,...,D$) and $\max ( |Y^{\alpha} (x)| )< \epsilon$ ($\alpha=1,2$) are satisfied simultaneously with a reasonably small value of~$\epsilon$. 
 
\par
\begin{figure}[!t] 
\centering\includegraphics[width=5cm]{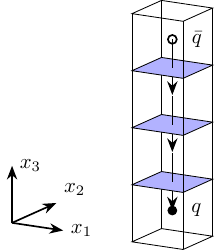} 
\caption{How to put the nonzero $\Sigma_{\mu\nu}$ for the quark-antiquark ($q\bar{q}$) system in $D=3$ dimensions. This example is for the $R=3$ case, where the $q\bar{q}$ axis is taken along the $x_{3}$-axis. The shaded part of $\Sigma_{12}$ is set to be $-N_{q}$.} 
\label{fig:dstring2} 
\end{figure}

 
The most elegant point in the dual lattice formulation is that we can simply treat the Dirac string $\Sigma_{\mu\nu}$ in the dual field strength as integer values, which allows us to obtain the solution for various kinds of flux-tube systems without caring about the detailed boundary condition of the field variables. For instance, if a quark-antiquark system with the charges $N_{q}$ and $-N_{q}$ with the distance $R$ along the $x_{3}$-axis in $D=3$ dimensions is considered ($N_{q}=1$ for the fundamental representation), we only set a connected stack of $N_{q}$ as 
\be 
\Sigma_{12}(0,0,l) = - N_{q} ~~~(l=1,2,...,R)\;, 
\label{eqn:qqbar-sample-setting} 
\ee 
and set zero for the rest. In this setting, the locations of the quark and antiquark correspond to $\vec{x}=(0,0,0) \to (1/2)\hat{1}+(1/2)\hat{2}+(1/2)\hat{3}$ and $\vec{x}=(0,0,R)\to (1/2)\hat{1}+(1/2)\hat{2}+(1/2+R)\hat{3}$. A schematic image is illustrated in Fig.~\ref{fig:dstring2} for the $R=3$ case. The dual Bianchi identity for the dual field strength is violated as $(1/2) \epsilon_{\mu\nu\rho} \del_{\mu}{}^{*\!} F_{\nu\rho} = + 2\pi N_{q}~(-2\pi N_{q})$ at the location of the cube where the quark $q$ (antiquark $\bar{q}$) is assumed to be inside, providing the total electric charge density in the system 
\be 
j_{0}(x) =N_{q}\delta_{x_{1}0}\delta_{x_{2}0}(\delta_{x_{3}0}-\delta_{x_{3}R}) \;. 
\ee 
 
\par 
Let us demonstrate what the numerical solution looks like for the quark-antiquark system in $D=3$ dimensions. We set here \eq{\hat{m}_{B}=\hat{m}_{\chi}=0.50} (\eq{\kappa =1.0}) and employ a $L^{3}=32^{3}$ lattice. After setting the Dirac string as in Fig.~\ref{fig:dstring2} with the trivial initial condition, we apply the Newton-Raphson method. To avoid direction dependence during the update process, we divide the lattice sites and links into odd and even parts, and the field variables on each are updated in turn. 
 
\par
\begin{figure}[!t] 
\centering\includegraphics[width=\figonecolumn]{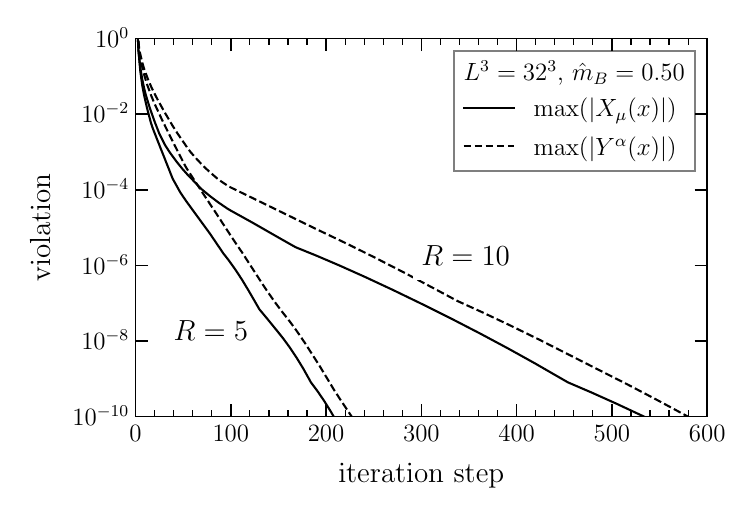} 
\caption{Typical histories of the maximum violation \eq{\max (|X_{\mu}(s)|)} and \eq{\max (|Y^{\alpha}(s)|)} for $R=5$ and $R=10$ for $\kappa =1.0$ 
as a function of iteration step.} 
\label{fig:iter_vs_violation} 
\vspace{0.5cm} 
\centering\includegraphics[width=\figonecolumn]{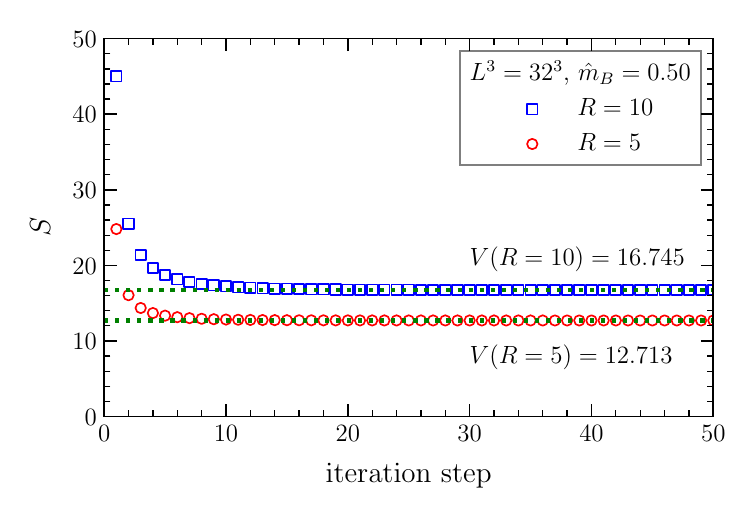} 
\caption{Typical histories of the value of $S$ in Eq.~\eqref{eqn:action} for $R=5$ and $R=10$ with $\kappa =1.0$ as a function of iteration step.} 
\label{fig:iter_vs_action} 
\end{figure}

 
Typical histories of the maximum violation \eq{\max (|X_{\mu}(x)|)} and \eq{\max (|Y^{\alpha}(x)|)} for $R=5$ and $R=10$ are shown in Fig.~\ref{fig:iter_vs_violation}. We find that the maximum violation is getting reduced exponentially as we increase the number of iteration steps. As a general tendency, a large value of $R$ requires more iteration steps. During the iteration, we also monitor the value of $S$ in Eq.~\eqref{eqn:action}, which finally converges to the potential energy in $D=3$ dimensions, $S \to V$. We find that the approximate value of the potential energy can be obtained only after several tens of iteration steps, as shown in Fig.~\ref{fig:iter_vs_action}, although the maximum violation is yet of $O(10^{-2})$ at this stage. 
 
\subsection{The Hodge decomposition} 
\label{sec:hodge-decomposition} 
 
\par 
The Hodge decomposition can also be applied to the dual gauge field on the lattice as in Eq.~\eqref{eqn:hodge-decomposition}. The decomposition is achieved by using the massless lattice Green function $G_{L}(x)$ in a finite volume $L^{D}$, which satisfies the relation 
\be 
\Delta_{L} G_{L}(x-x') = 
\del_{\mu}\del_{\mu}' G_{L}(x-x') =-\delta_{xx'}+\frac{1}{L^{D}} \;, 
\label{eqn:green-finite-volume-effect} 
\ee 
where $\del_{\mu}$ and $\del_{\mu}'$ are forward and backward differences, respectively. We summarize some basics of the lattice Green functions in Appendix~\ref{sect:app-lattice-green}. The decomposed parts of the dual gauge field are then given by 
\be 
B_{\mu}^{\rm reg} (x) = 
\sum_{x'}G_{L}(x-x') 
{\del_{\nu}'} {}^{*\!} F_{\mu\nu}(x')\;, 
\ee 
\be 
B_{\mu}^{\rm sing} (x) = 2\pi \sum_{x'}G_{L}(x-x') 
\del_{\nu}' \Sigma_{\mu\nu}(x')\;, 
\ee 
\be 
B_{\mu}^{\rm red} (x) =\sum_{x'}G_{L}(x-x')\del_{\mu} (\del_{\nu}' B_{\nu}(x')) \;. 
\ee 
 
\par 
The rotation of $B_{\mu}^{\rm reg}$ leads to 
\be 
(\del \wedge B^{\rm reg} )_{\mu\nu} (x) = {}^{*\!} F_{\mu\nu}(x) +\frac{2\pi \xi_{\mu\nu}}{L^{D}} - 2\pi C_{\mu\nu}(x ) \;, 
\label{eqn:dgf-reg} 
\ee 
where we denote 
\bea 
&& 
\xi_{\mu\nu} =\sum_{x} \Sigma_{\mu\nu}(x) \;,\\ 
&& C_{\mu\nu}(x ) = - \epsilon_{\mu\nu\sigma} \del_{\sigma}'\sum_{x'}G_{L}(x-x') j_{0}(x') \;. 
\label{eqn:coulomb-cmunu} 
\eea 
The correction term $2\pi \xi_{\mu\nu}/L^{D}$ appears due to the $1/L^{D}$ term in Eq.~\eqref{eqn:green-finite-volume-effect}. The $C_{\mu\nu}$ term represents the Coulombic electric field. The rotation of $B_{\mu}^{\rm sing}$ becomes 
\be 
(\del \wedge B^{\rm sing})_{\mu\nu}(x) = 2\pi \Sigma_{\mu\nu}(x )- \frac{2\pi \xi_{\mu\nu}}{L^{D}} +2\pi C_{\mu\nu}(x ) 
\; , 
\ee 
which also has the correction term with the opposite sign. Since the redundant part has no contribution to the dual field strength as $(\del \wedge {B}^{\rm red})_{\mu\nu} (x)=0$, we recover the relation in Eq.~\eqref{eqn:dualgauge-fieldstrength}, 
\bea 
(\del \wedge B)_{\mu\nu}(x) &=& (\del \wedge B^{\rm reg})_{\mu\nu}(x) +(\del \wedge B^{\rm sing})_{\mu\nu}(x) \nn\\ &=& {}^{*\!}F_{\mu\nu}(x) + 2\pi \Sigma_{\mu\nu}(x ) \; , 
\eea 
which has no explicit volume correction. Therefore, when the regular and the singular parts of the dual gauge field are evaluated separately, it is possible to improve them by removing the $1/L^{D}$ term in advance as 
\be 
(\del \wedge B^{\rm reg})_{\mu\nu}(x) \to (\del \wedge B^{\rm reg})_{\mu\nu}(x) - \frac{2\pi \xi_{\mu\nu}}{L^{D}} \;, 
\label{eqn:breg-correction} 
\ee 
\be 
(\del \wedge B^{\rm sing})_{\mu\nu}(x) \to (\del \wedge B^{\rm sing})_{\mu\nu}(x) +\frac{2\pi \xi_{\mu\nu}}{L^{D}}\;, 
\ee 
where this procedure has no visible effect if the volume is already large enough. Note that these improvements were not taken into account in our previous work~\cite{Koma:2000hw}. With this improvement, the field strength becomes 
\be 
{}^{*\!} F_{\mu\nu} = ( \del \wedge B^{\rm reg})_{\mu\nu}(x) +2\pi C_{\mu\nu} (x)\;. 
\label{eqn:fs-reg-c} 
\ee 
 
\par 
Since the contribution from the cross term of $(\del \wedge B^{\rm reg})_{\mu\nu}$ and $C_{\mu\nu}$ vanishes because of the relation $\epsilon_{\lambda\mu\nu}\del_{\lambda}(\del \wedge B^{\rm reg})_{\mu\nu}=0$, the action can also be decomposed into two parts as 
\bea 
S = S_{\mathrm{coul}} +S_{\mathrm{sole}} \;, 
\eea 
where 
\bea 
S_{\rm coul} & \equiv & 
\beta_{g} 
\sum_{x} \frac{1}{4} [2\pi C_{\mu\nu} (x)]^{2} \;, 
\label{eqn:action-coulomb} 
\\ 
S_{\rm sole} &\equiv &\beta_{g} \sum_{x}\Biggl [\frac{1}{4} [ (\del \wedge B^{\rm reg})_{\mu\nu}(x)]^{2} + 
\frac{\hat{m}_B^2}{2} 
(D_{\mu}\chi^{\alpha}(x) )^{2} 
\nn\\ 
&& + 
\frac{\hat{m}_B^2 \hat{m}_\chi^2}{8} \left ( {\chi^{\alpha}(x)}^2 - 1 \right )^2 
\Biggr ] \; . 
\label{eqn:action-solenoidal} 
\eea 
The labels ``coul'' and ``sole'' mean the Coulombic and the solenoidal parts, respectively, reflecting the structure of the resulting electric field profile (we will demonstrate this below). 
 
\begin{figure}[!t] 
\centering\includegraphics[width=\figonecolumn]{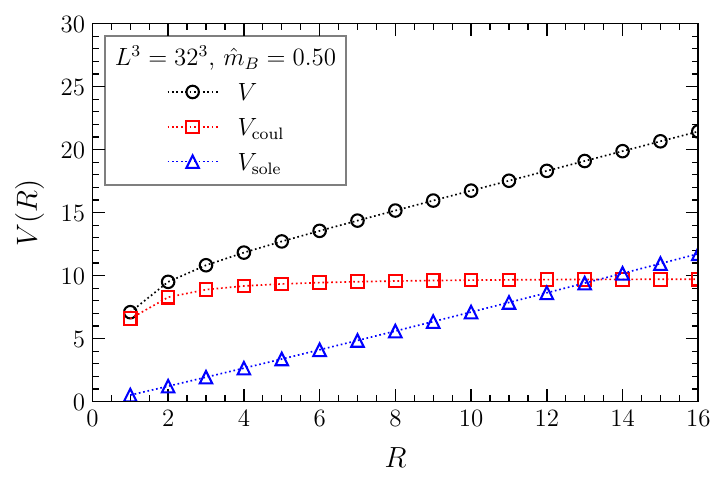} 
\caption{The interquark potential with the Hodge decomposition $V=V_{\mathrm{coul}}+V_{\mathrm{sole}}$ for $\kappa =1.0$ ($\hat{m}_{B}=\hat{m}_{\chi}=0.50$).} 
\label{fig:demo3d_potential} 
\end{figure} 
 
\par 
In Fig.~\ref{fig:demo3d_potential}, we present the potential energy $V$ (the converged value of $S$) with the Hodge decomposition, $V_{\rm coul}$ and $V_{\rm sole}$, for various distances $R$, where the lattice setting is the same as before (on the $L^{3}=32^{3}$ lattice with $\hat{m}_{B}=\hat{m}_{\chi}=0.50$, $N_{q}=1$), and the iteration process is terminated when the maximum violation becomes smaller than $\epsilon =10^{-6}$. One may immediately notice that the Coulombic part of the potential $V_{\rm coul}(R)$ becomes constant at large distances, while the solenoidal part $V_{\rm sole}(R)$ is linearly increasing.\footnote{The quark-antiquark potential in the lattice QCD simulations with the Abelian projection in the maximally Abelian (MA) gauge~\cite{tHooft:1981bkw} also exhibits a similar composed structure of the potential, where the ``photon'' and ``monopole'' parts in the MA gauge corresponds to the ``Coulombic'' and ``solenoidal'' parts in the DGL theory. This correspondence has been exploited in our previous investigation on the type of the dual superconducting vacuum~\cite{Koma:2003gq, Koma:2003hv, Koma:2003gi}. } 
 
\begin{figure*}[!t] 
\centering\includegraphics[width=\figtwocolumn]{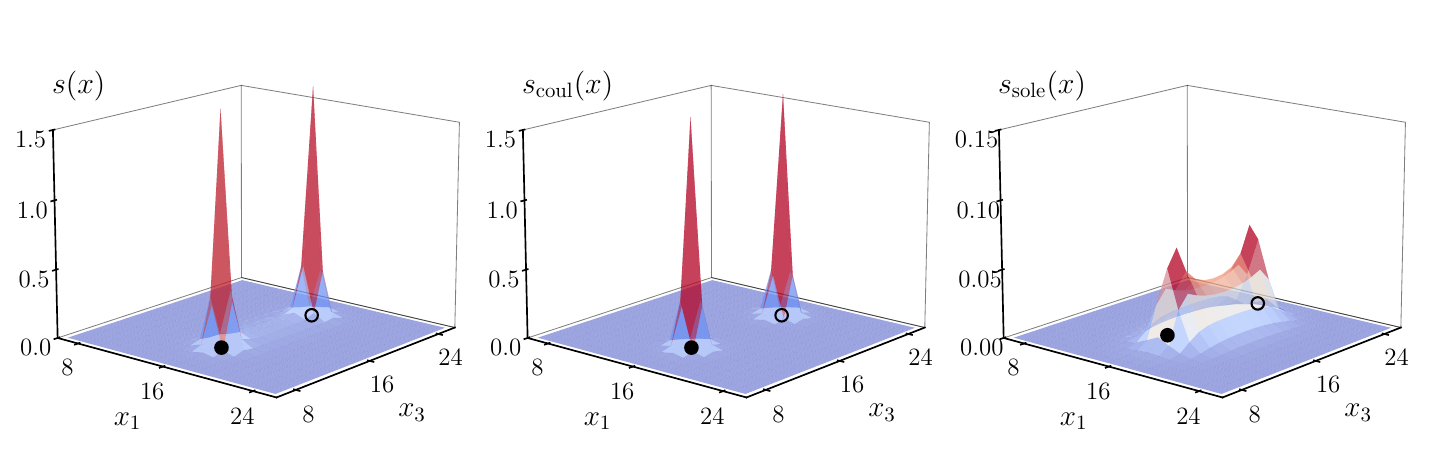} 
\caption{The profile of the full action density $s$ (left) with the Hodge decomposition, the Coulombic part $s_{\mathrm{coul}}$ (middle), and the solenoidal part $s_{\mathrm{sole}}$ (right), just on the median plane that the quark (the black-filled circle) and antiquark (the black-open circle) are placed for $\kappa =1.0$ ($\hat{m}_{B}=\hat{m}_{\chi}=0.50$) on the $L^{3}=32^{3}$ lattice, where the interquark distance is $R=10$. Note that the vertical axis scale of the solenoidal part is one-tenth smaller than the others.} 
\label{fig:profile_actd} 
\end{figure*} 
 
\begin{figure*}[!t] 
\centering\includegraphics[width=\figtwocolumn]{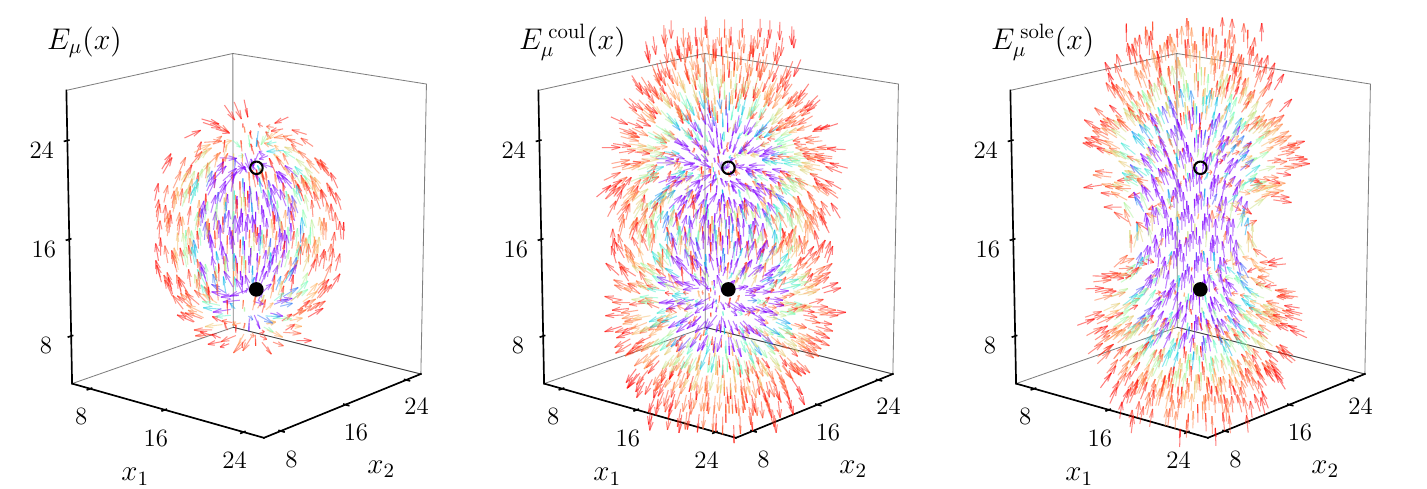} 
\caption{The profile of the full electric field $E_{\mu}$ (left) with the Hodge decomposition, the Coulombic part $E_{\mu}^{\mathrm{coul}}$ (middle), and the solenoidal part $E_{\mu}^{\mathrm{sole}}$ (right), for $\kappa =1.0$ ($\hat{m}_{B}=\hat{m}_{\chi}=0.50$) on the $L^{3}=32^{3}$ lattice, where the interquark distance is $R=10$, and the locations of the quark and antiquark are specified by the black-filled and black-open circles, respectively. The values for $|E_{\mu}| \geq 0.004$ are displayed, and the strength is represented by a color scale, with violet indicating higher magnitudes.} 
\label{fig:profile_ele} 
\end{figure*} 
 
\par 
In fact, $V_{\rm coul}(R)$ exactly coincides with the lattice Coulombic potential of the functional form 
\be 
V_{\rm coul}(R) = 4 \pi^{2} \beta_{g} N_{q}^{2} [G_{L}(0,0,0)- G_{L}(0,0,R) ] \; . 
\label{eqn:lattice-coulombic-pot} 
\ee 
In the current setting, $G_{L}$ is the massless lattice Green funtion on the $L^{3}=32^{3}$ lattice.\footnote{For instance, $G_{L}(0,0,0)=0.2456765511$, $G_{L}(0,0,1)=0.0790149708$, $G_{L}(0,0,2)=0.0358552903$, on the $L^{3}=32^{3}$ lattice.} Note that Eq.~\eqref{eqn:lattice-coulombic-pot} is derived by inserting an explicit form of 
\be 
C_{\mu\nu} (x) = -N_{q} 
\epsilon_{\mu\nu\sigma}{\del_{\sigma}'} 
[ G_{L}(x)-G_{L}(x_{1},x_{2},x_{3}-R) ] \; 
\ee 
into Eq.~\eqref{eqn:action-coulomb}, with the relation 
\be 
\sum_{x} (\del_{\mu}' G_{L}(x) )^{2} = - \sum_{x}G_{L} (x) \Delta_{L} G_{L}(x) 
\ee 
and Eq.~\eqref{eqn:green-finite-volume-effect}, where the finite volume correction terms $2\pi \xi_{\mu\nu}/L^{3}$ from the quark and the antiquark cancel each other. The partial cubic rotational symmetry of $C_{\mu\nu}$ around the $x_{3}$-axis such as $C_{23} (x) =C_{31} (x)$ is also taken into account, which appears when $L_1 = L_2$ is satisfied or both $L_1$ and $L_2$ are sufficiently large. In the continuum and the large volume limit, Eq.~\eqref{eqn:lattice-coulombic-pot} reduces to 
\be 
V_{\rm coul}(R) \to -\frac{\pi \beta_{g} N_{q}^{2}}{R} + const. \;, 
\ee 
since $G_{L}(0,0,R)$ will approach $1/(4\pi R)$. An important fact is that the Coulombic part is always independent of the mass parameters. 
 
\par 
On the other hand, $V_{\rm sole}(R)$ seems to be proportional to the distance $R$ with a positive slope as $V_{\rm sole}(R) \propto R$, which reflects approximate translational invariance of the solenoidal part along the $x_{3}$-axis. As already explained in Sec.~\ref{sec:dgl}, if the length of the initial Dirac string is straight and infinitely long, the system becomes $D=2$ dimensional. Since the string tension is known to be $\sigma = \pi \beta_{g} | N_{q} | m_{B}^{2}$ [Eq.~\eqref{eqn:stringtension-bb}] for $\kappa=1$, we expect 
\be 
V_{\rm sole}(R) \sim \sigma R = \pi \hat{m}_{B}^{2} R \approx 0.785 R \;. 
\ee 
With the resolution in Fig.~\ref{fig:demo3d_potential}, this expectation seems to be realized from short distances. In general, the string tension $\sigma$ depends on $\kappa$, and numerical methods are needed to obtain the value if $\kappa \ne 1$. 
 
\par 
The functional form of the potential has a direct connection to the profile of the action density $s(x)$, where $V=\sum_{x}s(x)$ in $D=3$ dimensions. We present an example of the profile in Fig.~\ref{fig:profile_actd} for the $R=10$ case just on the median plane of the flux tube. The action density $s(x)$ can be decomposed into two parts $s_{\mathrm{coul}}(x)$ and $s_{\mathrm{sole}}(x)$ as in the potential. We observe two sharp peaks at the location of the quark and antiquark in the Coulombic part, while a ridge structure between the quark and antiquark in the solenoidal part. Then, the sum of these two profiles gives the full action density. 
 
\par 
The height of the two sharp peaks themselves is irrelevant to the interaction, but just reflects the self-energies of the quark and antiquark. The interaction occurs when the action density is {\em continuously connected} by positive values. When the quark and antiquark are close to each other, the tails of the two shape peaks overlap, yielding the Coulombic potential. As the quark and antiquark are separated, the overlap of the tails is reduced, and the ridge structure comes out of hiding. Since the height and width of the ridge are almost constant, which persists even when the quark and antiquark are separated further, there appears a linearly rising potential with a constant positive slope. 
 
\par 
The behavior of the action density is closely related to the profile of the electric field, 
\be 
E_{\mu} = 
\frac{1}{2}\epsilon_{\mu\nu\rho}{}^{*\!}F_{\nu\rho} 
\;, 
\ee 
since the square of this quantity is a part of the action density. The electric field can also be decomposed into the Coulombic and the solenoidal parts. The Coulombic part of the electric field should behave as that around the electric dipole in electromagnetism. The solenoidal part is expected to behave similarly (but dual) to the magnetic field in and around a finite-length solenoid coil, which is the origin of its naming. In our case, the monopole supercurrent, 
\bea 
k_{\mu} &=&- 
\hat{m}_{B}^{2} 
[ \chi^{\alpha}(x) \chi^{\alpha}(x+\hat{\mu})\sin B_{\mu}(x) \nn\\* &&+ \epsilon_{\alpha\beta}\chi^{\alpha}(x) \chi^{\beta}(x+\hat{\mu}) \cos B_{\mu}(x) ] 
\;, 
\eea 
acts as the circular current. 
 
\begin{figure}[!t] 
\centering\includegraphics[width=\figonecolumnsmall]{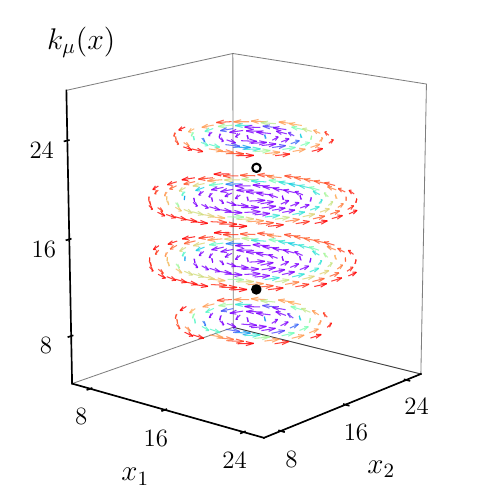} 
\caption{The profiles of the monopole supercurrent $k_{\mu}(x)$ for $\kappa =1.0$ ($\hat{m}_{B}=\hat{m}_{\chi}=0.50$) on the $L^{3}=32^{3}$ lattice, where the interquark distance is $R=10$, and the locations of the quark and antiquark are specified by the black-filled and black-open circles, respectively. The values with $|k_{\mu}| \geq 0.001$ on the four selected planes are displayed, and the strength is represented by a color scale, with violet indicating higher magnitudes.} 
\label{fig:profile_k} 
\centering\includegraphics[width=\figonecolumnsmall]{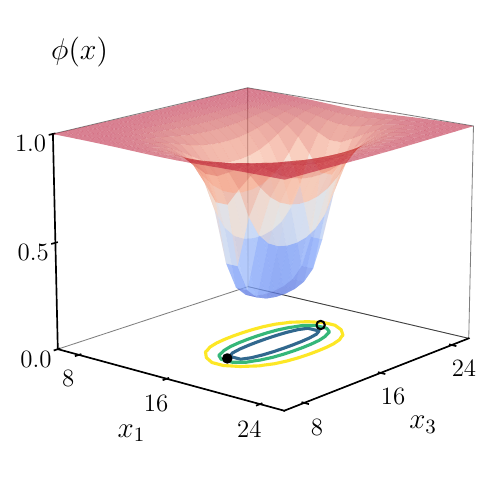} 
\caption{The profiles of the monopole field $\phi(x)=|\chi (x)| $ for $\kappa =1.0$ ($\hat{m}_{B}=\hat{m}_{\chi}=0.50$) on the $L^{3}=32^{3}$ lattice, where the interquark distance is $R=10$, and the locations of the quark and antiquark are specified by the black-filled and black-open circles, respectively. } 
\label{fig:profile_phi} 
\end{figure} 
 
\par 
In Fig.~\ref{fig:profile_ele}, we show the electric field $E_{\mu}(x)$ with the Hodge decomposition. Since $R=10$ corresponds to $5.0\;\mathrm{[1/m_{B}]}$ for $\hat{m}_{B}=0.50$, one may naively expect a tubelike structure with the ratio, $\mathrm{length}: \mathrm{diameter}=5:2$, however, the profile of the full electric field in Fig.~\ref{fig:profile_ele} (left) rather looks like a sphere. In this sense, the naming, such as the ``flux tube'' which evokes translational invariance of the profile, may not always be appropriate, though we use this term to express the solution throughout this paper. To observe a thin tubelike structure for the electric field, the $\kappa$ should be larger, or the length $R$ should be unrealistically longer compared to the penetration depth. 
 
\par 
In fact, the penetration depth alone does not characterize the width of the electric-field profile properly when $\kappa$ is finite. Since the electric field is getting suppressed exponentially beyond the distance of coherence length, it is reasonable to think that the width of the electric-field profile is roughly given by the sum of the coherence length and the penetration depth. For the $\kappa=1$ case, the width will then be twice the penetration depth, so that the above naive ratio should be modified to $\mathrm{length}: \mathrm{diameter}=5:4$, which may be consistent with the appearance (see Fig.~\ref{fig:fieldprofile}, where the electric field at $\rho a \sim 2\;\mathrm{[1/m_{B}]}$ becomes about $e^{-1}\sim 1/3$ of the maximum value at the center). 
 
\par 
Even if the full electric-field profile does not show a tubelike structure, it is not the same as the Coulombic field, since the full electric field is given by the sum of the real Coulombic field in Fig.~\ref{fig:profile_ele} (middle) and the solenoidal field in Fig.~\ref{fig:profile_ele} (right). The solenoidal part plays a role in canceling the Coulombic field in the outer region, which, at the same time, enhances the electric field in the inner region around the Dirac string. In Figs.~\ref{fig:profile_k} and~\ref{fig:profile_phi}, we show the profiles of the circulating monopole supercurrent $k_{\mu}(x)$ and the monopole field $\phi(x) =| \chi (x) |$. In the region where the electric field has a large value, $\phi(x)$ necessarily takes a value smaller than one, indicating the melting of vacuum condensate. 
 
\begin{figure*}[!t] 
\centering\includegraphics[width=\figtwocolumn]{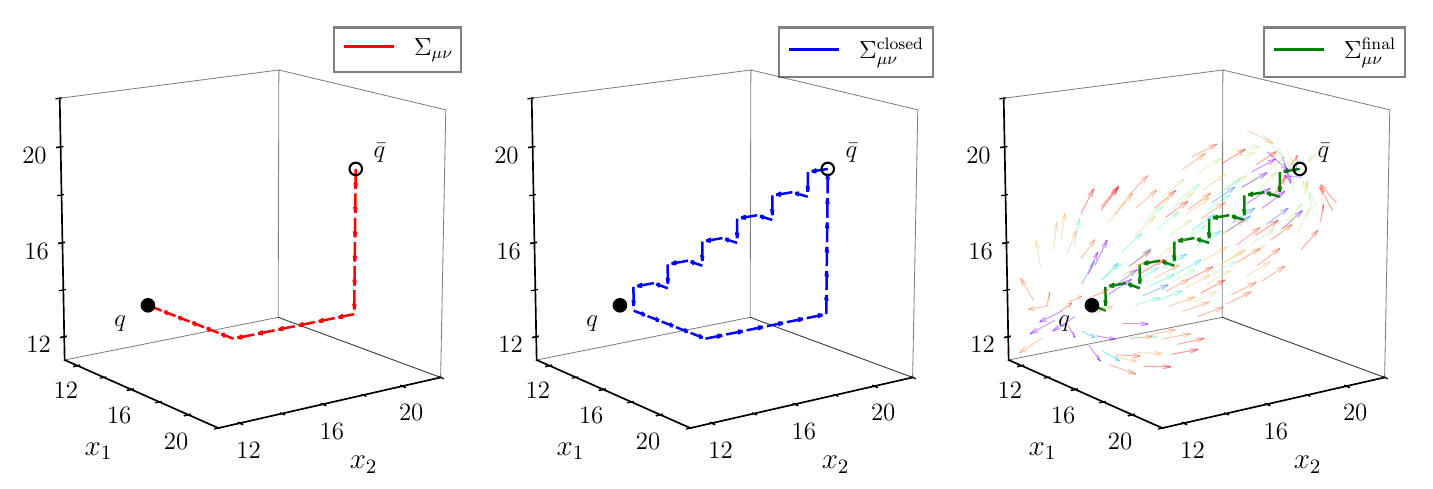} 
\caption{The vector representation of the original Dirac string $\Sigma_{\mu\nu}$ (left), the generated closed Dirac string $\Sigma_{\mu\nu}^{\rm closed}$ (middle), and the final Dirac string $\Sigma_{\mu\nu}^{\rm final}$ (right), where the ending and starting points of the string, $(x_{1},x_{2},x_{3})=(13.5,13.5,13.5)$ and $(19.5,19.5,19.5)$, correspond to the quark source and the antiquark sink, respectively. The profile of the electric field $E_{\mu}$ is also plotted with $\Sigma_{\mu\nu}^{\rm final}$.} 
\label{fig:demo3d-ds} 
\centering\includegraphics[width=\figtwocolumn]{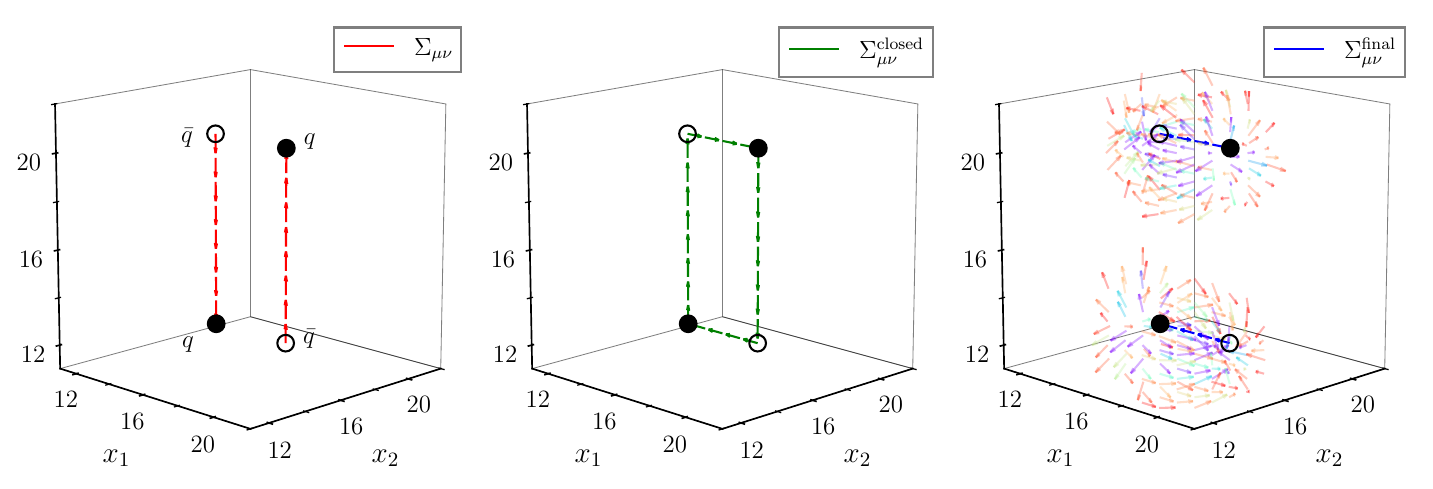} 
\caption{The similar plots as in Fig.~\ref{fig:demo3d-ds} for the two-flux-tube system, where the two original Dirac strings are put along the $x_{3}$-axis from $(x_{1},x_{2},x_{3})=(14.5,16.5,12.5)$ to $(14.5, 16.5, 20.5)$ and from $(18.5, 16.5, 20.5)$ to $(18.5, 16.5, 12.5)$.} 
\label{fig:demo3d-ds-flip} 
\end{figure*} 
 
\subsection{The final location of the Dirac string} 
\label{subsec:fluxtubecore} 

The shape of the flux-tube solution may depend on the path of how the Dirac string is put. As we will show below, however, the iteration process always tries to find a minimum energy configuration of the fields, so that the final path of the Dirac string is not always the same as the initial one. During the iteration process of the Newton-Raphson method, the phase of the monopole field or the redundant part of the dual gauge field can become multivalued, which generates a new set of Dirac strings. This happens when the flux tube along the original Dirac string $\Sigma_{\mu\nu}$ is energetically unstable. The sum of the original and the generated Dirac strings then determines the final Dirac string that is responsible for the flux-tube solution. 
 
\par 
In practice, the location of the final Dirac string in $D=3$ dimensions can be identified by the following procedure. First, we identify the total winding number of the dual gauge field, including the contribution from the phase of the monopole field $\eta (x) = \tan^{-1}(\chi^{2}(x)/\chi^{1}(x)) \in [-\pi,\pi)$ by 
\be 
n_{\mu}(x) \equiv {\rm floor} (\frac{ B_{\mu} (x)+\del_{\mu}\eta (x)+5\pi}{2\pi} )-2 \;, 
\ee 
where the shift of $5\pi$ is just for technical reasons, and the function ${\rm floor}(arg)$ is assumed to return the greatest integer value less than or equal to $arg$. When $n_{\mu} \ne 0$ the closed Dirac string is generated by 
\be 
\Sigma_{\mu\nu}^{\rm closed}(x) = 
(\del \wedge n)_{\mu\nu}(x) 
\;, 
\label{eqn:sigma-closed} 
\ee 
which satisfies $\sum_{x}\Sigma_{\mu\nu}^{\rm closed}(x)=0$ regardless of the choice of $\mu\nu$ planes. Second, we extract the remnant of the dual gauge field by 
\be 
B_{\mu}^{\rm rem} (x) =B_{\mu} (x) - 2 \pi n_{\mu} (x) \;, 
\ee 
which allows us to compute the final location of the Dirac string by 
\bea 
\Sigma_{\mu\nu}^{\rm final}(x) 
&=& {\rm floor} ( \frac{ (\del \wedge B^{\rm rem})_{\mu\nu} (x) - 2\pi C_{\mu\nu}(x) +5\pi}{2\pi} ) -2 \nn\\ &=& 
\Sigma_{\mu\nu}(x) + \Sigma_{\mu\nu}^{\rm closed}(x) \;. 
\label{eqn:sigma-final} 
\eea 
where the subtraction of the Coulombic electric field $C_{\mu\nu}$ is crucial when a finite-length flux tube with higher charges $N_{q}\geq 2$ is investigated, since the large value of the electric field proportional to $N_{q}$ often leads to the wrong integer value which is not related to the string singularity. Note that the location of the Dirac string identified by 
\be 
{\rm floor}( \frac{ (\del \wedge B)_{\mu\nu}(x) - 2\pi C_{\mu\nu}(x) +5\pi}{2\pi}) -2 \;, 
\label{eqn:sigma-bsing} 
\ee 
is the same as the original Dirac string $\Sigma_{\mu\nu}$, so that the cancellation of the Dirac string occurs in the field strength, leaving the Coulombic term as in Eq.~\eqref{eqn:fs-reg-c}. 
 
\par 
Let us see a typical example of the generation of the closed string on a $32^{3}$ lattice with $\hat{m}_{B} =\hat{m}_{\chi} =0.50$. We first put the Dirac string along the side of a cube as shown in Fig.~\ref{fig:demo3d-ds} (left) and start the Newton-Raphson method, where the ending and starting points of the string, $(x_{1},x_{2},x_{3})=(13.5,13.5,13.5)$ and $(19.5,19.5,19.5)$, correspond to the quark source and the antiquark sink, respectively. The convergence criterion is $\epsilon = 10^{-6}$. Note that we always compute $\frac{1}{2}\epsilon_{\mu\nu\rho}\Sigma_{\nu\rho}(x)$ etc. for identifying the string in the $\mu$ direction. Once the field equations are satisfied, we evaluate Eq.~\eqref{eqn:sigma-closed} and obtain the string as shown in Fig.~\ref{fig:demo3d-ds} (middle). The final string in Eq.~\eqref{eqn:sigma-final} has the structure as shown in Fig.~\ref{fig:demo3d-ds} (right), where the corresponding electric field $E_{\mu}(x)$ is also plotted. Clearly, the location of the Dirac string is different from that of the original one, and the flux tube is formed along the final Dirac string. 
 
\par 
It may also be interesting to see what happens if two open Dirac strings are placed in parallel with the {\em opposite} direction. With the same lattice volume and mass parameters as above, we put the two original Dirac strings along the $x_{3}$-axis from $(x_{1},x_{2},x_{3})=(14.5,16.5,12.5)$ to $(14.5, 16.5, 20.5)$ and from $(18.5, 16.5, 20.5)$ to $(18.5, 16.5, 12.5)$. We then obtain the result as shown in Fig.~\ref{fig:demo3d-ds-flip}, where the original Dirac string (left), the generated closed Dirac string (middle), and the final Dirac string (right) are plotted. We find that the sum of the original Dirac string and the generated closed Dirac string leads to the final Dirac string. If we only look at the original and the final Dirac strings, we may interpret this result as an occurrence of the {\em string flip}. We will investigate the case that two open Dirac strings are placed in parallel but in the {\em same} direction in the next section in detail. 
 
\subsection{The control of systematic lattice effects} 
\label{subsec:result-latticeeffects} 
 
\begin{figure}[!t] 
\centering\includegraphics[width=\figonecolumn]{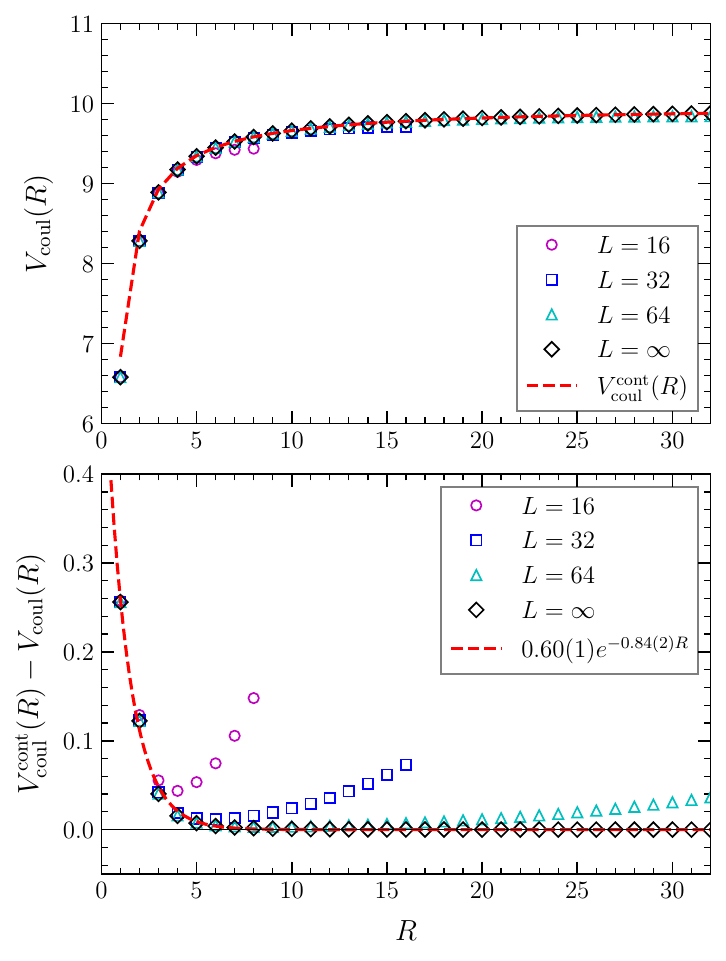} 
\caption{The upper panel shows the Coulombic part of the potential $V_{\mathrm{coul}}$ in Eq.~\eqref{eqn:lattice-coulombic-pot} for cubic lattice volumes $L^{3}$ with $L=16$, $32$, $64$, $\infty$, which are compared with that of its continuum limit counterpart $V_{\mathrm{coul}}^{\mathrm{cont}}$ in Eq.~\eqref{eqn:lattice-coulombic-pot-cont}. The lower panel shows the difference between $V_{\mathrm{coul}}^{\mathrm{cont}}$ and $V_{\mathrm{coul}}$. There is a characteristic difference at very short distances quantified approximately by an exponential function.} 
\label{fig:pot_coul_vdep} 
\end{figure} 
 
\par 
Our method is based on solving the field equations defined on the dual lattice, therefore the solution can be affected by the systematic lattice effects due to the finite cutoff and finite volume. To close this section, we consider a way of controlling such lattice effects in the flux-tube solution, and propose optimal sets of parameters for $\kappa=0.50$ (type I), $1.0$ (border), and $1.5$ (type II). 
 
\begin{figure}[!t] 
\centering\includegraphics[width=\figonecolumn]{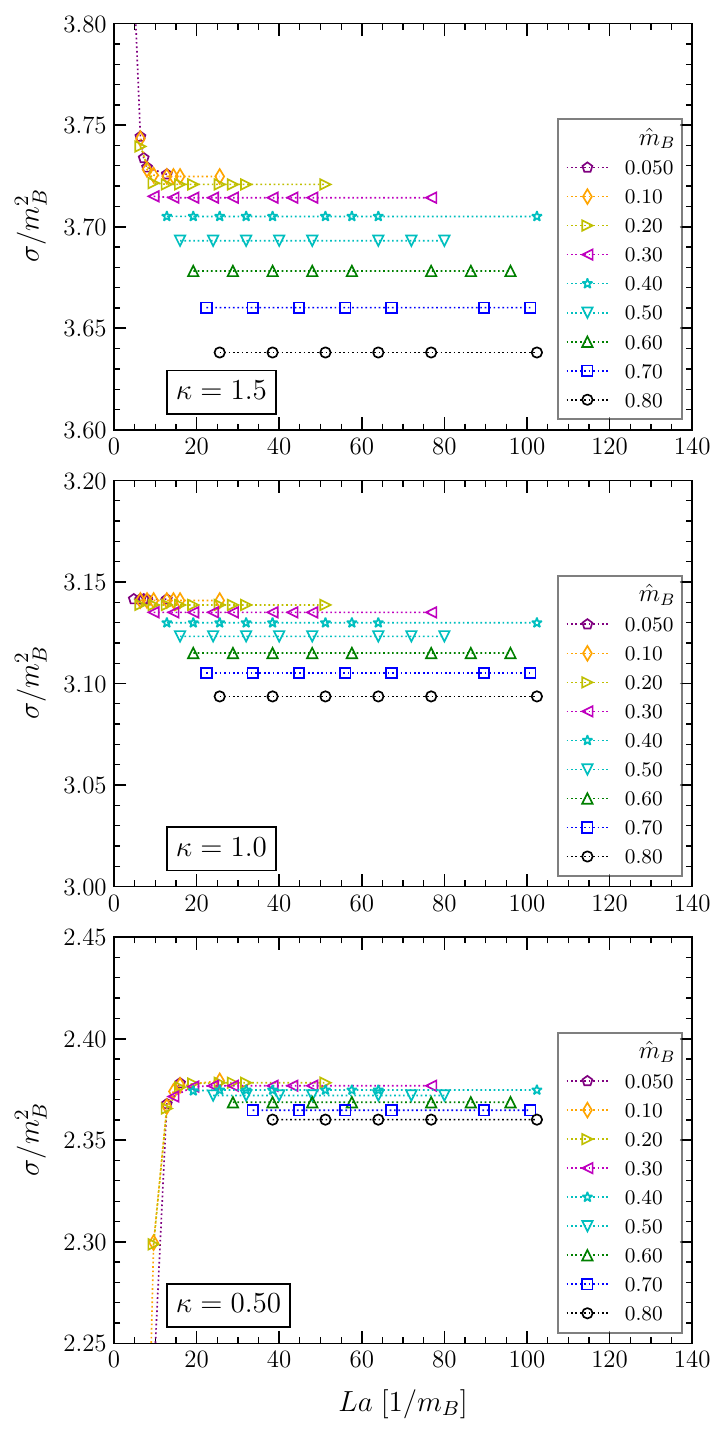} 
\caption{The $L$-dependence on the string tension $\sigma$ for $ \kappa=0.50$ (lower), 1.0 (middle), and 1.5 (upper) with various $\hat{m}_{B}$. The span of the vertical axis for the three figures is taken to be the same as $\Delta \sigma/m_{B}^{2}=0.20$ to observe how the string tension is sensitive to $\hat{m}_{B}$ among three $\kappa$ values.} 
\label{fig:sigma-ldep} 
\end{figure} 
\begin{figure}[!t] 
\centering\includegraphics[width=\figonecolumn]{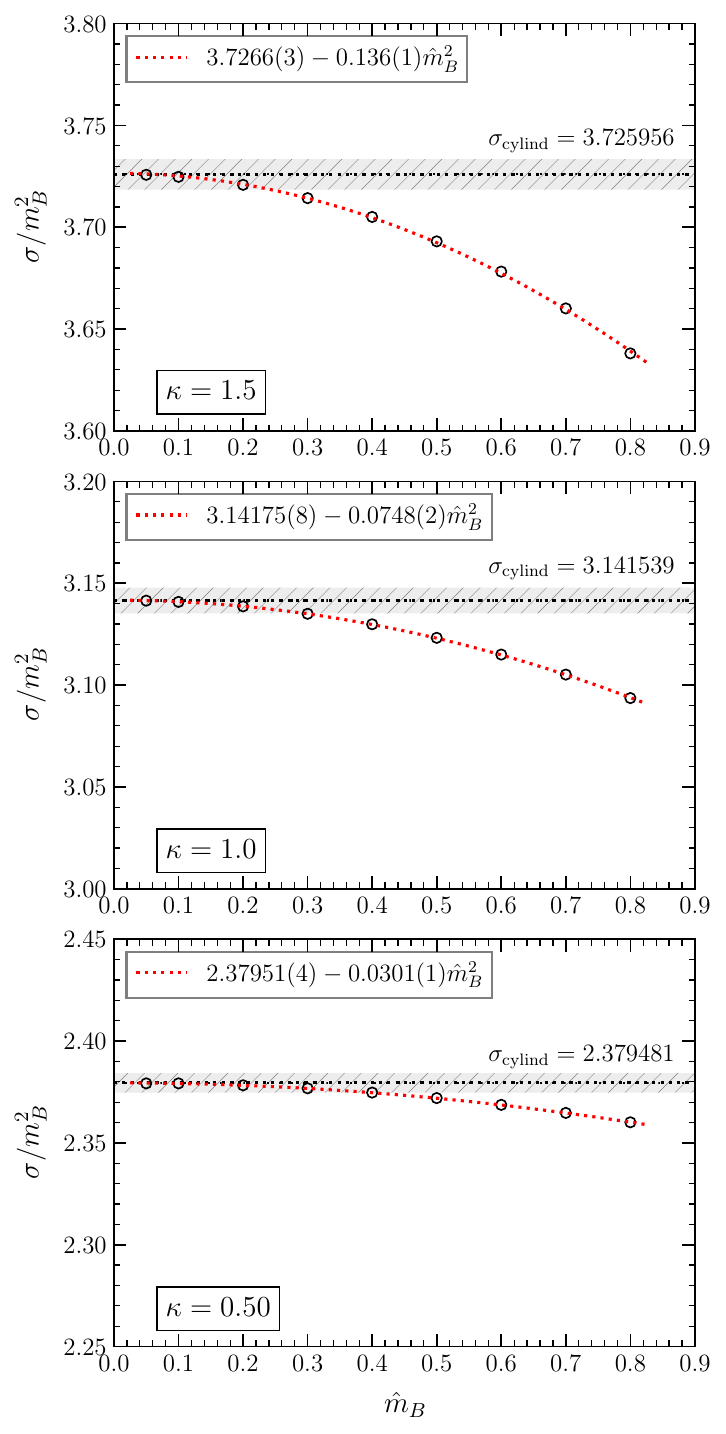} 
\caption{The $\hat{m}_{B}$-dependence on the string tension $\sigma$ (after taking $L\to \infty$ limit) for $\kappa=0.50$ (lower), 1.0 (middle), and 1.5 (upper) with the quadratic fitting functions. For a given $\hat{m}_{B}$ the lattice spacing corresponds to $a=\hat{m}_{B}\;\mathrm{[1/m_{B}]}$. The shaded regions represent the detailed value of $\sigma_{\rm cylind}$ in Table~\ref{tbl:comparison1D2D} within 0.2 percent of relative error.} 
\label{fig:sigma-adep} 
\end{figure} 
 
\par 
As discussed in Sec.~\ref{sec:dgl-cylind-numerical}, the physical scale of a dimensionless distance can be recovered by multiplying it by $a=\hat{m}_{B}/m_{B}$ in Eq.~\eqref{eqn:physical-lat-a}, which means that any physical distances can be expressed in units of the physical penetration depth $1/m_{B}$. Then, the lattice cutoff effect can be investigated by changing $\hat{m}_{B}$ by assuming the existence of a unique physical value of $m_{B}$, and the finite volume effect by changing the lattice size $L_{\mu}$, where the physical size is given by $L_{\mu} a =L_{\mu} \hat{m}_{B}\;\mathrm{[1/m_{B}]}$. As the flux-tube solution consists of the Coulombic and the solenoidal parts as in Eqs.~\eqref{eqn:action-coulomb} and ~\eqref{eqn:action-solenoidal}, it will be useful to look at the lattice effect separately. 
 
\par 
First, we look at the Coulombic part of the potential in Eq.~\eqref{eqn:action-coulomb}, which is expressed by using the massless lattice Green function $G_{L}$ in a finite volume up to the overall factor $\beta_{g}$ and the electric charge $N_{q}$ as in Eq.~\eqref{eqn:lattice-coulombic-pot}. Thus, the lattice effect for this contribution can be investigated by looking at how $V_{\mathrm{coul}}(R)$ is far from the continuum Coulombic potential in infinite volume, $V_{\mathrm{coul}}^{\mathrm{cont}}(R) =- \pi \beta_{g} N_{q}^{2}/R +const.$. However, one cannot directly compare $V_{\mathrm{coul}}(R)$ and $V_{\rm coul}^{\mathrm{cont}}(R)$ due to the presence of the unspecified constant term from the self-energy. Thus, we do this comparison with the help of the massless lattice Green function in infinite volume $G_{\infty}$, which has the value $G_{\infty}(0)=0.25273100985866$ at the origin~\cite{Necco:2001xg} and is known to behave as the continuum function $G_{\infty}(R) \sim 1/(4\pi R)$ at large $R$. As the relation $G_{L}(0) - G_{L}(R) \simeq G_{\infty}(0) -G_{\infty}(R)$ is held for large lattice volume, it makes sense to compare $G_{L}(0) - G_{L}(R)$ with $G_{\infty}(0) -1/(4\pi R)$, which requires no constant shift. 
 
\par 
In the upper panel in Fig.~\ref{fig:pot_coul_vdep}, we plot the Coulombic part of the potential in Eq.~\eqref{eqn:lattice-coulombic-pot} for various cubic lattice volumes $L^{3}$ with $L=16$, $32$, $64$, $\infty$, where $N_{q}=1$, which are compared with that of its continuum limit counterpart 
\be 
V_{\mathrm{coul}}^{\mathrm{cont}}(R) =4 \pi^{2} \beta_{g} N_{q}^{2} (G_{\infty}(0)-\frac{1}{4\pi R})\;. 
\label{eqn:lattice-coulombic-pot-cont} 
\ee 
The lower panel in Fig.~\ref{fig:pot_coul_vdep} is the difference between $V_{\mathrm{coul}}^{\mathrm{cont}}$ and $V_{\mathrm{coul}}$, that is, $\delta V_{\mathrm{coul}} =V_{\mathrm{coul}}^{\mathrm{cont}}-V_{\mathrm{coul}}$. These figures indicate that the finite volume effect for larger $R$ can be cured simply by increasing the lattice size $L$, while the short distance behavior, especially for $R < 5$, never coincides with the continuum one. Based on a fitting analysis, the characteristic difference at short distances can be quantified approximately by an exponential function, $\delta V_{\mathrm{coul}} =0.60(1) \exp ( - 0.84(2)R )$. To reconcile this short-distance lattice effect with the continuum theory, we may redefine the distance $R$ by $\tilde{R}\equiv (4 \pi G_{\infty}(R))^{-1}$ as often used in the lattice QCD simulations with the name of tree-level improvement~\cite{Necco:2001xg}. Note that $\tilde{R} \equiv (4 \pi (G_{L}(R)-G_{L}(0)+G_{\infty}(0)))^{-1}$ can also be applied if $R \ll L$. 
 
\par 
Second, we look at the solenoidal part of the potential in Eq.~\eqref{eqn:action-solenoidal}, which is responsible for the string tension of the flux tube. Thus, the main lattice effects in this part can be investigated by computing the infinitely long flux-tube solution without terminals. This is possible by putting the Dirac string along the whole $x_{3}$-axis, which is nothing but the $D=2$ dimensional system. Hence, it is economical to perform the computation on the $D=2$ dual lattice. In $D=2$ dimensions, the converged value of $S$ in Eq.~\eqref{eqn:action} directly corresponds to the string tension, $S \to \sigma$. The convergence is faster than the $D=3$ case, and it is also simple to put the Dirac string; we just assign $\Sigma_{12}(x) = - N_{q}\delta_{x0}$ for a single flux tube at $x=0$. 
 
\par 
In Fig.~\ref{fig:sigma-ldep}, we show the $L$-dependence of the string tension $\sigma$ for $\kappa=0.50, 1.0, 1.5$ on the squared lattice of the size $L^{2}$, where various values of $\hat{m}_{B}$, that is, various lattice spacings are examined. The convergence criterion is set to be $\epsilon =10^{-4}$ here, which is already small enough to guarantee the convergence of $\sigma$. As a common feature for all values of $\kappa$, we find that $\sigma$ becomes constant at large $L$ for each value of $\hat{m}_{B}$. At small $L$ we observe that $\sigma$ shifts {\em upward} for $\kappa >1$ and {\em downward} for $\kappa <1$, while it remains {\em flat} for $\kappa=1$. As we will clarify later in Sec.~\ref{subsect:fluxtube-interaction}, this tendency reflects the effect of interaction among the flux tubes: attractive for $\kappa <1$, no interaction for $\kappa =1$, repulsive for $\kappa >1$. 
 
\begin{table}[!t] 
\caption{The continuum and large volume limits of the normalized string tension $\sigma/m_{B}^{2}$ on the $D=2$ dimensional lattice with $N_{q}=1$ for three $\kappa$ values. The physical size corresponds to $L a =25.6 \;\mathrm{[1/m_{B}]}$. They are compared to the string tension of the cylindrical flux-tube solution, $\sigma_{\rm cylind}$, computed by using the method described in Sec.~\ref{sec:dgl}, where we set $\rho_{\rm max} a =40 \;\mathrm{[1/m_{B}]}$ and $\hat{m}_{B}=0.01$ ($n_{\rm max} =4000$) with $\epsilon =10^{-9}$.} 
\label{tbl:comparison1D2D} 
\begin{tabular}{lcll} 
\hline\hline 
$\kappa$ & $\sigma /m_{B}^{2}$ & $\sigma_{\rm cylind}/m_{B}^{2}$ \\ 
\hline 
$0.50 $ &2.37951(4) & 2.379481 \\ $1.0 $ & 3.14175(8)& 3.141539 \\ $1.5$ &3.7266(3) & 3.725956 \\ 
\hline\hline 
\end{tabular} 
\caption{The summary of the GL parameter $\kappa = \hat{m}_{\chi}/\hat{m}_{B}$, the mass parameters $\hat{m}_{B}$ and $\hat{m}_{\chi}$, and the lattice volume with $La = 128 \cdot 0.20/m_{B}=25.6 \;\mathrm{[1/m_{B}]}$ and $L_{3}a = 256 \cdot 0.20/m_{B}= 51.2 \;\mathrm{[1/m_{B}]}$.} 
\begin{tabular}{cccc} 
\hline 
\hline 
$\kappa$ & $\hat{m}_{B}$ & $\hat{m}_{\chi}$ & $L^{2} L_{3}$ \\ 
\hline 
0.50 & 0.20 & 0.10& $128^{2} 256$ \\ 1.0 &0.20 & 0.20 & $128^{2} 256$\\ 1.5 & 0.20 & 0.30 &$128^{2} 256$\\ 
\hline 
\hline 
\end{tabular} 
\label{tbl:massparameters} 
\end{table} 
 
\par 
We then extract $\sigma$ for each $\hat{m}_{B}$ in the large $L$ limit, and plot them as a function of $\hat{m}_{B}$ in Fig.~\ref{fig:sigma-adep}, which are compared to the string tension of the cylindrically symmetric flux-tube solution $\sigma_{\mathrm{cylind}}$ (see, Table~\protect{\ref{tbl:comparison1D2D}}). We find that $\sigma$ approaches $\sigma_{\rm cylind}$ as $\hat{m}_{B} \to 0$, where the $\hat{m}_{B}$-dependence of $\sigma$ is well described by the quadratic functions. If $\hat{m}_{B}<0.20$, it is possible to obtain the same string tension as $\sigma_{\rm cylind}$ within 0.2 percent of accuracy, indicating that the rotational symmetry around $\Sigma_{12}$ is recovered properly. 
 
\par 
From these experiences, we conclude that the lattice effect can be controlled by the following method; one investigates the lattice volume dependence of the observable for various $\hat{m}_{B}$ and finds the $L$-independent results at first, and extracts the value in the continuum limit from them by taking $\hat{m}_{B} \to 0$ next. As a common problem in the lattice formulation, it takes much computation time for larger lattice volumes with smaller lattice spacings. Thus, it is economical to choose the mass parameters depending on the desired accuracy; if one requires about 0.2 percent of accuracy for the wide range of the potential, it may be necessary to take $\hat{m}_{B} \lesssim 0.2$ and $L \gtrsim O(10^{2})$. This is how we arrive at the lattice setting and mass parameters as in Table~\ref{tbl:massparameters}. As for the convergence condition $\epsilon$, we find that if only the potential energy and the field profiles are of interest, it seems to suffice to take $\epsilon = 10^{-4}$, while if the \textit{derivative} (corresponding to the difference on the lattice) of the potential is of interest, it should be $\epsilon \leq 10^{-6}$. Thus, in most cases of the following computation, we adopt the convergence criterion $\epsilon = 10^{-6}$.

\section{Numerical results} 
\label{sec:results} 
 
\begin{figure*}[t] 
\centering\includegraphics[width=\figtwocolumn]{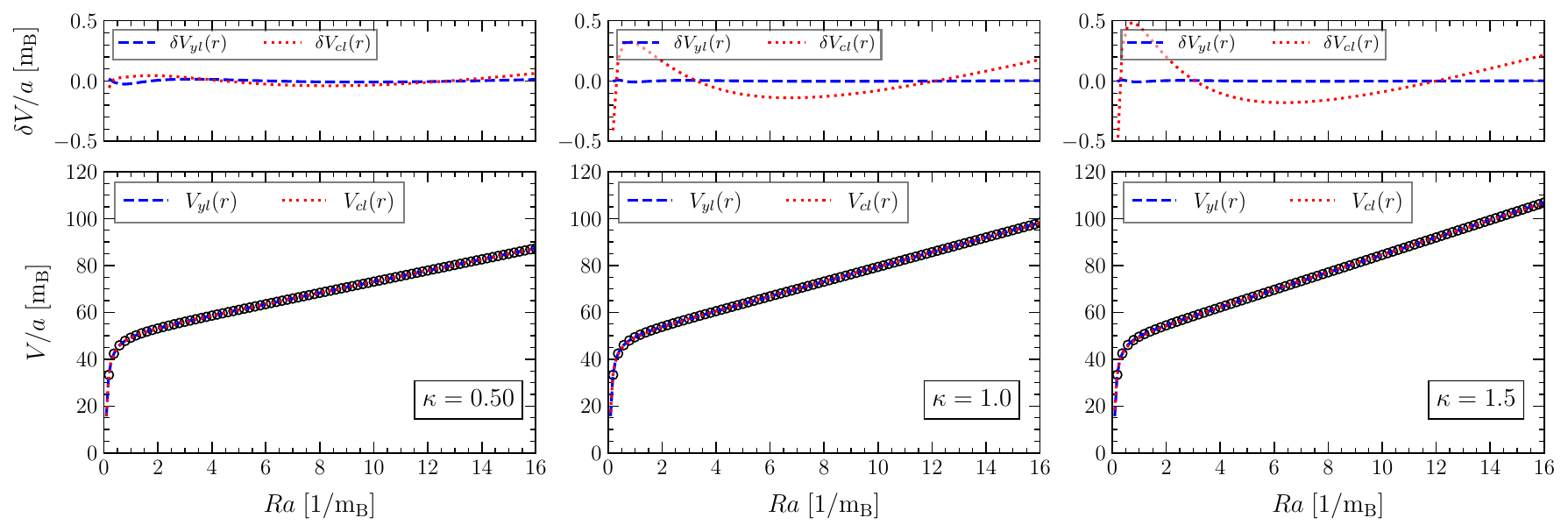} 
\caption{The interquark potential $V(r=Ra)/a$ with the fitting functions $V_{yl}(r)$ in Eq.~\eqref{eqn:fit-yl} and $V_{cl}(r)$ in Eq.~\eqref{eqn:fit-cl} (lower three panels) and the fitting residue (upper three panels) for $\kappa =0.50, 1.0, 1.5$ (from left to right). The maximum value of $Ra =16 \;\mathrm{[1/m_{B}]}$ corresponds to $R=80$, which is about $1/3$ of $L_{3}=256$.} 
\label{fig:result-potential} 
\end{figure*} 

Based on the optimal parameter set in Table~\ref{tbl:massparameters}, we compute the interquark potential from the finite-length flux-tube solution, and analyze its functional form in detail. Using also the twice coarse-grained parameter set of Table~\ref{tbl:massparameters}, we investigate the width of the finite-length flux-tube, and the interaction properties of flux tubes in the two-flux-tube and multiflux-tube systems. 
 
\subsection{The interquark potential from the finite-length flux tube} 

We compute the interquark potential for the three values of the GL parameters $\kappa=0.50, 1.0, 1.5$ as in Table~\ref{tbl:massparameters} on the $L^{2}L_{3}=128^{2}256$ lattice. The length of the flux tube is varied from $R=1$ to $80$, where $R=80$ is about $1/3$ of $L_{3}=256$. We keep the location of the quark at the origin and shift that of the antiquark along the $x_{3}$-axis. The initial field variables for the $R=1$ case are set as trivial such that $B_{\mu}=0$, $\chi^{1}=1$, and $\chi^{2}=0$. For computing the $R \geq 2$ cases, we use the solution of the $R-1$ case, which helps to reduce the number of iteration steps toward convergence as it is closer to the final solution than the trivial setting. 
 
\begin{figure*}[!t] 
\centering\includegraphics[width=\figtwocolumn]{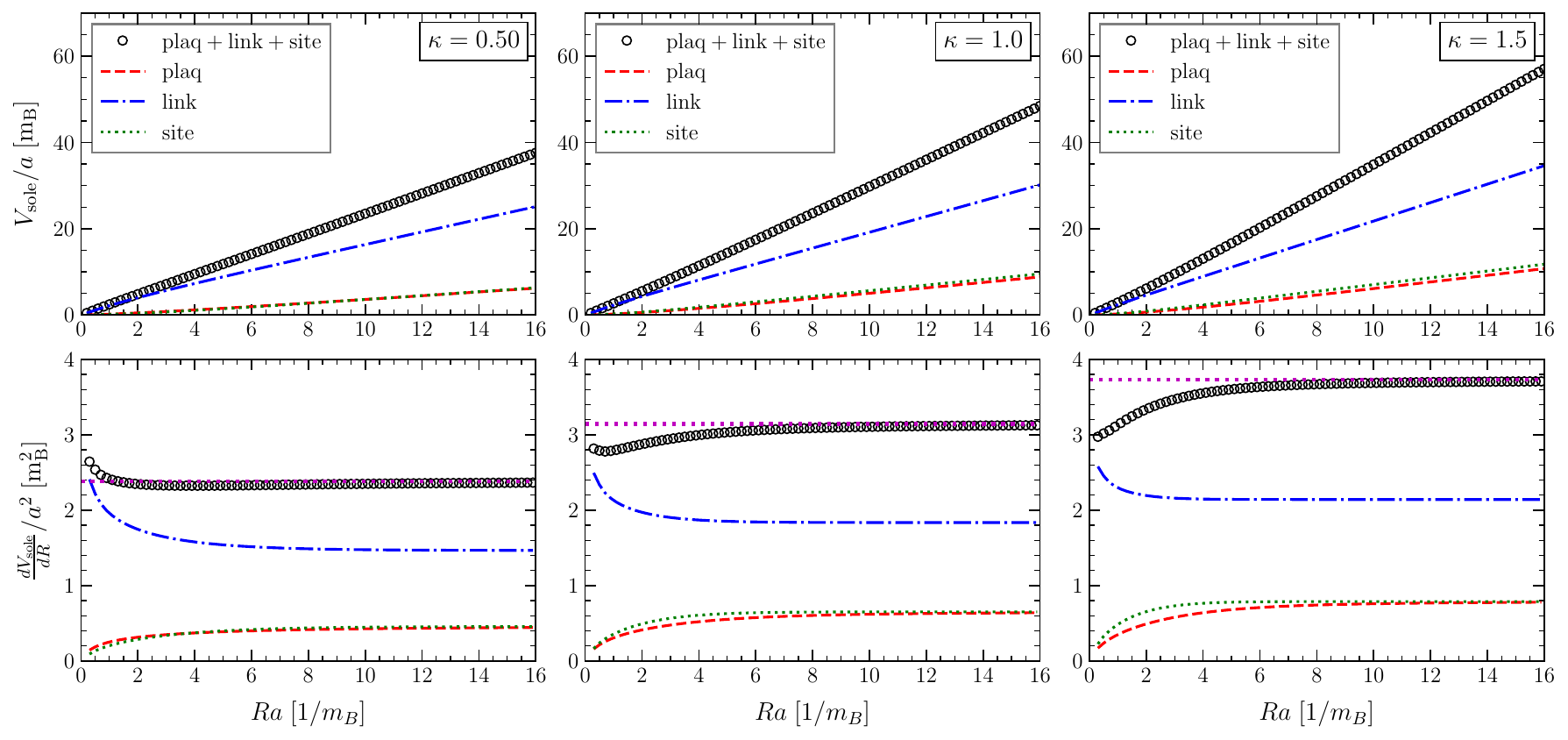} 
\caption{The solenoidal part of the interquark potential $V_{\mathrm{sole}}(r=Ra)/a$ (upper three panels) and its derivative with respect to the distance $(dV_{\mathrm{sole}}/dR)/ a^{2}$ (lower three panels) for $\kappa=0.50, 1.0, 1.5$ (from left to right). The horizontal dotted lines in the lower panels denote the values of the string tension $\sigma/m_{B}^{2}=2.37951, 3.14175, 3.7266$ for $\kappa=0.50,1.0,1.5$, computed in the $D=2$ dimensions (see, Fig.~\ref{fig:sigma-adep} and Table~\ref{tbl:comparison1D2D}). The definitions of ``plaq,'' ``link,'' and ``site'' are given in Eqs.~\eqref{eqn:sole-p},~\eqref{eqn:sole-l}, and \eqref{eqn:sole-s}, respectively.} 
\label{fig:pot-sole} 
\end{figure*} 
 
\par 
In Fig.~\ref{fig:result-potential}, we plot the potential with the physical scale $V/a \;\mathrm{[m_{B}]}$ as a function of the physical distance, $r \equiv Ra \;\mathrm{[1/m_{B}]}$, where two functional forms are examined. One is the Yukawa plus linear function 
\be 
V_{yl}(r)=-\frac{A_{y}}{r} e^{-m_{y} r} +\sigma_{y} r +\mu_{y} \;, 
\label{eqn:fit-yl} 
\ee 
which is motivated by the theoretical expectation of the superconductor in the type-II limit as in Ref.~\cite{Nambu:1974zg, Suganuma:1993ps}. The parameters $A_{y}$, $\sigma_{y}$, $\mu_{y}$ denote a constant to control the strength of the Yukawa interaction of the mass $m_{y}$, the string tension, and a cutoff-dependent constant with mass dimension, respectively. The other is the Coulomb plus linear function, as often used in the lattice QCD analyses, 
\be 
V_{cl}(r) =-\frac{A_{c}}{r} + \sigma_{c} r + \mu_{c} \; , 
\label{eqn:fit-cl} 
\ee 
where the parameters $A_{c}$, $\sigma_{c}$, $\mu_{c}$ have similar meaning as above, but when the Coulombic interaction is assumed at short distances. To use these continuum functional forms, we apply the tree-level improvement to the lattice distance as discussed in Sec.~\ref{subsec:result-latticeeffects}, with the massless lattice Green functions in infinite volume. We take into account all data (80 data points for each $\kappa$) in the fitting, and summarize the result of the fitting parameters in Tables~\ref{tbl:potfit-yl} and \ref{tbl:potfit-cl}. 
 
\par 
We find that both functional forms $V_{yl}$ and $V_{cl}$ seem to explain the behavior of the potential quite well for all $\kappa$ cases, only if the fitting curves are looked at. However, the residues, $\delta V_{yl}\equiv V-V_{yl}$ and $\delta V_{cl}\equiv V-V_{cl}$, shown in the panels above the potential plots, indicate that $V_{yl}$ is clearly better than $V_{cl}$, especially for the cases $\kappa \geq 1$. In fact, $\sigma_{y}$ for all three $\kappa$ values are consistent with the string tensions computed directly in the $D=2$ setting, and $A_{y}$ is stable compared to $A_{c}$. 
 
\par 
It may be interesting to notice that the masslike quantity $m_{y}$ in $V_{yl}$, which we call the effective mass, is not common but dependent on $\kappa$, which is monotonically increasing with the growth of $\kappa$. Since the distance $r$ is measured in units of $1/m_{B}$, $m_{y}$ itself corresponds to the effective ratio of the interaction range to $1/m_{B}$. Hence, if the value is smaller (larger) than one, it means that the effective interaction range of the dual gauge field is longer (shorter) than $1/m_{B}$. We will consider the meaning of this tendency later. 
 
\begin{table} 
\caption{The fit results of the interquark potential in Fig.~\ref{fig:result-potential} to $V_{yl}=-A_{y} e^{-m_{y} r}/r + \sigma_{y} r + \mu_{y}$ in Eq.~\eqref{eqn:fit-yl}. The errors of the fit parameters are estimated by assuming $\delta \chi^{2}/N_{\rm df}=1$. The potentials for $\kappa \geq 2$ are not presented explicitly in this paper, as their behaviors are qualitatively similar to that for $\kappa = 1.5$. The potential in the type-II limit $\kappa \to \infty$ is shown in Appendix~\ref{sec:app-london} (* is for the case $\hat{m}_{B}=0.10$ to examine the ultraviolet cutoff effect).} 
\begin{tabular}{lccccc} 
\hline\hline 
$\kappa$ & $A_{y}$ & $m_{y}$ & $\sigma_{y}$ & $\mu_{y}$ & fit range $R$\\ 
\hline 
0.50 &3.190(2) &0.229(4) & 2.3716(6) & 49.393(8)&1--80 \\ 1.0 &3.1678(1) & 0.629(1)& 3.1376(1) & 48.000(1)&1--80\\ 1.5 & 3.1687(9) & 0.852(1) & 3.71998(8)& 47.2876(9)&1--80\\ 2.0 &3.1700(9) &1.005(1) &4.20273(7) & 46.8059(7)&1--80\\ 3.0 &3.169(1)&1.201(1) &4.98335(7) & 46.1646(7)&1--80\\ 4.0& 3.164(2) &1.321(2)& 5.60424(9) & 45.7426(9)&1--80\\ $\infty$ & 3.084(6) & 1.668(9) & 9.5344(3) & 43.987(3) &1--80\\ $\infty$* & 3.081(4) & 1.66(1) & 11.686(2) & 93.651(8) &1--80\\ 
\hline\hline 
\end{tabular} 
\label{tbl:potfit-yl} 
\end{table} 
 
\begin{table} 
\caption{The same as in Table~\ref{tbl:potfit-yl} but to $V_{cl}=-A_{c} /r + \sigma_{c} r + \mu_{c}$ in Eq.~\eqref{eqn:fit-cl}. The fit parameters for the $\kappa = \infty$ case are not included as the fitting residue got even worse than the other cases. } 
\begin{tabular}{lcccc} 
\hline\hline 
$\kappa$ & $A_{c} $ & $\sigma_{c} $& $\mu_{c} $ & fit range $R$\\ 
\hline 
0.50 & 3.151(6) & 2.3439(9) & 49.977(9) &1--80 \\ 1.0 &2.92(3)& 3.077(4)& 48.97(4) &1--80 \\ 1.5 & 2.79(3)& 3.652(5)& 48.33(5) &1--80 \\ 2.0&2.70(4)& 4.132(6)&47.88(6) &1--80 \\ 3.0&2.58(5)& 4.910(7)&47.25(7) &1--80 \\ 4.0& 2.51(5)& 5.531(7) &46.83(7) &1--80 \\ 
\hline\hline 
\end{tabular} 
\label{tbl:potfit-cl} 
\end{table} 
 
\par 
By applying the Hodge decomposition, the potential can be decomposed as $V=V_\mathrm{coul}+V_\mathrm{sole}$. The functional form of $V_\mathrm{coul}$ is exactly described by using the massless lattice Green function in Eq.~\eqref{eqn:lattice-coulombic-pot} on the $128^{2}256$ lattice.\footnote{For instance, $G_{L}(0,0,0)=0.25160834912$, $G_{L}(0,0,1)=0.08494163028$, $G_{L}(0,0,2)=0.04176644516$, on the $L^{2}L_{z}=128^{2}256$ lattice.} In other words, the values of $V_\mathrm{coul}$ are perfectly under control. On the other hand, the functional form of $V_\mathrm{sole}$ is slightly nontrivial, which cannot be exactly linear, as the fitting results show that $V_{yl}$ is favorable than $V_{cl}$; $V_\mathrm{sole}$ should contain nontrivial $r$-dependence at short distances that modifies the pure Coulombic term to the Yukawa term with the effective mass $m_{y}$. 
 
\par 
To investigate the detailed feature of $V_\mathrm{sole}$, we further decompose it into three terms as $V_\mathrm{sole}=V_\mathrm{sole}^{\mathrm{(plaq)}}+V_\mathrm{sole}^{\mathrm{(link)}}+V_\mathrm{sole}^{\mathrm{(site)}}$ with 
\bea 
V_\mathrm{sole}^{\mathrm{(plaq)}}&\equiv & 
\beta_{g} \sum_{x}\Biggl [\frac{1}{4} [ (\del \wedge B^{\rm reg})_{\mu\nu}(x)]^{2} 
\Biggr ] \; , 
\label{eqn:sole-p}\\ 
V_\mathrm{sole}^{\mathrm{(link)}}&\equiv & 
\beta_{g} \sum_{x}\Biggl [ 
\frac{\hat{m}_B^2}{2} (D_{\mu}\chi^{\alpha}(x) )^{2} 
\Biggr ] \; , 
\label{eqn:sole-l}\\ 
V_\mathrm{sole}^{\mathrm{(site)}} &\equiv & 
\beta_{g} \sum_{x}\Biggl [ 
\frac{\hat{m}_B^2 \hat{m}_\chi^2}{8} \left ( \chi^{\alpha}(x)^2 - 1 \right )^2 
\Biggr ] \; , 
\label{eqn:sole-s} 
\eea 
where the additional superscripts of $V_\mathrm{sole}$ are just named after the geometric shapes of the contribution to the potential, from plaquettes, links, and sites on the dual lattice. The results are shown in Fig.~\ref{fig:pot-sole}, where the derivatives of these terms with respect to $r$ are also computed. 
 
\par 
Let us first look at the result for $\kappa=0.50$, the left panels of Fig.~\ref{fig:pot-sole}. We find that all the contributions $V_\mathrm{sole}^{\mathrm{(plaq)}}$, $V_\mathrm{sole}^{\mathrm{(link)}}$, $V_\mathrm{sole}^{\mathrm{(site)}}$ seem to behave linearly, where the largest contribution is of $V_\mathrm{sole}^{\mathrm{(link)}}$, and the next is provided by $V_\mathrm{sole}^{\mathrm{(plaq)}}$ and $V_\mathrm{sole}^{\mathrm{(site)}}$ almost identically. Although each component looks linear, its derivative exhibits nontrivial behavior as $r \to 0$. Typically, the slopes of $V_\mathrm{sole}^{\mathrm{(plaq)}}$ and $V_\mathrm{sole}^{\mathrm{(site)}}$ approach smoothly zero for $r \to 0$. In contrast, the slope of $V_\mathrm{sole}^{\mathrm{(link)}}$ increases as $r \to 0$ up to a certain finite value, which gives a dominant contribution to $V_{\rm sole}$ in this limit. The slope of $V_{\rm sole}$ for large $r$ approaches the string tension $\sigma/m_{B}^{2}=2.37951$ obtained on the $L^{2}=128^{2}$ lattice. The contribution to the string tension from each term for larger $r$ is, $\mathrm{plaq:link:site}=0.19: 0.62: 0.19\sim 1:3:1$. 
 
\par 
We then look at the results for $\kappa=1.0$ and $1.5$, the middle and right panels in Fig.~\ref{fig:pot-sole}. We find that the slope of each component becomes relatively steeper as $\kappa$ increases, which finally leads to the larger string tension. The qualitative behavior of the derivatives is similar to that of $\kappa=0.50$, but as $\kappa$ increases, the difference between $V_\mathrm{sole}^{\mathrm{(plaq)}}$ and $V_\mathrm{sole}^{\mathrm{(site)}}$ becomes noticeable for smaller $r$. It is interesting to note that the ratio of the contribution to the string tension of each term is almost the same as that for $\kappa=0.50$. We find the ratio is given by $\mathrm{plaq:link:site}=0.20: 0.59: 0.21$ for $\kappa=1.0$, and $0.21: 0.58: 0.21$ for $\kappa=1.5$, both approximately $\sim 1:3:1$. 
 
\par 
It may be remarkable that all the values of $\lim_{r\to 0}V_{\rm sole}$ for three $\kappa$ seem to be around $\pi$ due to the contribution of $V_\mathrm{sole}^{\mathrm{(link)}}$, implying a topological origin rather than the dynamical effects of interaction. In fact, we observed that the value becomes just twice for the $N_{q}=2$ system. This means that the effect of the quantized Dirac string is present even at short distances in the DGL theory, though its effect on the flux-tube dynamics is probably hidden by the contribution of $V_\mathrm{coul}$. 
 
\begin{figure}[!t] 
\centering\includegraphics[width=\figonecolumn]{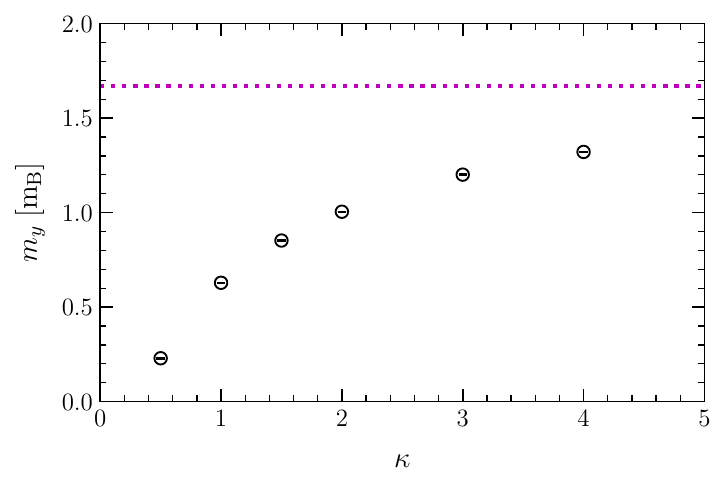} 
\caption{The $\kappa$-dependence of the effective mass $m_{y}$ in the fitting function $V_{yl}$ in Eq.~\eqref{eqn:fit-yl}. The data are from Table~\ref{tbl:potfit-yl}, where the horizontal dotted line corresponds to $m_{y} =1.668\;\mathrm{[m_{B}]}$ from the potential in the type-II limit.} 
\label{fig:kappa_fitmass} 
\end{figure} 
 
\par 
Finally, we make some comments on the behavior of the effective mass $m_{y}$ in $V_{yl}$. According to \cite{Nambu:1974zg, Suganuma:1993ps}, $m_{y}$ is expected to approach $m_{y} \to 1$ in the type-II limit, which is based on an ansatz such that the monopole field prefers to take the value of condensate $v$ everywhere in the type-II limit. Then, the dual gauge field acquires the mass $m_{B}$ regardless of the direction of propagation, which results in the interaction to be $\propto -e^{-m_{B}r}/r$. In fact, our fitting result on $m_{y}$ as summarized in Table~\ref{tbl:potfit-yl} for three $\kappa$ values seems to support this expectation; $m_{y}$ tends to increase monotonically from the value $m_{y} <1$ against $\kappa$. However, our further investigation with larger $\kappa$ values such as $\kappa=2.0,~3.0,~4.0$, and the computation in the type-II limit provides us with a different view as shown in Fig.~\ref{fig:kappa_fitmass}, where the fitting results are also included in Table~\ref{tbl:potfit-yl}. The result exhibits that $m_{y}$ can be larger than one beyond $\kappa=2.0$, and approaches a constant value around $m_{y} = 1.668\;\mathrm{[m_{B}]}$. This indicates that the suppression effect of the interaction at short distances in the type-II limit is stronger than previously expected. It may be interesting to explore theoretically the origin of this behavior~\cite{Koma:2001pz}. 
 
\subsection{The width of the finite-length flux tube} 

As mentioned in the Introduction, the width of the flux-tube profiles, especially for the quark-antiquark system, is often investigated by lattice QCD simulations to identify the type of dual superconducting phase~\cite{Singh:1993jj, Cea:1995zt, Bali:1997cp, Gubarev:1999yp, Koma:2003gq, Koma:2003hv, Haymaker:2005py, Chernodub:2005gz, DAlessandro:2006hfn, Sekido:2007mp, Suzuki:2007jp, Suzuki:2009xy, Cea:2012qw, Shibata:2012ae, Kato:2014nka, Cea:2014uja, Nishino:2019bzb, Battelli:2019lkz}. However, most previous studies seem to assume translational invariance of the flux-tube profile along the quark-antiquark axis. We focus on whether such an assumption is valid or not for a finite-length flux tube. We compute the profile of the electric field, the monopole supercurrent, and the action density, and investigate how they behave as a function of the flux-tube length. 
 
\begin{figure*} 
\centering\includegraphics[width=\figtwocolumn]{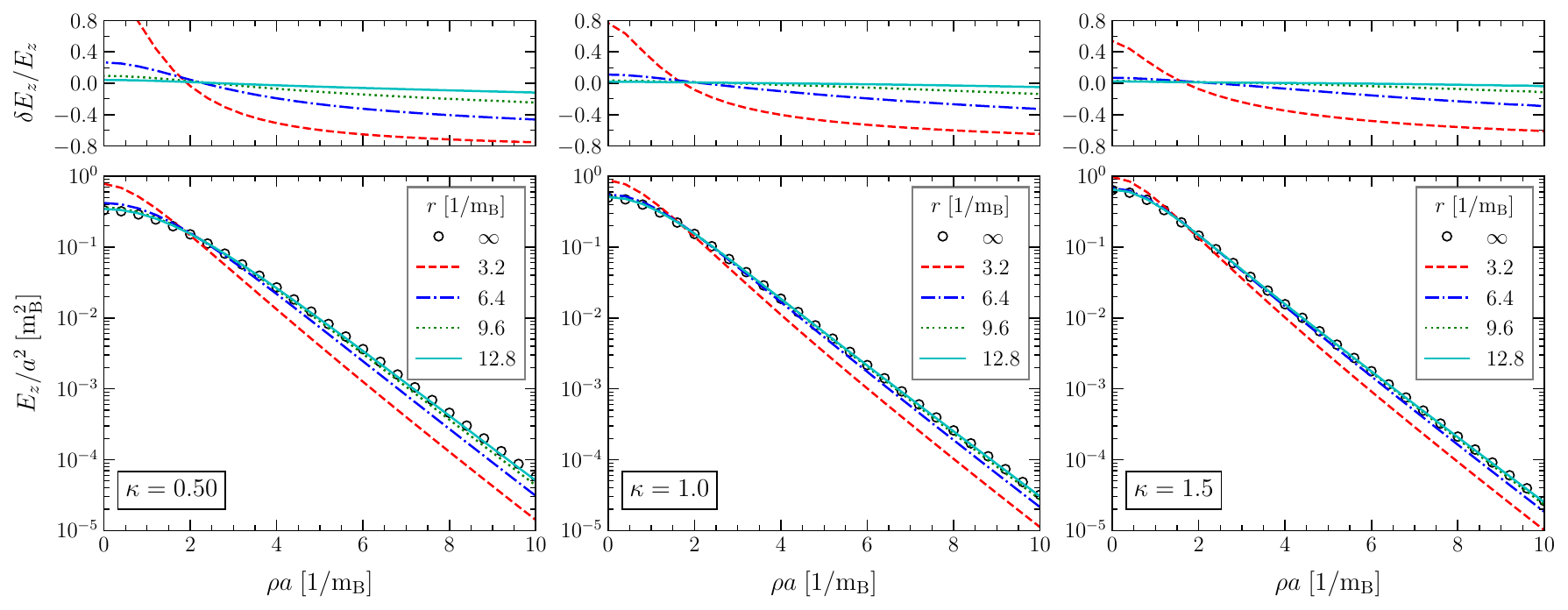} 
\caption{The profile of the electric field on the midtransverse plane of the finite-length flux tube in comparison with the $D=2$ profile denoted as $r=\infty$ (lower) and the relative error from the $D=2$ profile (upper) as a function of the distance from the flux-tube center $\rho$ for $\kappa=0.50,1.0,1.5$ (from left to right). The lattice size is $L^{2}L_{3}=64^{2} 128$ and the flux-tube lengths are $R=8,16,24,32$, which correspond to $r\equiv Ra=3.2,6.4,9.6,12.8\;\mathrm{[1/m_{B}]}$ for $\hat{m}_{B}=0.40$.} 
\label{fig:profile_ez_lx64_lz128} 
\end{figure*} 
 
\begin{figure*} 
\centering\includegraphics[width=\figtwocolumn]{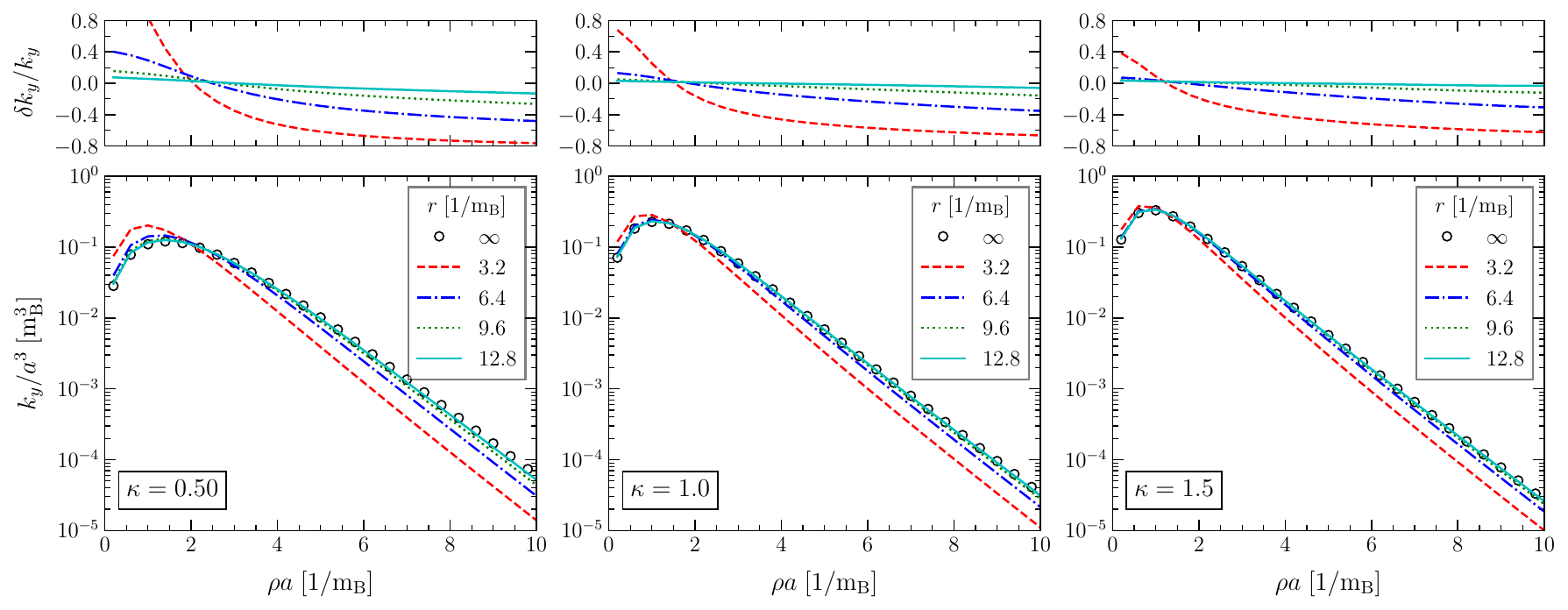} 
\caption{The similar plot as in Fig.~\ref{fig:profile_ez_lx64_lz128} for the monopole supercurrent.} \label{fig:profile_ky_lx64_lz128} 
\end{figure*} 
 
\par 
For economical reasons of visualization of the field profiles, we use twice the coarser lattice than that used in the potential study; we set $\hat{m}_{B}=0.40$ ($N_{q}=1$, and $\epsilon =10^{-6}$) and $L^{2}L_{3}=64^{2} 128$. This coarse-graining may affect the quantitative argument; however, the effect on the string tension of a single flux tube is at most 0.6 percent for $\kappa=1.5$ according to Fig.~\ref{fig:sigma-adep}, and is further smaller for $\kappa=0.50$ and $\kappa=1.0$. We compute the profiles of flux tubes with the length $R=8,16,24,32$ along the $x_{3}$-axis, and compare them with the $R=\infty$ profiles computed by the $D=2$ setting on the $L^{2}=64^{2}$ lattice. To provide a quantitative sense, we introduce a physical scale with respect to the penetration depth $1/m_{B}$, hence, the lengths $R=8,16,24,32$ correspond to $r \equiv Ra = 3.2, 6.4, 9.6,12.8\;\mathrm{[1/m_{B}]}$ for $\hat{m}_{B}=0.40$. Just like the potential study, we also examine the $\kappa$-dependence by setting $\hat{m}_{\chi}=0.20, 0.40, 0.60$, corresponding to $\kappa=0.50,1.0,1.5$. 
 
\par 
In Figs.~\ref{fig:profile_ez_lx64_lz128} and \ref{fig:profile_ky_lx64_lz128}, we show the results of the electric field and the monopole supercurrent with the semilog scale, where each field is normalized by $a^{2}$ and $a^{3}$, respectively, and the profiles along the $x_{1}$-axis on the midtransverse plane are plotted as a function of the distance from the flux-tube center using the radial coordinate $\rho$ as in Fig.~\ref{fig:fieldprofile}. As a common feature of the profiles for all $\kappa$ values, the width of the flux tube depends on $r$ as long as $r$ is finite; the profiles decrease exponentially for large $\rho$, have a narrow structure when $r$ is small, tend to be broad with increasing $r$, and finally approach the $D=2$ result. One can see how the $D=3$ profiles converge to the $D=2$ ones by looking at the relative errors (not taking the absolute value) plotted on top of the profiles. 
 
\begin{figure*}[!t] 
\centering\includegraphics[width=\figtwocolumn]{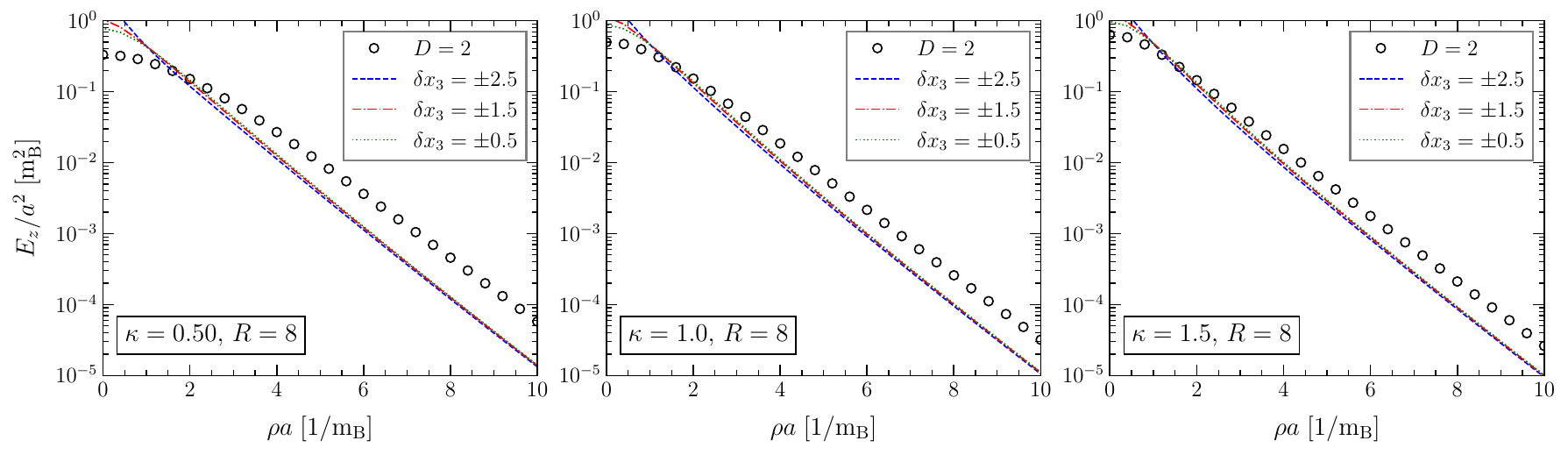} 
\centering\includegraphics[width=\figtwocolumn]{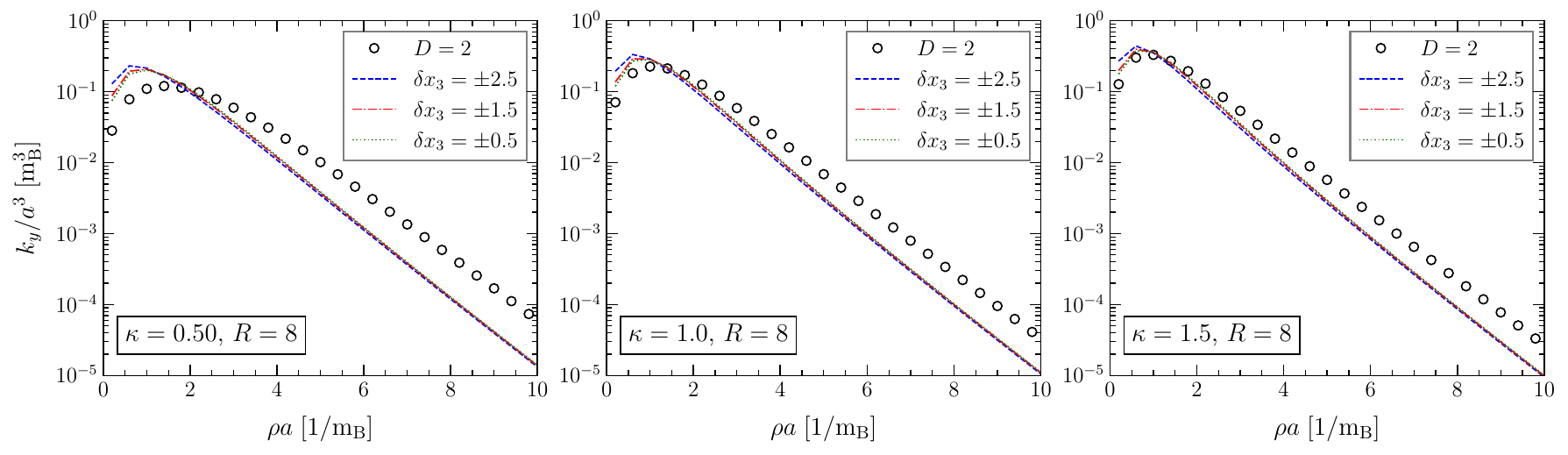} 
\caption{The profile of the electric field (upper three panels) and the monopole supercurrent (lower three panels) for $r=3.2\;\mathrm{[1/m_{B}]}$ ($R=8$) {\em around} the midtransverse plane, which are compared to the $D=2$ profile. The value of $\delta x_{3}=\pm 0.5$ corresponds to the midtransverse plane for the even length $R=8$, and $\delta x_{3}=\pm 1.5$ corresponds on the planes just shifted a lattice spacing $a$ along the $x_{3}$ axis above and below the center, and so on. The location of the terminals is $\delta x_{3}=\pm 4$.} 
\label{fig:profile_ek_slices_r8_kdep} 
\end{figure*} 
 
\par 
If one does not know how these profiles are obtained, one may speculate that the broadening of the profile is caused by a roughening effect of the flux tube~\cite{Luscher:1980iy}. However, this is not the case as the profiles are completely based on the classical solution. We would like to point out that the broadeninglike behavior reflects the change of contribution to the total profile from the Coulombic part and the solenoidal part for the increase of $r$ under the {\em flux quantization condition}; if one part of the electric-field profile is enhanced, the other part must be reduced, and vice versa, since the sum of the electric field on a plane is always quantized to be $2\pi N_{q}$ if the plane is pierced by the Dirac string of the charge $N_{q}$. 

For smaller $r$, the electric field has a larger value at small $\rho$ due to a significant contribution from the Coulombic part, and then, it must take a smaller value at large $\rho$ instead. On the other hand, for larger $r$, the electric field has a smaller value at small $\rho$ as the Coulombic contribution is suppressed, and then, it must take a larger value at large $\rho$ by enhancing the solenoidal contribution. When $r$ becomes large enough at the Coulombic contribution is completely negligible, the solenoidal part of the profile should cover the total electric flux $2\pi N_{q}$, which then coincides with the $D=2$ result. The $r$-dependence of the monopole supercurrent is directly related to that of the solenoidal part of the electric field via the dual Amp\`ere law. 
 
\par 
An important lesson from this result may be that when one investigates the flux-tube profile in the quark-antiquark system similarly in lattice QCD simulations, one must keep in mind that this kind of broadening effect can affect the determination of the width, and thus, of the penetration depth. One difficulty in the analysis may be that one does not know a priori how the examined quark-antiquark distance is close to the limit $r \to \infty$, where the use of the $D=2$ profile in the analysis makes sense. Then, one usually investigates translational invariance of the field profiles in advance. In Fig.~\ref{fig:profile_ek_slices_r8_kdep}, we show the profiles of the electric field and the monopole supercurrent for $r=3.2\;\mathrm{[1/m_{B}]}$ ($R=8$) {\em around} the midtransverse plane, which are compared to the $D=2$ profiles as in Figs.~\ref{fig:profile_ez_lx64_lz128} and \ref{fig:profile_ky_lx64_lz128}. We find that the $D=3$ profiles exhibit a translational-invariant behavior except for the region near the center of the flux tube, but they are clearly different from the $D=2$ profile. This means that observing the translational invariance in the field profile only from a limited distance of $r$ does not always guarantee the realization of the $D=2$ system. 
 
\begin{figure*}[!t] 
\centering\includegraphics[width=\figtwocolumn]{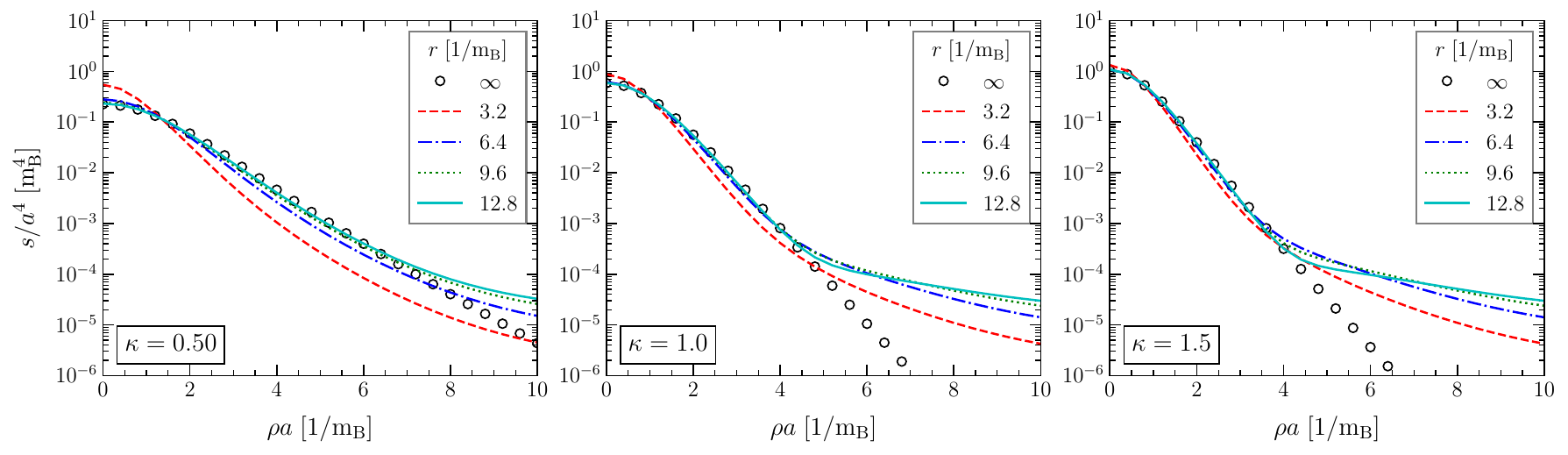} 
\caption{The similar plot as in Fig.~\ref{fig:profile_ez_lx64_lz128} for the action density.} \label{fig:profile_actd_lx64_lz128} 
\centering\includegraphics[width=\figtwocolumn]{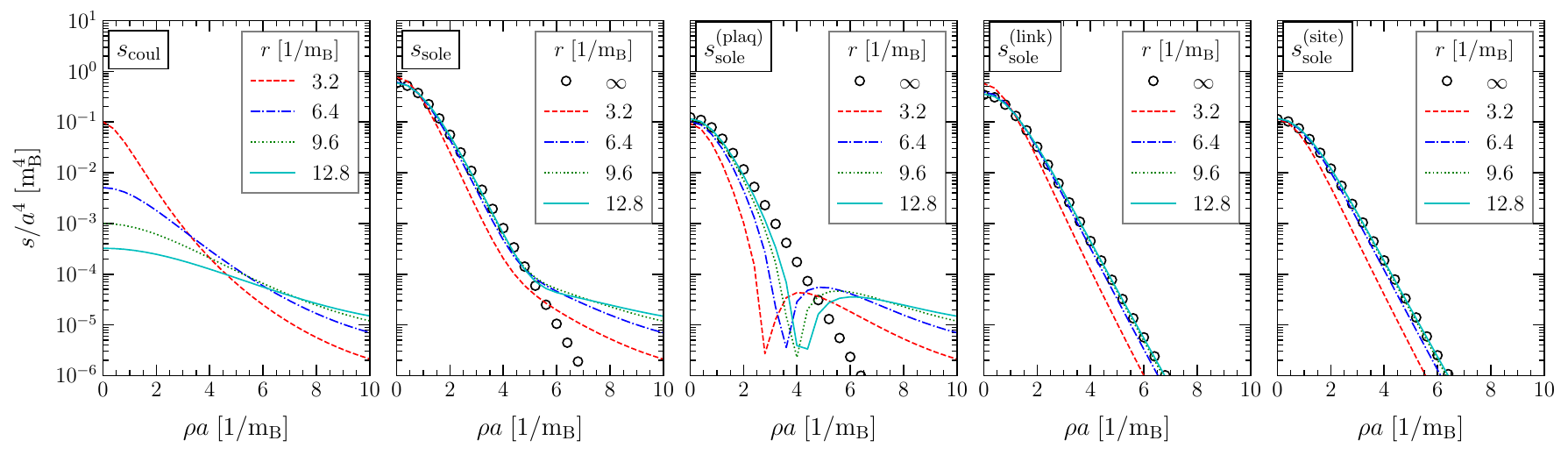} 
\caption{The ingredient of the action density for $\kappa =1.0$ in Fig.~\ref{fig:profile_actd_lx64_lz128} (middle), where $s =s_{\mathrm{coul}} + s_{\mathrm{sole}}$ and $s_{\mathrm{sole}}=s_{\mathrm{sole}}^{\mathrm{(plaq)}}+s_{\mathrm{sole}}^{\mathrm{(link)}}+s_{\mathrm{sole}}^{\mathrm{(site)}}$.} 
\label{fig:profile_actd_lx64_lz128_k10_ingredient} 
\end{figure*} 
 
\par 
In Fig.~\ref{fig:profile_actd_lx64_lz128}, we next show the profiles of the action density along the $x_{1}$-axis on the midtransverse plane (the same location as the field profiles). As in the field profiles, the action densities have a narrow structure when $r$ is small, and tend to be broad with increasing $r$. In contrast to the field profiles, however, the action densities do not show a convergence behavior. Rather, they tend to spread and deviate from the $D=2$ profile, especially at large $\rho$. Since the contribution to the action density is not only from the electric field but also from the monopole field, the interpretation of the behavior may not be straightforward. 
 
\par 
To understand this peculiar broadening behavior, it is useful to go back to the definition of the Coulombic electric field $C_{\mu\nu}$ in Eq.~\eqref{eqn:coulomb-cmunu} and $S_{\mathrm{coul}}$ in Eq.~\eqref{eqn:action-coulomb}. By using the continuum Green function in $D=3$ dimensions, $G(x)=1/(4\pi \sqrt{x_{1}^{2}+x_{2}^{2}+x_{3}^{2}})$, we can estimate the action density for $S_{\mathrm{coul}}$ along the $x_{1}$-axis on the midtransverse plane as 
\bea 
s_{\mathrm{coul}}(\rho a,0,0) &=& 2 \pi^{2} \beta_{g} C_{12}(\rho a,0,0)^{2} \nn\\* &=& \frac{ \beta_{g} N_{q}^{2} r^{2}}{ 8 ((\rho a)^{2} +r^{2}/4)^{3} }\hat{m}_{B}^{4} \;. 
\label{eqn:s_coul} 
\eea 
This indicates that it is hard to reproduce the $D=2$ result from the $D=3$ setting, since this Coulombic contribution can only be suppressed by a power of $r^{-4}$ for a fixed $\rho$ and is independent of $\kappa$ (for a fixed $r$, it also behaves as $\rho^{-6}$ at large $\rho$). At least, the maximum length $r=12.8\;\mathrm{[1/m_{B}]}$ ($R=32$) in Fig.~\ref{fig:profile_actd_lx64_lz128} is yet far from the $r \to \infty$ limit. 
 
\par 
In Fig.~\ref{fig:profile_actd_lx64_lz128_k10_ingredient}, we reveal the ingredients of the action density for $\kappa=1.0$ with the decomposition by $s=s_{\mathrm{coul}}+s_{\mathrm{sole}}$ and $s_{\mathrm{sole}}=s_{\mathrm{sole}}^{\mathrm{(plaq)}}+s_{\mathrm{sole}}^{\mathrm{(link)}}+s_{\mathrm{sole}}^{\mathrm{(site)}}$. The behavior of $s_{\mathrm{coul}}$ can be explained by Eq.~\eqref{eqn:s_coul}, which is responsible for half of the tail behavior of the action density. We find that another half is supplied by $s_{\mathrm{sole}}^{\mathrm{(plaq)}}$ in $s_{\mathrm{sole}}$. This is quite reasonable as the solenoidal part $(\del \wedge B^{\mathrm{reg}})_{\mu\nu}$ plays a role in canceling the Coulombic electric field $C_{\mu\nu}$ in the large $\rho$ region to form a flux tube. One may find a cusp structure in $s_{\mathrm{sole}}^{\mathrm{(plaq)}}$ at a certain $\rho$ for each $r$, which is just the radius that the solenoidal electric field changes its direction: the same (opposite) direction with the Coulombic electric field inside (outside) the cusp. On the other hand, we observe that $s_{\mathrm{sole}}^{\mathrm{(link)}}$ and $s_{\mathrm{sole}}^{\mathrm{(site)}}$ approach the $D=2$ result when $r$ becomes large, as in the field profiles. 
 
\par 
The profile of the action density is also often used to evaluate the width of the flux tube in the lattice QCD simulation. Our result here indicates that a naive fitting with a sort of exponential function may be problematic; one may extract a widthlike quantity, but it can potentially depend on the flux-tube length $r$ and the fitting range of $\rho$. Thus, the profile of the action density may not be suitable to evaluate the width, at least as it is. Even if one could observe a translational invariance of the action density around the midtransverse plane in the lattice QCD simulation, the profile can be different from the $D=2$ one as long as the Coulombic contribution persists. The Coulombic contribution cannot be removed easily, as it also affects the behavior of the solenoidal part. 
 
\subsection{The flux-tube interaction in the two-flux-tube system} 
 
\begin{figure*}[!t] 
\centering\includegraphics[width=\figtwocolumn]{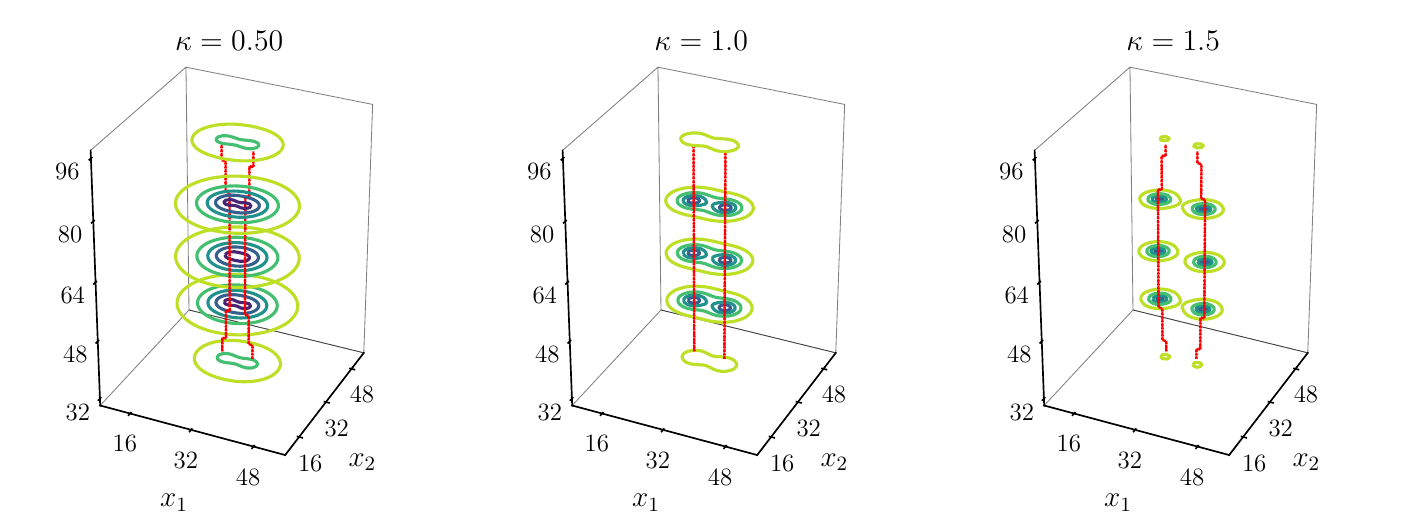} 
\caption{The profile of the final Dirac string for the two-flux-tube system ($R=56$, $X=8$) with the contours of the monopole field $\phi =|\chi |$ at $x_{3}=35,51,64,78,94$. The values of the contour lines from outer to inner are $\phi = 0.9, 0.7, 0.5, 0.3, 0.1$.} 
\label{fig:ff-3d} 
\centering\includegraphics[width=\figtwocolumn]{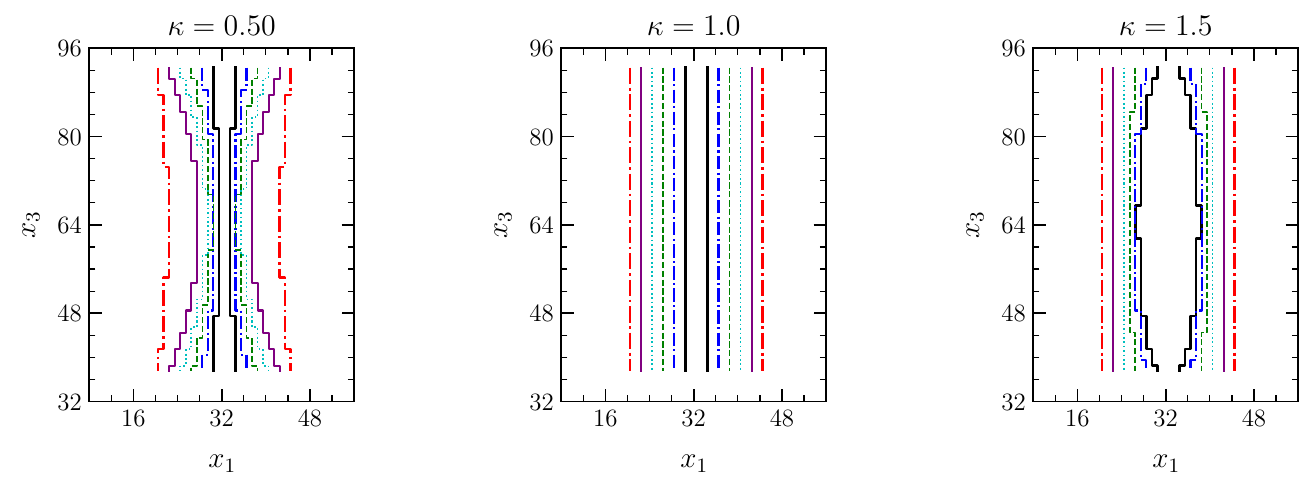} 
\caption{The profile of the final Dirac strings for the two-flux-tube system with $R=56$ with various original separations $X=4, 8,12,16,20,24$, where the same line-type and color are the pair for each $X$ ($X=24$ is the outermost pair). The 3D profile for $X=8$ is shown in Fig.~\ref{fig:ff-3d}.} 
\label{fig:ff-bending-r56} 
\end{figure*} 
 
\begin{figure*}[!t] 
\centering\includegraphics[width=\figtwocolumn]{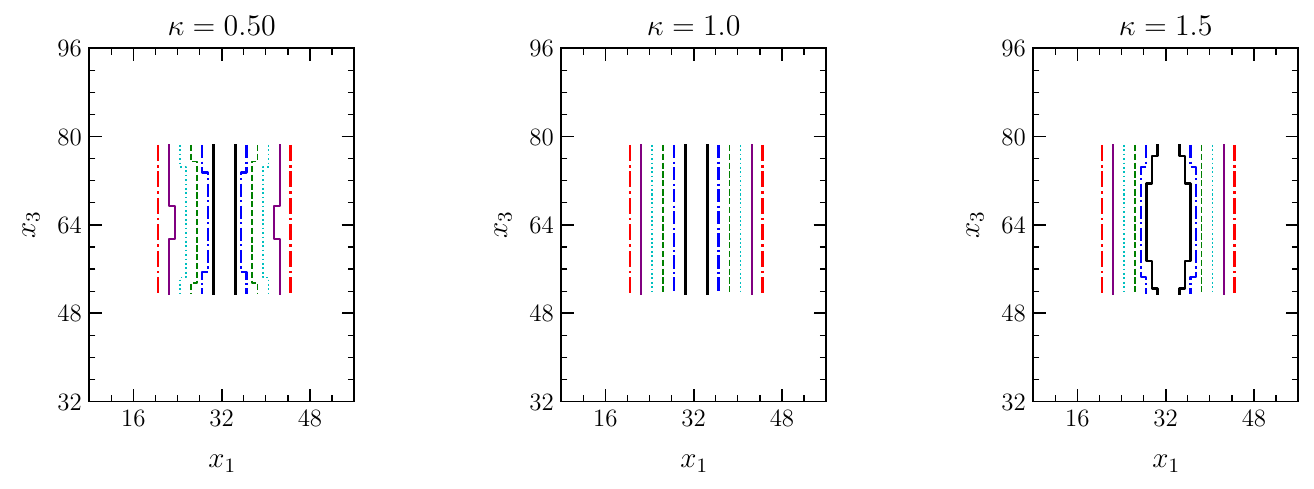} 
\caption{The same as in Fig.~\ref{fig:ff-bending-r56} for $R=28$.} \label{fig:ff-bending-r28} 
\end{figure*} 
 
\par 
In our method, all required before computation is to set a connected stack of integer values $\Sigma_{\mu\nu}$ along the Dirac string for a flux tube as in Fig.~\ref{fig:dstring2}. This feature then enables us to investigate the two-flux-tube system quite easily, as demonstrated in Sec.~\ref{subsec:fluxtubecore}. As already argued, the interaction properties of the flux tube can depend on $\kappa$; it is expected that the attractive or repulsive force works for $\kappa<1$ (type I) or $\kappa>1$ (type II). 
 
\subsubsection{The parallel case} 

We first investigate the two-flux-tube system in which the two straight Dirac strings are placed in parallel with the {\em same} direction. Here, to simplify the visualization, we set $\hat{m}_{B} =0.40$ and $L^{2}L_{3}=64^{2}128$, the same as the study of the flux-tube width. 
 
\begin{figure*}[!t] 
\centering\includegraphics[width=\figtwocolumn]{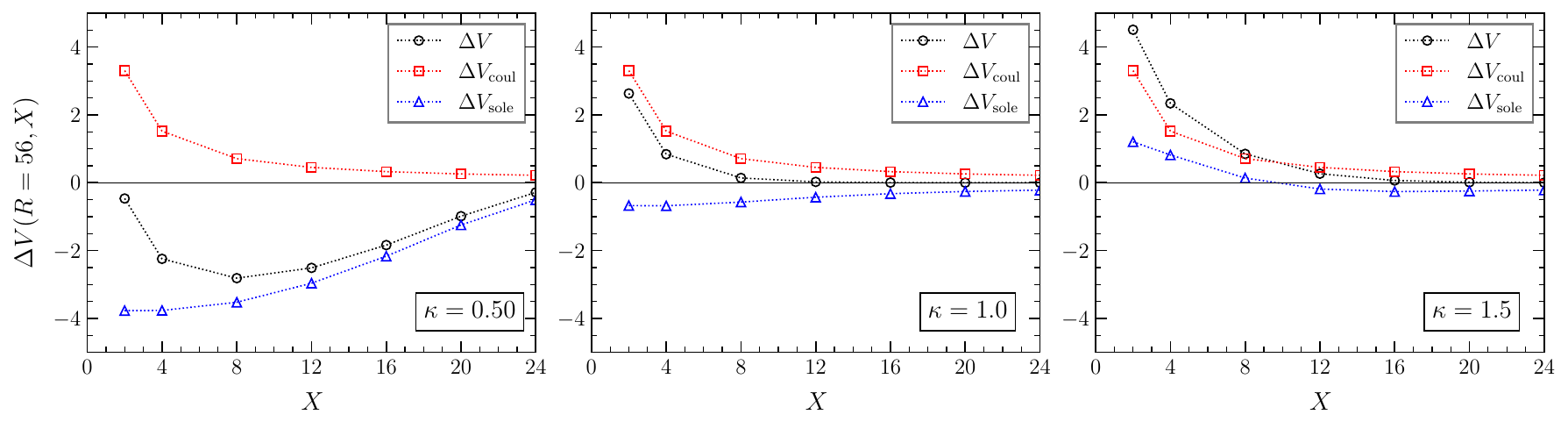} 
\caption{The interaction energy between two flux tubes defined by Eq.~\eqref{eqn:intenergy} for $R=56$ as a function of the distance between two original Dirac strings $X$, where the Hodge decomposition is applied.} 
\label{fig:ff-intenergy-decomose-r56} 
\centering\includegraphics[width=\figtwocolumn]{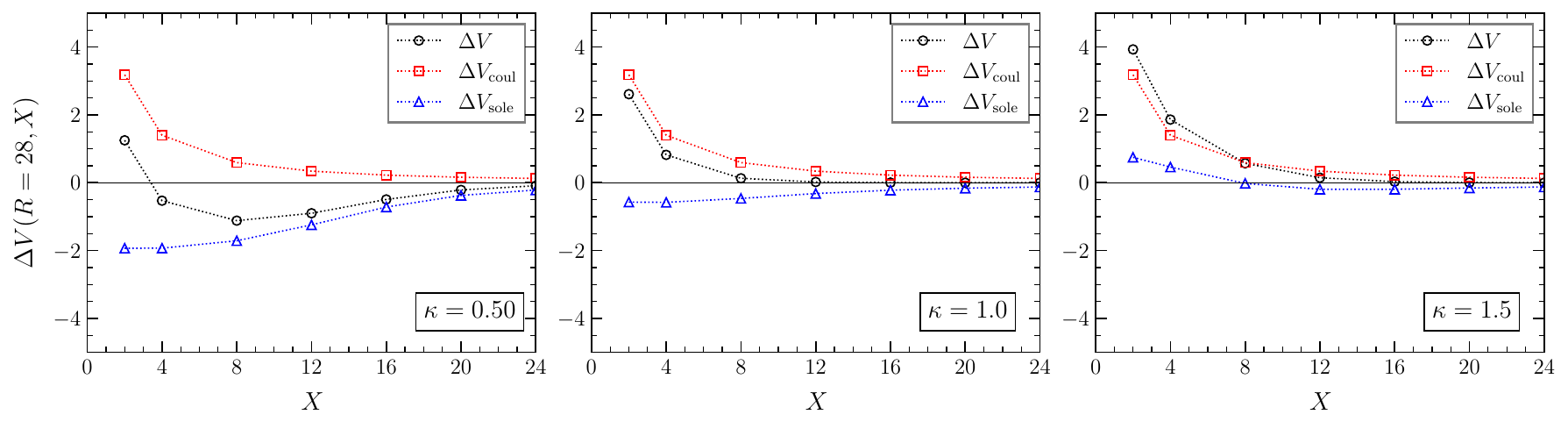} 
\caption{The same plot as in Fig.~\ref{fig:ff-intenergy-decomose-r56} but for $R=28$.} \label{fig:ff-intenergy-decomose-r28} 
\end{figure*} 
 
\par 
In Fig.~\ref{fig:ff-3d}, we show the final profile of the Dirac string as a representative of the flux-tube solution for $R=56$ ($22.4\;\mathrm{[1/m_{B}]}$) with the distance between two original Dirac strings $X=8$ ($3.2\;\mathrm{[1/m_{B}]}$), where some selected cross sections of the monopole field $\phi =|\chi |$ are also presented. We find that the two Dirac strings are bent inward with each other for $\kappa =0.50$ (left), and outward for $\kappa =1.5$ (right), which clearly indicates the appearance of attractive and repulsive forces between the two flux tubes, respectively. On the other hand, the two Dirac strings remain straight as the original setting for $\kappa =1.0$ (middle), as if there is no interaction. From the contour of the monopole field, we may learn that the total energy cost for digging holes in the monopole field along the path of the Dirac string is related to the final shape of the Dirac string, such that a single hole is preferable for $\kappa =0.50$ even though the hole diameter is enlarged, while double holes are preferable for $\kappa =1.5$. There is no such preference for $\kappa =1.0$ in terms of the energy cost, so that the original Dirac strings can remain intact. 
 
\par 
The interaction of the two flux tubes can further be investigated by changing the length $R$ and the distance $X$ as shown in Figs.~\ref{fig:ff-bending-r56} and~\ref{fig:ff-bending-r28}. These plots indicate that for $\kappa=0.50$ the two flux tubes are eager to be combined to create a single flux tube as they are closer, while for $\kappa=1.5$ the two flux tubes are eager to be apart from each other and can be independent once they are separated enough. For $\kappa=1.0$, it is clear that the two Dirac strings remain straight regardless of the length $R$ and the distance $X$. A shorter flux tube ($R=28$) seems to resist bending more than a longer one ($R=56$), which suggests that the flux tube has a certain rigidity. Another feature of the flux-tube interaction suggested from these figures is that a flux tube has a finite interaction range depending on $\kappa$; the final shape of the flux tubes might not be determined solely by the minimum energy condition of the whole system. 
 
\par 
The interaction energy between two flux tubes may be quantified by looking at the difference of the energy between the two-flux-tube system $V^{(2)}(R,X)$ and twice the single flux-tube system $V(R)$ (the interquark potential) as 
\be 
\Delta V (R,X) =V^{(2)}(R,X) - 2 V(R) \; . 
\label{eqn:intenergy} 
\ee 
This definition is similar to that used for investigating the inter-meson potential \cite{Kodama:1997zc}, but what we can learn from this definition for the finite-length flux-tube system in this dual lattice formulation is not the energy before the interaction, but after or intermediate. In other words, it corresponds to a remnant of the interaction energy that is not liberated yet because of the fixed terminals. Note that in the case without fixed terminals in the $D=2$ setting, the flux tubes can move freely until all the interaction energy is liberated. 
 
\par 
In Fig.~\ref{fig:ff-intenergy-decomose-r56}, we plot $\Delta V (R=56,X)$ against the distance $X$ (corresponding to Fig.~\ref{fig:ff-bending-r56}) for $\kappa=0.50, 1.0, 1.5$, where the Hodge decomposition is applied to see the breakdown of the contribution as $\Delta V=\Delta V_{\mathrm{coul}}+\Delta V_{\mathrm{sole}}$. As already argued, the contribution from the Coulombic part does not depend on the $\kappa$ values. In the current setting, it is always positive for small $X$ because repulsive interaction between the electric charges of the same sign gives a dominant contribution to the energy for $X < R$ as 
\be 
\Delta V_{\mathrm{coul}}(R,X) \sim \frac{2\pi \beta_{g} N_{q}^{2}}{X} \; . 
\label{eqn:ff-intenergy-coulomb} 
\ee 
On the other hand, the contribution from the solenoidal part exhibits a qualitatively different behavior below and above $\kappa=1.0$. It would be clear that the negative or positive value indicates the appearance of an attractive or repulsive interaction, which causes the bending of the Dirac strings as in Fig.~\ref{fig:ff-bending-r56}. Similarly, $\Delta V (R=28,X)$ (corresponding to Fig.~\ref{fig:ff-bending-r28}) is plotted in Fig.~\ref{fig:ff-intenergy-decomose-r28}. We find that the behavior of the Coulombic part is almost the same as for $R=56$, which is quite reasonable since Eq.~\eqref{eqn:ff-intenergy-coulomb} is still dominant for small $X$. The solenoidal parts for $\kappa=0.50$ and $\kappa=1.5$ are approximately half of the $R=56$ case, just reflecting the difference in the flux-tube length. In contrast, for $\kappa=1.0$, the solenoidal part of the $R=28$ case seems to be identical to the $R=56$ case, which may indicate that the energy of the middle part of the flux tube is completely canceled in Eq.~\eqref{eqn:intenergy}, leaving only the remnant of the terminal contribution in the solenoidal part. 
 
\par 
In general, the flux-tube interaction depends not only on the $\kappa$ value but also on the initial geometric configuration of the two flux tubes. Thus, if we shift one of the two flux tubes along the $x_3$-axis, or change the length of one of them, we will obtain another final configuration and interaction energy. What is different from particle interaction is that a nonlocal interaction occurs between the middle part of the two flux tubes. In particular, the attractive interaction for $\kappa <1$ may be remarkable in the sense that, if we regard a flux tube as a hadronic state, it might produce a new type of bound state of hadrons, such as an exotic four-quark (tetraquark) state or a mesonic molecule. 
 
\subsubsection{The nonparallel case} 
 
\begin{figure*}[!t] 
\centering\includegraphics[width=\figtwocolumnsmall]{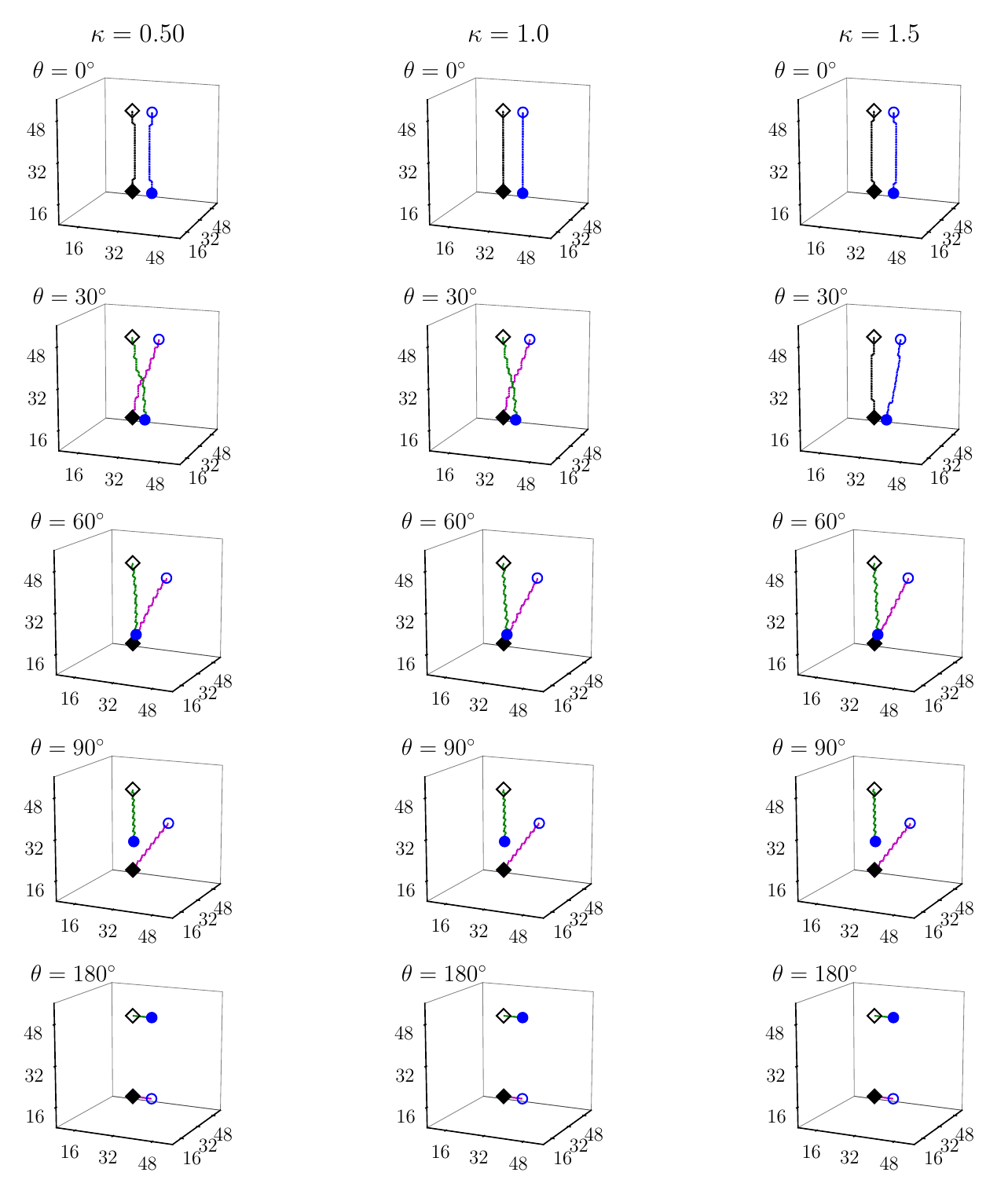} 
\caption{The final Dirac string profile for $\kappa=0.50$ (left), $1.0$ (middle), and $1.5$ (right) in the two-flux-tube system with various relative angles. Note that some of the two Dirac strings appear to intersect from this view angle, but they do not. The string flip occurs around $\theta=30^{\circ}$ for $\kappa=0.50$ and $1.0$, which around $\theta=60^{\circ}$ for $\kappa=1.50$. The colors of connected strings are changed before and after the string flip.} 
\label{fig:ff-rotation-ds} 
\end{figure*} 

If we regard the finite-length flux tube as a quark-antiquark system, the interaction of two flux tubes does not always happen in parallel. It is then interesting to investigate how the interaction depends on the initial relative angles between two flux tubes. We employ here a cubic lattice of the size $L^{3}=64^{3}$ with $\hat{m}_{B}=0.40$ and take $R=32$ and $X=8$ as an example. For the two points $\vec{x}_{c1}=(\frac{L}{2} - \frac{X}{2}, \frac{L}{2} ,\frac{L}{2} )$ and $\vec{x}_{c2}= ( \frac{L}{2} + \frac{X}{2}, \frac{L}{2} ,\frac{L}{2} )$, we put the 1st Dirac string from $\vec{x}_{q}= \vec{x}_{c1} - \frac{R}{2} \vec{e}_{z}$ to $\vec{x}_{\bar{q}}= \vec{x}_{c1} + \frac{R}{2} \vec{e}_{z}$, and the 2nd one from $\vec{x}_{q}= \vec{x}_{c2} - (\frac{R}{2}\sin \theta) \vec{e}_{y} - (\frac{R}{2}\cos \theta) \vec{e}_{z}$ to $\vec{x}_{\bar{q}}= \vec{x}_{c2} + (\frac{R}{2}\sin \theta) \vec{e}_{y} + (\frac{R}{2}\cos \theta) \vec{e}_{z}$, where $\theta$ varies from the same direction ($\theta=0^{\circ}$) to the opposite ($\theta=180^{\circ}$). 
 
\begin{figure*}[!t] 
\centering\includegraphics[width=\figtwocolumn]{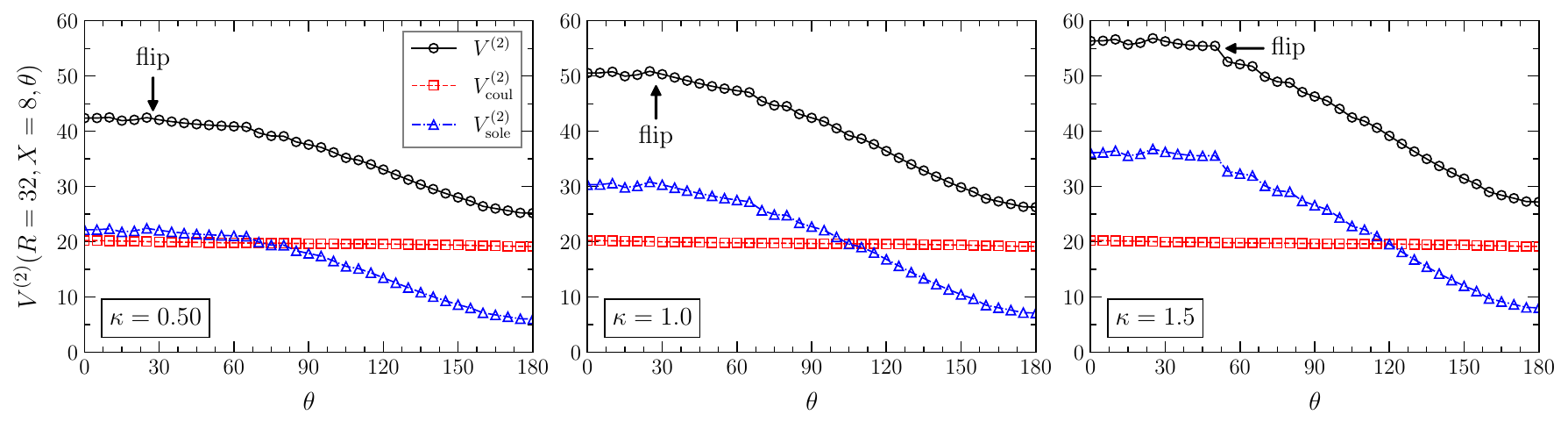} 
\caption{The potential energy in the two-flux-tube system $V^{(2)}$ as a function of relative initial angles of the Dirac strings for $\kappa=0.50$ (left), $\kappa=1.0$ (center), and $\kappa=1.5$ (right). The ``flip'' indicates the angle at which we observe the string flip by monitoring the profile of the Dirac strings as in Fig.~\ref{fig:ff-rotation-ds}.} 
\label{fig:ff-rotation-act} 
\end{figure*} 
 
\par 
In Fig.~\ref{fig:ff-rotation-ds}, we show the final Dirac-string profile of the flux-tube solution for $\kappa=0.50, 1.0, 1.5$ against $\theta$. The change of the Dirac-string shape from the original straight one indicates the appearance of an attractive or repulsive force. As $\theta$ increases from $0^{\circ}$ to $180^{\circ}$, we find that the 1st and 2nd Dirac-string shapes are distorted and the string flip occurs beyond a certain angle. Within the plotted profiles, the string flip is already seen at $\theta = 30^{\circ}$ for $\kappa=0.50$ and $1.0$, while at $\theta =60^{\circ}$ for $\kappa=1.5$, where connected strings are displayed with the same color. A remarkable feature is that the interaction between the middle part of the flux tubes occurs even for $\kappa=1.0$ once two flux tubes are originally placed with $\theta > 0$, which is attractive. 
 
\par 
For the relative angle $\theta$, the total energy $V^{(2)}(R=32,X=8,\theta)$ is found to behave as in Fig.~\ref{fig:ff-rotation-act}. We find that the Coulombic part of the energy is almost constant for $\theta$ compared to the change of the solenoidal part, implying rotational symmetry around the axis that pierces the center of the original 1st and 2nd Dirac strings in the current geometrical setting, and thus, the change of the total energy is mostly governed by that of the solenoidal part. The qualitative behavior seems to be common to all $\kappa$ values; the energy starts to decrease gradually at around $\theta = 30^{\circ}$. Although the energy may possess a small fluctuation against $\theta$, this is mostly due to the lattice discretization effect. Only an exception is around the angle between $\theta=50^{\circ}$ and $55^{\circ}$ for $\kappa=1.5$, which indicates a jump of the energy due to the string flip (the corresponding figure is not displayed in Fig.~\ref{fig:ff-rotation-ds}, but we have examined the final Dirac-string profiles with these energy data as well). 
 
\subsection{The flux-tube interaction in the multiflux-tube system} 
\label{subsect:fluxtube-interaction} 

So far, we have regarded the finite volume effect as an undesired lattice artifact to be removed. However, this effect can rather be exploited to investigate the interaction properties among the {\em multiflux} tubes~\cite{Ichie:1996jr} simply by putting a single flux tube in a small box with the periodic boundary condition. In fact, the result of the string tension on the $D=2$ lattices presented in Fig.~\ref{fig:sigma-ldep} of Sec.~\ref{subsec:result-latticeeffects} already contains this effect especially when the size of $L^{2}$ is small, such that the string tension decreases for $\kappa=0.50$ or increases for $\kappa =1.5$ as the size $L$ becomes small, while it remains constant for $\kappa=1.0$. 
 
\par 
The question is then what happens if the lattice size becomes further smaller. Since the increase of the flux-tube density causes a reduction of the spatial region for monopole condensation, a phase transition is expected to occur from the superconducting phase $\phi \ne 0$ to the normal one $\phi \simeq 0$ everywhere when it exceeds a critical density. In such a normal phase, the value of $S$ in Eq.~\eqref{eqn:action} in $D=2$ dimensions is reduced to 
\be 
S\to \sigma = \beta_{g} \left [ \frac{2\pi^{2} N_{q}^{2}}{L^{2}} + L^{2} \frac{\hat{m}_{B}^{2}\hat{m}_{\chi}^{2}}{8} \right ] 
\;, 
\label{eqn:action-smallbox} 
\ee 
regardless of the $\kappa$ values. The situation is, so to speak, an appearance of a uniformly distributed electric field between the metal plates with the electric charges $N_{q}$ and $-N_{q}$, like in a capacitor. Hence, $\sigma$ is not the string tension of the flux tube anymore, but simply means the energy per unit length. If we regard Eq.~\eqref{eqn:action-smallbox} as a function of the density $1/(La)^{2} \; \mathrm{[m_{B}^{2}]}$, this takes a minimum value $\sigma_{\rm min}/m_{B}^{2} = \pi \beta_{g} \kappa | N_{q} |$ when $1/(La)^{2} = \kappa / (4\pi | N_{q} | ) \; \mathrm{[m_{B}^{2}]}$. 
 
\par
\begin{figure}[!t] 
\centering\includegraphics[width=\figonecolumn]{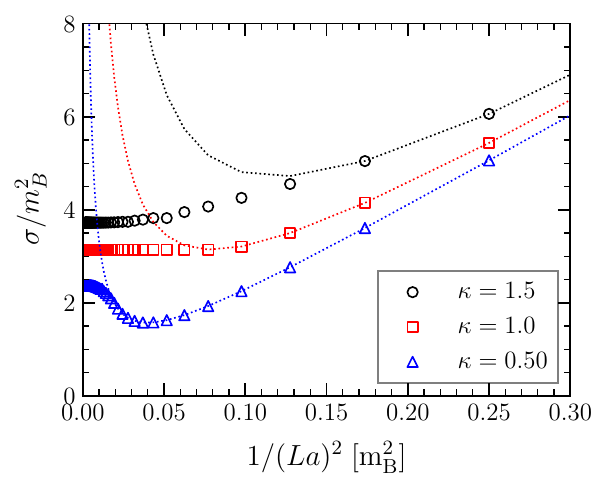} 
\caption{The string tension or the energy per unit length $\sigma$ as a function of the density. The dotted lines are Eq.~\eqref{eqn:action-smallbox} for each $\kappa$ value, and their minimum values correspond to $\pi \kappa$ at $1/(La)^{2} =\kappa/(4\pi )\;\mathrm{[m_{B}^{2}]}$.} 
\label{fig:sigma2d_smallbox} 
\end{figure}

 
In Fig.~\ref{fig:sigma2d_smallbox}, we show the value of $\sigma/m_{B}^{2}$ for $N_{q}=1$ against the density $1/(La)^{2}$ for $\hat{m}_{B}=0.20$. As the density increases, the value of $\sigma$ for the solution starts to fall into the expectation in Eq.~\eqref{eqn:action-smallbox}, indicating the disappearance of the superconducting phase. The melting point of the normal phase [the point at which the numerical data starts to follow Eq.~\eqref{eqn:action-smallbox}] seems to depend on the $\kappa$ value, where the tendency is such that the superconducting phase of smaller $\kappa$ can melt to the normal phase with a lower density. 
 
\begin{figure}[!t] 
\centering\includegraphics[width=\figonecolumn]{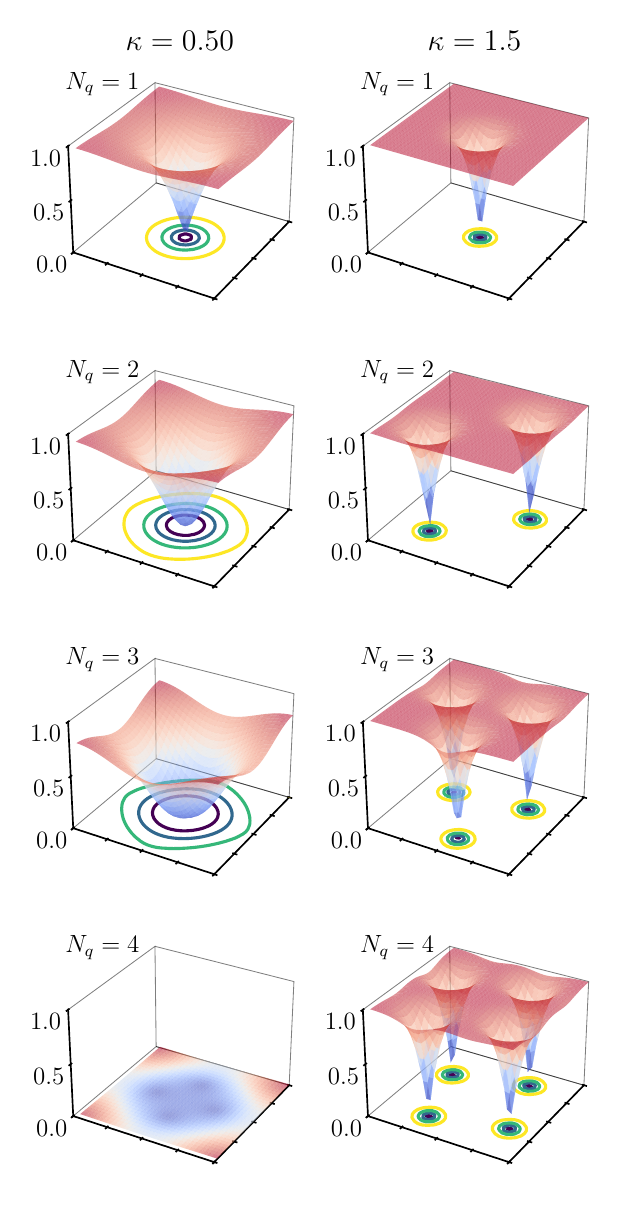} 
\caption{The profiles of monopole field for various charges $N_{q}$ on the $L^{2}=32^{2}$ lattice, where the left panels are for $\kappa=0.50$ and the right for $\kappa=1.5$. The range of the vertical axis is from 0 to 1. Note that the initial Dirac string of $N_{q}$ charges is always put at the center of the lattice $(x_{1},x_{2})=(16,16)$, but it can split into $N_{q}$ strings of single charge for $\kappa =1.5$ due to internal repulsive forces.} 
\label{fig:monopolefield_smallbox} 
\end{figure} 
 
\par 
The flux-tube density can also be increased for a fixed lattice size by increasing the charge $N_{q}$ instead. We show how the monopole field $\phi =|\chi |$ is melting for the increase of $N_{q}$ on a fixed $L^{2}=32^{2}$ lattice in Fig.~\ref{fig:monopolefield_smallbox}. In this computation, we put a Dirac string with the charge $N_{q}$ at the center of the lattice and solve the field equations for $\hat{m}_{B}=0.40$ with $\epsilon=10^{-4}$. The flux-tube density will be given by $|N_{q}|/(L a)^{2} \sim 0.006 |N_{q}| \;\mathrm{[m_{B}^{2}]}$. For $\kappa =0.50$, we observe the signs of melting already for $N_{q}=2$. As for $N_{q}=4$, we find $\phi \sim 0$ in the whole region of the lattice, where the corresponding density is $0.024 \;\mathrm{[m_{B}^{2}]}$, which is already in the region that the data point falls into Eq.~\eqref{eqn:action-smallbox} in Fig.~\ref{fig:sigma2d_smallbox}. On the other hand, for $\kappa=1.5$ we find that the $N_{q}$-charged flux tube splits into $N_{q}$ single flux tubes with a characteristic distribution depending on the number of $N_{q}$. We note that even for $N_{q}=10$ the monopole field keeps condensation $\phi \sim 1$ except at the cores of the flux tubes, which is a similar situation to the Abrikosov vortex lattice in type-II superconductors. 
 
\begin{figure*}[!t] 
\centering\includegraphics[width=\figtwocolumn]{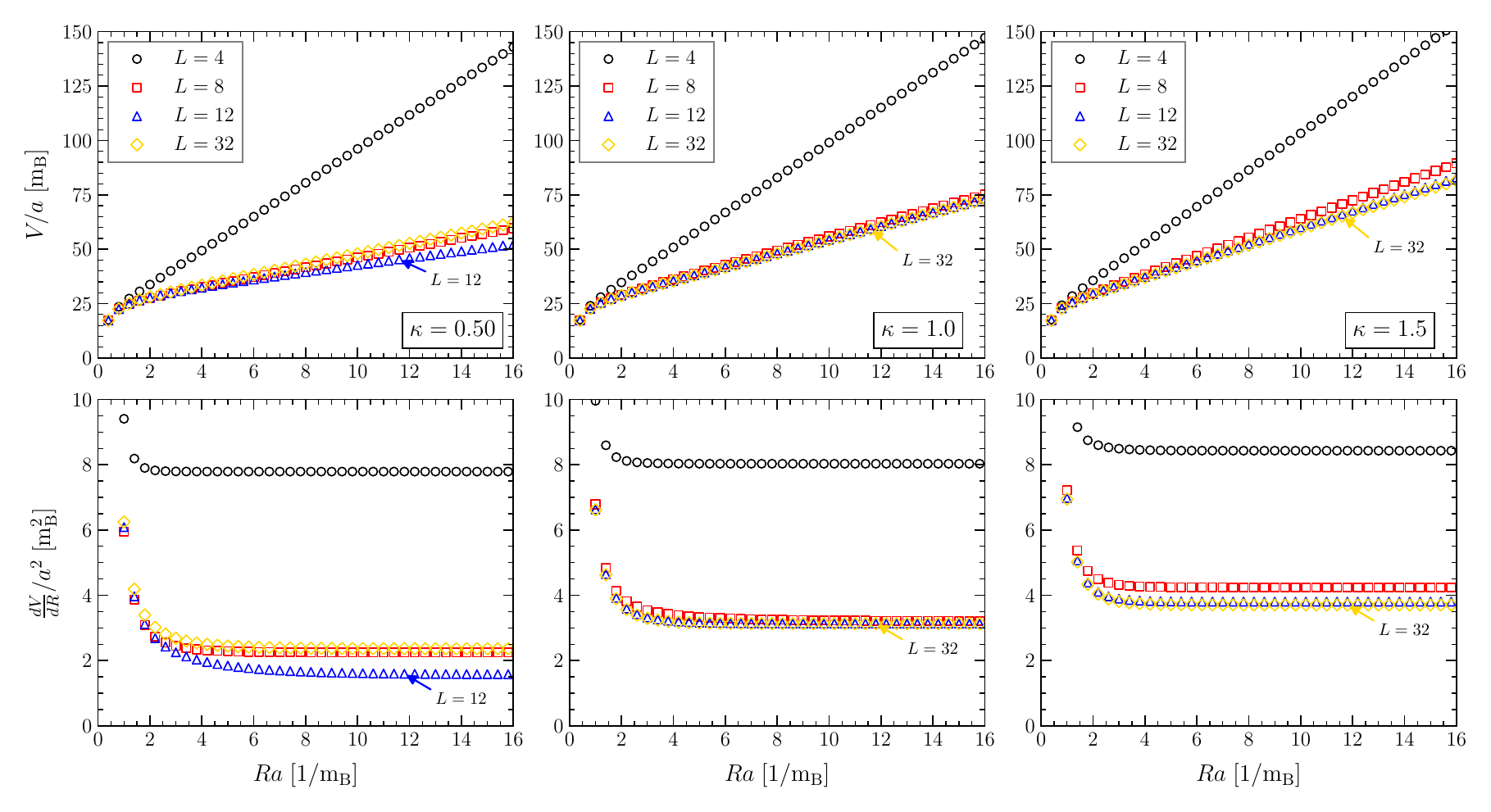} 
\caption{The interquark potential and its derivative with respect to the distance $r=Ra$ for $\kappa=0.50, 1.0, 1.5$ (with $\hat{m}_{B}=0.40$), where the interquark distance is taken along the $x_{3}$-axis on the $L^{2}L_{3}$ lattice with $L_{3}=128$. The lattice size perpendicular to the interquark axis is $L=4,8,12,32$; the corresponding flux-tube density is $1/(La)^{2}=0.39, 0.098, 0.043, 0.0061\;\mathrm{[m_{B}^{2}]}$ (cf. Fig.~\ref{fig:sigma2d_smallbox}). The maximum value of $Ra=16 \;\mathrm{[1/m_{B}]}$ corresponds to $R=40$, which is about $1/3$ of $L_{3}=128$.} 
\label{fig:pot-lxy-lz128} 
\end{figure*} 
 
\par 
It is also possible to investigate the density dependence of the multiflux-tube system in $D=3$ dimensions by putting the finite-length flux tube along the $x_{3}$-axis on a $L^{2}L_{3}$ lattice, where $L_{3}$ is taken to be large enough, while $L$ is changed to control the flux-tube density. In Fig.~\ref{fig:pot-lxy-lz128}, we plot the interquark potential and its derivative with respect to the distance for $\kappa=0.50, 1.0, 1.5$ (with $\hat{m}_{B}=0.40$), where the lattice volume is $L_{3}= 128$ with $L=4,8,12,32$. The lattice sizes $L=4,8,12,32$ correspond to the flux-tube density $1/(La)^{2}=0.39, 0.098, 0.043, 0.0061\;\mathrm{[m_{B}^{2}]}$, respectively. We find that the interquark potential contains a linearly rising part regardless of the density, and its slope tends to be steep as the density increases. Note that the slope extracted from the potential is consistent with the value of $\sigma$ in Fig.~\ref{fig:sigma2d_smallbox}. The steep slope for higher densities indicates a realization of a uniformly distributed electric field along the $x_{3}$-axis as in a capacitor. 
 
\par 
The difference from the ordinary capacitor may be that the electric charges are now discretely distributed, the interquark distance is not limited to short, and the slope yet has a dependence on $\kappa$. According to Eq.~\eqref{eqn:action-smallbox}, the $\kappa$-dependence will disappear in the high-density limit, $1/L^{2} \to \infty$, where the second term becomes negligible. We find an interesting feature of the slope for $\kappa =0.50$, as already expected from Fig~\ref{fig:sigma2d_smallbox}, such that the case for $L=12$ has the smallest slope within the plotted data, which means that it is not necessarily true that the higher density leads to a larger slope. This behavior is qualitatively different from that for $\kappa \geq 1$. 
 
\section{Summary and outlook} 
\label{sec:summary} 
 
\par 
For efficient use of the DGL theory toward understanding the dual superconducting properties of the QCD vacuum, we have demonstrated a powerful numerical method for solving the field equations in the DGL theory with U(1) dual gauge symmetry. An essential aspect of our method is to formulate the DGL theory on the dual lattice, which enables us to investigate any system composed of finite-length flux tubes in a systematic manner. 
 
\par 
The tasks of solving simultaneous nonlinear partial differential equations for the field variables remain the same as in the ordinary discretization methods, but we do not need to struggle to set complicated boundary conditions for the field variables. For the quark-antiquark system, all we have to care about is specifying the locations of a quark and an antiquark as the ends of an open Dirac string. Using the ordinary Newton-Raphson method, we always obtain the field configurations that minimize the energy of the system, regardless of the original path of the Dirac string. Moreover, the systematic lattice effects accompanied by the finite lattice spacing and the finite volume can be controlled depending on the desired accuracy of the numerical results. 
 
\par 
Using this formulation, we have obtained the finite-length flux-tube solution for three types of the dual superconducting phases of $\kappa=0.50$ (type I), $\kappa=1.0$ (the border), and $\kappa=1.5$ (type II) as templates, and have investigated (i) the interquark potential, (ii) the width of the finite-length flux tube, and (iii, iv) the interaction properties of flux tubes in the two-flux-tube and multiflux-tube systems. Let us summarize our findings on these topics briefly. 
 
\begin{enumerate}[(i)] 
\itemsep=0cm 
 
\item 
The functional form of the interquark potential for the distance $r$ is well-described by the Yukawa term $-e^{-m_{y} r}/r$ plus linear function $\sigma_{y} r$, where the $\kappa$-dependence of the potential can be successfully captured by the effective mass $m_{y} (\kappa)$, and the string tension $\sigma_{y} (\kappa)$. $m_{y} (\kappa)$ grows monotonically from a value smaller than the dual gauge boson mass $m_{B}$ to the value about $1.668\, m_{B}$ in the type-II limit. 
 
\item 
The width of the finite-length flux tube on the midtransverse plane can grow with the increase of $r$, if we define it from the tail behavior of the profile of the electric field, or the monopole supercurrent, or the action density. The profile of the finite-length flux tube cannot be reproduced by the infinitely long flux-tube solution in $D=2$ dimensions as long as the Coulombic contribution from the flux-tube terminals remains, which does not go away easily, even at quite large $r$. 
 
\item 
The flux-tube interaction depends not only on the $\kappa$ value but also on the geometric configuration of flux tubes specified by the distance, the length, and the relative angles. If the original flux tubes are located within the range of interaction, an attractive force or a repulsive force between flux tubes can cause the change of the original shapes of the flux tubes, such as the bending or the string flip. An attractive force between flux tubes might produce a new type of hadronic state, such as an exotic four-quark (tetraquark) state or a mesonic molecule. 
 
\item 
A phase transition occurs from the dual superconducting phase to a normal phase if the flux-tube density exceeds a critical one, where the linearly rising behavior of the interquark potential persists with a different slope, indicating an appearance of a uniformly distributed electric field, as in a capacitor. The slope for $\kappa \geq 1$ tends to grow monotonically with increasing density, while the slope for $\kappa < 1$ once becomes smaller than that in the dual superconducting phase and then increases, which is qualitatively different from the behavior with $\kappa \geq 1$. 
 
\end{enumerate} 
 
\par 
In fact, some of the above findings may contain the features already known or speculated by approximate treatments. We would like to emphasize, what is new is that it is now possible to investigate these features quantitatively in a systematic manner with the dual lattice formulation of the DGL theory. Another useful feature of the dual lattice formulation may be that we can obtain the solution that contains systematic lattice effects on purpose. Since the results of lattice QCD simulations certainly contain the lattice effects, this will be helpful to isolate the effects from the results when the comparison is made. 

When analyzing the lattice QCD results of the flux-tube profiles in the quark-antiquark system, the finite-length flux-tube solution will be more effective than the infinitely long one for extracting quantitative values on the dual superconducting QCD vacuum. In QCD, there also exists a three-quark state corresponding to a baryon. It is quite interesting to analyze our recent results of the three-quark potential in SU(3) lattice gauge theory~\cite{Koma:2017hcm} with the baryonic flux-tube solution in the DGL theory, which is possible by extending the dual gauge symmetry of the DGL theory to $\mathrm{U(1) \times U(1)}$~\cite{Kamizawa:1992np, Koma:2000hw}. It is also interesting to investigate the two-flux-tube and multiflux-tube systems by lattice QCD simulations, as we already have the solution for these systems. These analyses can provide new insights into the nonperturbative vacuum structure, which will be addressed in our future publications. 
 
\section*{Acknowledgments} 

The main idea of this work is based on our previous collaboration with E.-M. Ilgenfritz, T. Suzuki, and H. Toki~\cite{Koma:2000hw}. We would like to thank them for the valuable discussions at that time. 
 
\appendix 
 
\section{The lattice Green function for the massless field} 
\label{sect:app-lattice-green} 

There are two types of the lattice Green function for the massless field, whether it is defined in an infinite volume or in a finite volume. In an infinite volume, the lattice Green function in the coordinate representation $G_{\infty}(x)$ is defined by the equation 
\be 
\Delta_{L} G_{\infty}(x) = - \delta_{x0} \;, 
\label{eqn:green-def} 
\ee 
where $\Delta_{L}$ denotes the lattice Laplacian. In the $D$-dimensional Euclidean space, the left-hand side of Eq.~\eqref{eqn:green-def} is written as 
\bea 
\Delta_{L} G_{\infty}(x) 
&= & \del_{\mu}\del_{\mu}' G_{\infty}(x)\nn\\ &=& 
\sum_{\mu =1}^{D} [ G_{\infty}(x + \hat{\mu}) + G_{\infty}(x - \hat{\mu}) - 2G_{\infty}(x) ]\;, 
\nn\\ 
\eea 
where $\del_{\mu}$ and $\del_{\mu}'$ are forward and backward differences to a direction $\mu$, respectively. 
 
\par 
The solution of Eq.~\eqref{eqn:green-def} is formally obtained by performing the Fourier transformation of the Green function in the momentum representation. By using the definition of the momentum 
\be 
\hat{p}_{\mu}=2 \sin \frac{p_{\mu}}{2} \quad ( p_{\mu} \in [-\pi,\pi] )\;, 
\label{eqn:mom-domain} 
\ee 
it is given by 
\be 
\tilde{G}_{\infty}(p) 
= \frac{1}{\hat{p}^{2}} = \frac{1}{2D-2 \sum_{\mu}\cos p_{\mu}}\;, 
\ee 
and the Fourier transform is 
\be 
G_{\infty}(x) = 
\int_{-\pi}^{\pi} \frac{d^{D} p}{(2\pi)^{D}} e^{ip_{\mu} x_{\mu}}\tilde{G}_{\infty}(p) 
\; . 
\label{eqn:green-infinite} 
\ee 
 
\par 
In a finite volume with periodic boundary conditions in all directions, ${p}_{\mu}$ in Eq.~\eqref{eqn:mom-domain} is discretized as $p_{\mu} =2\pi n_{\mu}/L_{\mu}$ with integers $n_{\mu}=0,1,2,...,{L_{\mu}-1}$, so that the Green function in the momentum representation is given by 
\be 
\tilde{G}_{L}(p) 
= \frac{1}{\hat{p}^{2}} = \frac{1}{2D - 2 \sum_{\mu}\cos \frac{2\pi n_{\mu}}{L}} \;, 
\ee 
and the Fourier transform is 
\be 
G_{L}(x) =\frac{1}{L^{D}}\sum_{n_{\mu} \ne 0}e^{i \frac{2 \pi n_{\mu}}{L_{\mu}}x_{\mu}}\tilde{G}_{L}(p) \; , 
\label{eqn:green-finite} 
\ee 
where $L^{D} \equiv \prod_{\mu=1}^{D}L_{\mu}$. To avoid division by zero, one may exclude the zero momentum mode $n_{\mu}=0$ for all $\mu$ reluctantly, which yields a finite volume correction in Eq.~\eqref{eqn:green-def} as 
\be 
\Delta_{L} G_{L}(x) =-\delta_{x0}+\frac{1}{L^{D}} \; . 
\ee 
 
\par 
The Fourier transformation of $\tilde{G}_{L}(p)$ in Eq.~\eqref{eqn:green-finite} is easily achieved by using the FFT algorithm, while that of $\tilde{G}_{\infty}(p)$ in Eq.~\eqref{eqn:green-infinite} needs alternative approach for lower dimensions $D\leq 4$ as demonstrated in Refs.~\cite{Luscher:1995zz, Necco:2001xg, Shin:1997ib}, since the convergence of the numerical integration is quite slow. 
 
\begin{figure}[!t] 
\centering\includegraphics[width=\figonecolumn]{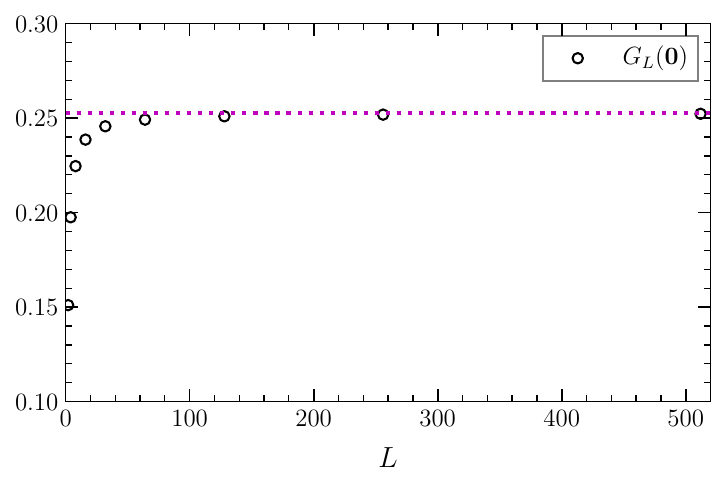} 
\caption{The behavior of $G_{L}(\mathbf{0})$ on the $L^{3}$ lattice, which is compared with $G_{\infty}(\mathbf{0})=0.25273100985866$ (dotted line).} 
\label{fig:g0const} 
\end{figure} 
 
\begin{figure}[!t] 
\centering 
\centering\includegraphics[width=\figonecolumn]{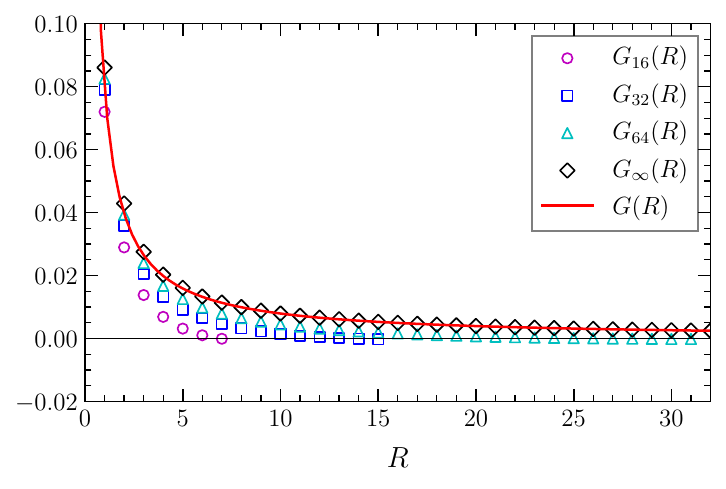} 
\caption{The behaviors of the lattice Green functions $G_{L}(R)$ and $G_{\infty}(R)$, which are compared with the continuum Green function $G(R)=1/(4\pi R)$.} 
\label{fig:gl-vdep} 
\end{figure} 
 
\par
\begin{figure}[!t] 
\centering\includegraphics[width=\figonecolumn]{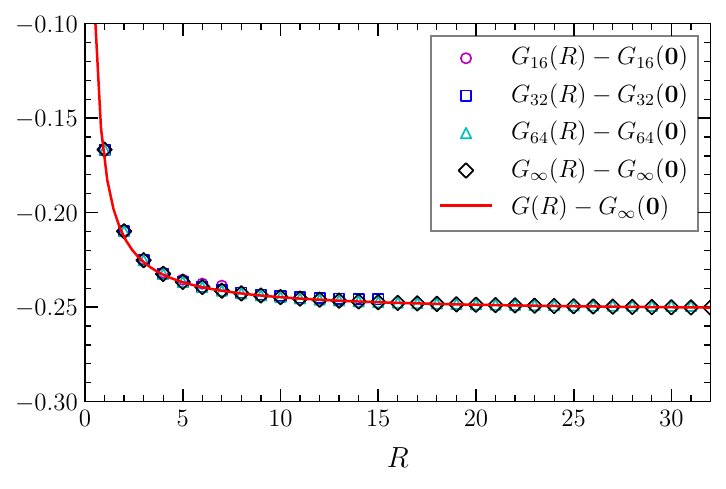} 
\caption{The behaviors of the normalized lattice Green functions, where the values at the origin are subtracted.} 
\label{fig:norm-gl} 
\centering\includegraphics[width=\figonecolumn]{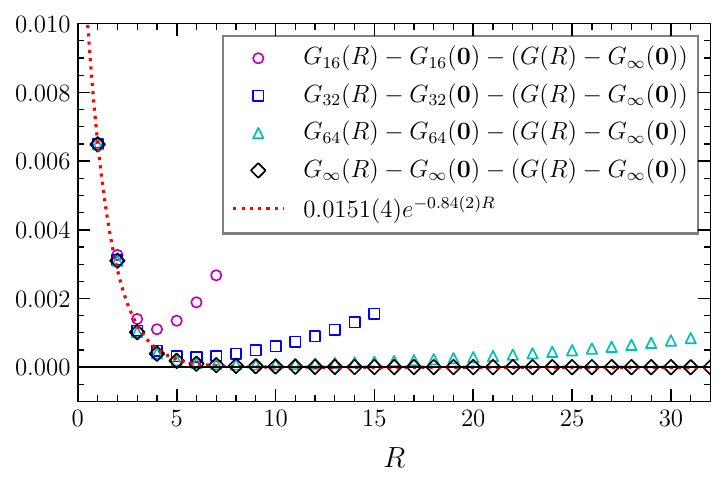} 
\caption{The difference between the normalized lattice Green functions and the normalized continuum Green function. The characteristic short-distance behavior is described by the exponential function (dotted curve). } 
\label{fig:norm-gl-diff} 
\end{figure}

 
We summarize the typical behavior of the lattice Green function in $D=3$ dimensions. In the continuum case, the Green function 
\be 
G(R)=\frac{1}{4\pi R} 
\label{eqn:green-continuum} 
\ee 
will diverge for $R\to 0$. In contrast, the lattice Green function at the origin is always finite. In Fig.~\ref{fig:g0const}, we plot the values of $G_{L}(\bvec{0})\equiv G_{L}(0,0,0)$ on the $L^{3}$ lattice as a function of~$L$, where the horizontal dotted line corresponds to $G_{\infty}(\bvec{0})\equiv G_{\infty}(0,0,0)$. We find that $G_{L}(\bvec{0})$ depends on the size $L$, which approaches $G_{\infty}(\bvec{0})$ from below as the size $L$ is increased. In Fig.~\ref{fig:gl-vdep}, we plot the on-axis lattice Green functions, $G_{L}(R)\equiv G_{L}(R,0,0)$ and $G_{\infty}(R)\equiv G_{\infty}(R,0,0)$, which are compared with the continuum Green function $G(R)$ in Eq.~\eqref{eqn:green-continuum}. $G_{\infty}(R)$ seems to be identical to $G(R)$ except for the distances near the origin, while $G_{L}(R)$ is clearly different from $G(R)$ and is dependent on the size $L$. Although $G_{L}(R)$ approaches $G(R)$ gradually with the increase of $L$, there still seems to be a constant gap even after taking the limit $L \to \infty$. As shown in Fig.~\ref{fig:norm-gl}, however, once the values at the origin are subtracted from the lattice Green functions, they can fall into one curve corresponding to $G(R) -G_{\infty}(\bvec{0})$. We look at the difference between them carefully in Fig.~\ref{fig:norm-gl-diff}, which clearly shows that the difference tends to disappear as the size $L$ is increased. Only at very short distances for $R \leq 5$, there is an inevitable difference. A chi-square fitting analysis for the infinite volume data $G_{\infty}(R) -G(R)$ indicates that the difference approximately obeys an exponential function $\delta G = 0.0151(3) e^{-0.84(2)R}$. Multiplying a factor $4\pi^{2}$ to the vertical axis of Fig.~\ref{fig:norm-gl-diff} leads to Fig.~\ref{fig:pot_coul_vdep} (lower) in Sec.~\ref{sec:results}. 
 
\section{The flux-tube solution in the large $\kappa$ limit} 
\label{sec:app-london} 

For a large $\kappa$, the monopole field prefers to be condensed everywhere $| \chi | =v$ to reduce the energy originating from the monopole self-interaction term. On the other hand, if this happens, the energy contribution from the monopole kinetic term, which contains interaction between the Dirac string and the monopole field via the singular part of the dual gauge field, requires large energy instead, which will diverge in the continuum limit. Thus, the monopole field should be zero at least just on the Dirac string, even in the type-II limit, $\kappa \to \infty$. 
 
\par 
On the dual lattice, however, such a potential ultraviolet divergence from the Dirac string is already regularized by admitting a finite size of $\Sigma_{\mu\nu}$, so that the monopole field can take nonzero values on all sites of the lattice. This feature then allows us to investigate the case of the type-II limit at any lattice spacing, which will help us to learn the numerical limitations of the dual lattice formulation. 
 
\par 
To set $| \chi | =v$ everywhere from the beginning, it is convenient to start from the polar coordinate representation of the monopole field $\chi (x)= \phi(x)e^{i\eta (x)}$ ($\phi,\;\eta \in \Re$), and write the (dimensionless) DGL action as 
\bea 
&&S = 
\beta_{g} 
\sum_{x} 
\Biggl [ 
\frac{1}{4} 
{}^{*\!}F_{\mu\nu} (x)^{2} 
\nn\\* 
&& +\hat{m}_{B}^2 
\sum_{\mu=1}^{D} 
\biggl ( \phi(x)^{2}- \phi (x) \phi (x+\hat{\mu}) \cos {B}_{\mu}(x) \biggr) 
\nn\\* 
&& +\frac{\hat{m}_{B}^2 \hat{m}_{\chi}^2}{8} \left ( \phi (x)^2 - 1 \right )^2 
\Biggr ], 
\eea 
where the dual gauge field is redefined as $B_{\mu} + \del_{\mu}\eta (x) \to B_{\mu}(x)$. Then, we set $\phi (x) =1$ everywhere and obtain the action in the type-II limit, 
\be 
S = 
\beta_{g} 
\sum_{x} 
\Biggl [ 
\frac{1}{4} 
{}^{*\!}F_{\mu\nu} (x)^{2} + \hat{m}_{B}^2 \sum_{\mu=1}^{D} (1 - \cos {B}_{\mu}(x) ) 
\Biggr ] \;. 
\label{eqn:dglaction-london} 
\ee 
The field equation for the dual gauge field is derived by 
\be 
\frac{\del S}{\del B_{\mu}(x) } = \beta_{g} X_{\mu}(x) =0 \;, 
\ee 
where 
\be 
X_{\mu}(x) = \sum_{\nu \ne\mu} [ F_{\mu \nu}(x)+ F_{\nu \mu}(x-\hat{\nu}) ] +\hat{m}_{B}^{2} \sin B_{\mu}(x) \; , 
\ee 
which is nothing but the lattice version of the London equation. To solve the field equation $X_{\mu}(x) =0$, the Newton-Raphson method is still applicable. The derivative of the field equation with respect to $B_{\mu}$ is further given by 
\be 
\delta_{B}X_{\mu} (x) \equiv 
\frac{\del X_{\mu}(x)}{\del B_{\mu}(x) } = 2(D-1)+ \hat{m}_{B}^{2} \cos B_{\mu}(x) \;, 
\ee 
and then, the update process can be performed by 
\be 
B_{\mu} (x) \to B_{\mu}^{\rm new} (x) = B_{\mu} (x) -\frac{X_{\mu}(x)}{\delta_{B}X_{\mu} (x) } \;. 
\ee 
 
\begin{figure}[!t] 
\centering\includegraphics[width=\figonecolumn]{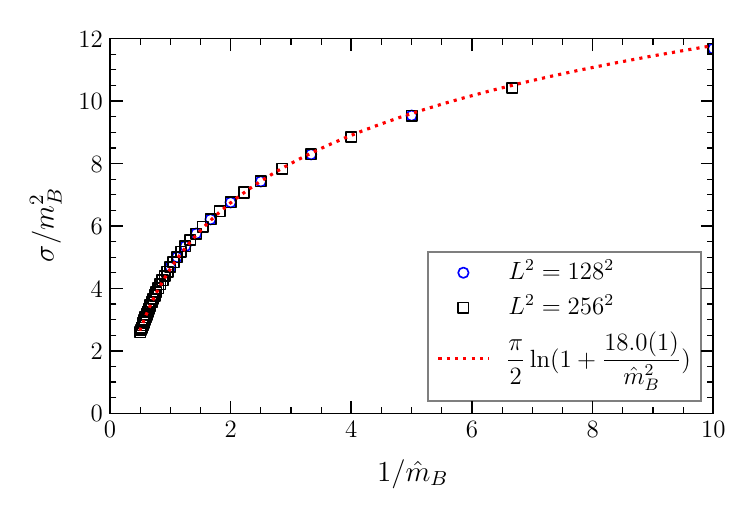} 
\caption{The string tension computed from Eq.~\eqref{eqn:dglaction-london} in $D=2$ as a function of $1/\hat{m}_{B}$ with the fitting function in Eq.~\eqref{eqn:london_sigma_fit}.} 
\label{fig:london_sigma} 
\end{figure} 
 
\begin{figure}[!t] 
\centering\includegraphics[width=\figonecolumn]{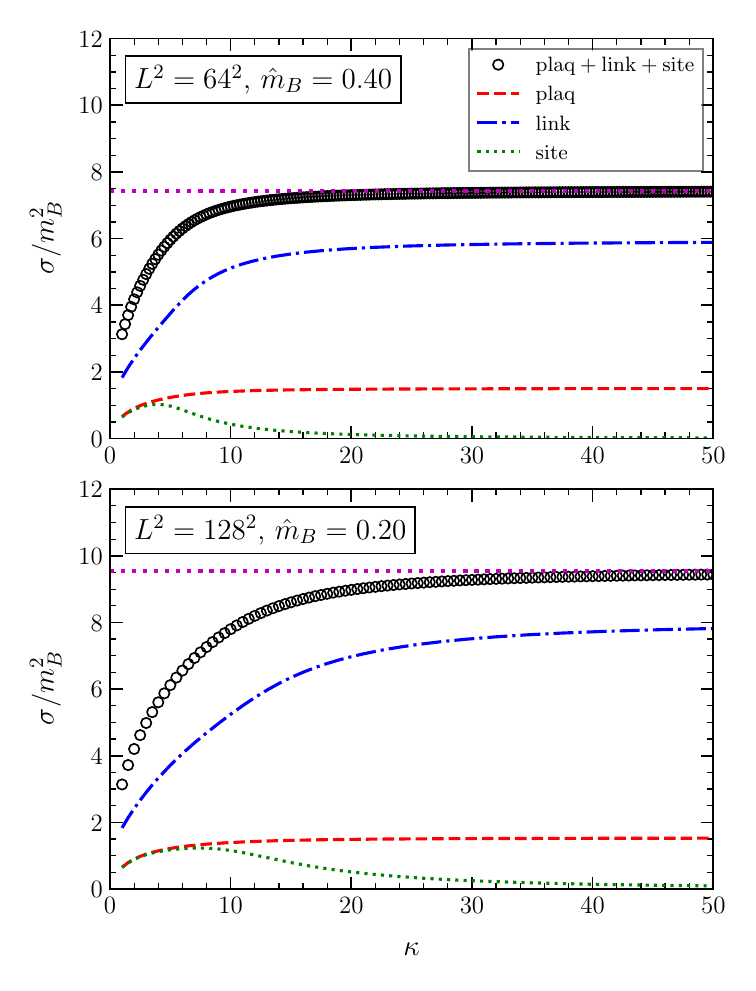} 
\caption{The $\kappa$-dependence of the string tension with the normal $D=2$ setting [Eq.~\eqref{eqn:action}] for $\hat{m}_{B}=0.40$ on the $L^{2}=64^{2}$ lattice (upper) and $\hat{m}_{B}=0.20$ on the $L^{2}=128^{2}$ lattice (lower). The labels, ``plaq,'' ``link,'' ``site'' are the separated contributions to the string tension as in Eqs.~\eqref{eqn:sole-p}-\eqref{eqn:sole-s}. The horizontal dotted lines correspond to the values in the type-II limit computation, $\sigma/m_{B}^{2} = 7.42778$ for $\hat{m}_{B}=0.40$ and $\sigma/m_{B}^{2} = 9.53481$ for $\hat{m}_{B}=0.20$.} 
\label{fig:london_sigma_kappadep} 
\end{figure} 
 
\par 
In the continuum theory, the ultraviolet cutoff of the momentum is usually assumed to be the mass of the monopole field $m_{\chi}$. The inverse of $m_{\chi}$ corresponds to the coherence length, which can be regarded as the thickness of the flux-tube core, where the magnitude of the monopole field changes drastically from the value of condensate $v$ (outer region) to zero (inside). The introduction of a sharp cutoff regularizes the string tension of the flux tube to be 
\be 
\sigma^{\mathrm{cont}}_{\mathrm{London}} = \frac{\beta_{g} \pi |N_{q}| m_{B}^{2}}{2} \ln (1+\frac{m_{\chi}^{2}}{m_{B}^{2}}) \;. 
\label{eqn:london_sigma} 
\ee 
 
\begin{figure}[!t] 
\centering\includegraphics[width=\figonecolumn]{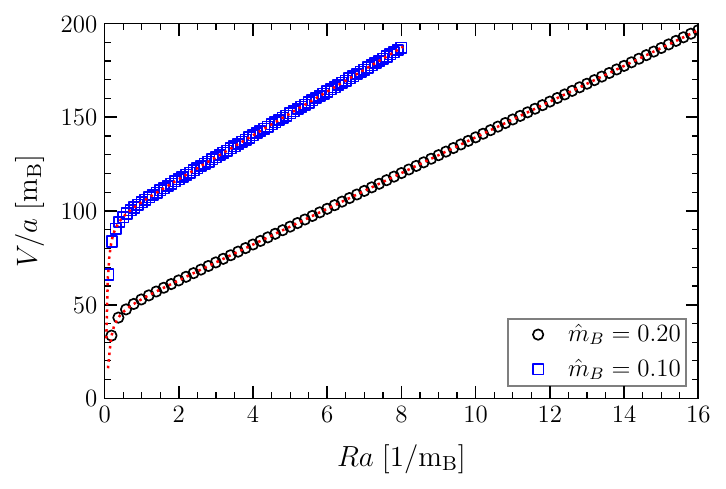} 
\caption{The interquark potential $V(r=Ra)$ in the type-II limit for $\hat{m}_{B}=0.20$ and $\hat{m}_{B}=0.10$ on the $L^{2}L_{3}=128^{2}256$ lattice, together with the fitting function $V_{yl}$ in Eq.~\eqref{eqn:fit-yl}. The fit parameters are summarized in Table~\ref{tbl:potfit-yl}.} 
\label{fig:london_pot} 
\end{figure} 
 
\par 
It is not clear how this expression will be modified in our lattice formulation, as the monopole mass $m_{\chi}$ is absent, and in addition, the lattice cutoff is not just the inverse of $m_{\chi}$. However, the numerical result presented in Fig.~\ref{fig:london_sigma} exhibits a quite similar behavior except for $m_{\chi}$, where the string tension computed from Eq.~\eqref{eqn:dglaction-london} in $D=2$ is plotted as a function of $1/\hat{m}_{B}$ on the $L^{2}=128^{2}$ and $256^{2}$ lattices (we have set $N_{q}=1$) with the one-parameter fitting function motivated by Eq.~\eqref{eqn:london_sigma}, 
\be 
f (\hat{m}_{B}) =\frac{\beta_{g} \pi |N_{q}| }{2} \ln (1+\frac{\gamma}{\hat{m}_{B}^{2}}) \;. 
\label{eqn:london_sigma_fit} 
\ee 
We find that this function excellently describes the behavior of the string tension. We may attempt to identify the fitting value of $\gamma = 18.0(1)$ as the square of the (dimensionless) monopole mass, such as $\hat{m}_{\chi}= \sqrt{18.0} \sim 4.24$. The GL parameter is then defined by $\kappa = 4.24/\hat{m}_{B}$. However, this correspondence slightly needs to be modified. In Fig.~\ref{fig:london_sigma_kappadep} we show the $\kappa$-dependence of the string tension with the normal $D=2$ setting for $\hat{m}_{B}=0.40$ on the $L^{2}=64^{2}$ lattice and $\hat{m}_{B}=0.20$ on the $L^{2}=128^{2}$ lattice and its breakdown as in Eqs.~\eqref{eqn:sole-p}-\eqref{eqn:sole-s}. In both cases, the site contribution due to the monopole self-interaction term starts to decrease beyond a certain value of $\kappa$, but the tail remains even at quite large $\kappa$. The relation, $\kappa = 4.24/\hat{m}_{B}$, leads to $\kappa = 10.6$ and $\kappa = 21.2$ for $\hat{m}_{B}=0.40$ and $\hat{m}_{B}=0.20$, respectively, but the value of the string tension has not yet reached the value in the type-II limit computation. To obtain the type-II limit result from the field equations in Eqs.~\eqref{eqn:feq-duallattice-b} and \eqref{eqn:feq-duallattice-chi}, the value of $\kappa$ at least should be twice as large as that estimated from the above fitting. 
 
\par 
It is possible to compute the interquark potential in the type-II limit with the $D=3$ setting. In Fig~\ref{fig:london_pot}, we show the result for $\hat{m}_{B}=0.20$ on the $L^{2}L_{3}=128^{2}256$ lattice, together with the fitting function $V_{yl}$ in Eq.~\eqref{eqn:fit-yl}. The linear slope is consistent with the string tension from the $D=2$ computation. An interesting observation may be that the effective mass in the Yukawa term is larger than one, $m_{y} =1.668(9)\,\mathrm{[m_{B}]}$. We note that the value of $m_{y}$ is the same even in the finer lattice with $\hat{m}_{B}=0.10$, while the slope of the linear term is described by Eq.~\eqref{eqn:london_sigma_fit}. This suggests that the Yukawa term in the type-II limit decays more quickly than previously thought.

\end{document}